%% file: main_arxiv.tex
\ifdefined\pdfobjcompresslevel
\fi
\PassOptionsToPackage{table,dvipsnames}{xcolor}
\documentclass{selfevolagent}

\usepackage{amsmath}
\usepackage{amssymb}
\usepackage{array}
\usepackage{makecell}
\usepackage{threeparttable}
\usepackage{subcaption}
\usepackage{pifont}
\usepackage{siunitx}
\usepackage{wrapfig}
\usepackage{xspace}     % so "\evoskillrec is" does not swallow the space
\usepackage{algorithm}
\usepackage{algorithmic}
\usepackage{tikz}
\usepackage{booktabs}
\usepackage{array}
\usepackage{xurl}
\usepackage{titletoc,placeins}

\usepackage{url}

\input{math_commands.tex}

\title{EvoSkillRec: Skill-Genome Evolution for\\
       Recommender Architecture Discovery}

\author[1,\dagger]{Xiaopeng Li}
\author[2]{Kuo Cai}
\author[2]{Bo Chen}
\author[1]{Wenlin Zhang}
\author[1]{Mengyang Ma}
\author[1]{Yingyi Zhang}
\author[1]{Zichuan Fu}
\author[1]{Yu Yang} 
\author[3]{Qidong Liu}
\author[2]{Yiyu Wang}
\author[2]{Ruiming Tang}
\author[2]{Wenwu Ou}
\author[3]{Jiang Wu}
\author[3]{Zhanbo Xu}
\author[1, *]{Xiangyu Zhao}

\affiliation[1]{City University of Hong Kong}
\affiliation[2]{Kuaishou Technology}
\affiliation[3]{Xi'an Jiaotong University}
\contribution[\dagger]{Work done during an internship at Kuaishou}
\contribution[*]{Corresponding author}

\newcommand{\evoskillrec}{\textsc{EvoSkillRec}\xspace}
\newcommand{\skill}{\mathcal{S}}
\newcommand{\library}{\mathcal{L}}
\newcommand{\model}{\mathcal{M}}
\newcommand{\eval}{\mathcal{E}}
\newcommand{\cmark}{\textcolor{ForestGreen}{\ding{51}}}
\newcommand{\xmark}{\textcolor{BrickRed}{\ding{55}}}
\DeclareUrlCommand\skillid{\urlstyle{tt}}

\abstract{

Modern recommender systems advance not only by scaling data and parameters, but also by encoding task-specific inductive biases through architecture, including sparse feature interactions for click-through rate (CTR) prediction, temporal attention for sequential recommendation, and expert routing for multi-task learning. However, these biases are typically human expert designed or searched within predefined operator spaces. Although Recent LLM-driven code evolution expands this space, unconstrained edits often produce invalid or ineffective architectures, underuse established architecture design knowledge, and fail to preserve successful innovations for reuse. We introduce \evoskillrec, a promotion-and-reuse framework for cumulative recommender architecture evolution. It first decomposes recommenders into atomic executable skills and represents architectures as typed skill genomes, with each skill equipped with input--output types, semantic annotations, and implementation code. We then evolve models with different tasks through two coupled spaces: a constrained skill--space that mutates, recombines, specializes, and reuses validated skills, and an open-ended code--space in which LLM planners and synthesizers invent new skill modules using prior evolution traces and accumulated experience. An autoresearch controller evaluates candidates, diagnoses failures, retrieves relevant skills, promotes validated innovations into the skill library, and adaptively allocates the proposal budget between the two spaces. Extensive experiments on CTR prediction, multi-task learning, and multi-domain learning, including resource-constrained co-optimization of predictive quality and model FLOPs utilization in generative ranking models, consistently demonstrate the effectiveness of our proposed \evoskillrec.\\
\\
\textbf{Code}: \url{https://github.com/Xiaopengli1/EvoSkill-Rec} \\
\textbf{Contact}: xiaopli2-c@my.cityu.edu.hk
}

\begin{document}
\maketitle

\input{1Intro}
\input{2RelatedWork}
\input{3Problem}
\input{4Method}

\input{5Experiment}
\input{6Conclusion}

\bibliography{references}
\bibliographystyle{plainnat}

\input{7Appendix}

\end{document}

%% file: math_commands.tex
\usepackage{amsmath,amsfonts,bm}

\def\1{\bm{1}}

\DeclareMathAlphabet{\mathsfit}{\encodingdefault}{\sfdefault}{m}{sl}
\SetMathAlphabet{\mathsfit}{bold}{\encodingdefault}{\sfdefault}{bx}{n}

\DeclareMathOperator*{\argmax}{arg\,max}

%% file: 1Intro.tex
\section{Introduction}
\label{sec:intro}

Recommender systems are large-scale prediction engines that estimate user preferences from
historical interactions and side information. Their backbones have changed substantially
alongside the platforms they serve, moving from content-based filtering to deep ranking
models and, more recently, to generative ranking models. 
Across every stage, however, many of the most consequential advances have stemmed from a narrower and more identifiable source:
\emph{\textbf{architectural inductive bias}}. In click-through-rate (CTR) prediction, the field
converged on the view that sparse categorical fields should interact in structured,
often high-order ways, motivating explicit factorization and feature-crossing modules such
as DeepFM \citep{guo2017deepfm} and DCN \citep{wang2017dcn}. In target-aware ranking, a consensus emerged that different candidate items should activate different parts
of a user's history, first realized through attention-based interest extraction in DIN
\citep{zhou2018din} and later extended by DIEN \citep{zhou2019dien}. In multi-task
recommendation, the shared insight is that tasks should share representations without
collapsing task-specific signal, as instantiated in MMoE \citep{ma2018mmoe} and PLE
\citep{tang2020ple}. These examples support a unifying view: a recommender is not merely a
stack of neural layers, but a composition of reusable inductive-bias units.

This view exposes a bottleneck. The first generation of strong recommender
architectures, such as Wide \& Deep \citep{cheng2016wide}, DeepFM \citep{guo2017deepfm}, DCN
\citep{wang2017dcn}, SASRec \citep{kang2018sasrec}, etc., demonstrates the value of expert-designed inductive bias, but the
design cycle behind each system is slow and largely manual. Moving to a new task, feature
distribution, or business scenario typically requires many rounds of human trial and error,
and the knowledge embedded in one model family is rarely expressed in a form another task
can reuse directly. 

Recommender AutoML and neural architecture search (NAS) automate part of this burden by searching over operators, connectivity patterns, or loss functions rather than hand-designing them. Automated neural interaction search discovers feature-interaction architectures for CTR prediction \citep{song2020autonid}; NASRec trains weight-sharing supernets over recommender architectures \citep{zhang2022nasrec}; AutoLossGen searches loss-function programs for recommendation objectives \citep{li2022autolossgen}; and recent surveys call for broader automated model design across the recommendation pipeline \citep{zhang2024automated}. These approaches are valuable but inherit a structural limitation: the search space itself is still specified by humans. A NAS system can efficiently select among candidate operators, but it rarely invents a new module, objective, or routing mechanism that lies outside the grammar it was given.

LLM-driven program evolution offers a possible escape from this fixed-space restriction. Program-search systems such as FunSearch \citep{romera2024funsearch} and AlphaEvolve \citep{novikov2025alphaevolve} use language models to propose code within a broad, largely unconstrained program space, relying on evaluators to select useful variants. Concurrent work has begun to apply this idea to recommender systems: Self-EvolveRec uses LLM-based directional feedback to iteratively edit recommender code \citep{kim2026selfevolverec}, and a related system for large-scale video recommendation coordinates offline and online LLM agents for autonomous optimization \citep{wang2026selfevolving}. These works support an important finding: model design can become an iterative, evaluator-driven research process, with LLMs acting as the source of proposals.

Direct LLM code evolution, however, is not yet sufficient for recommender architecture discovery, for
two reasons. First, open-ended code edits, unconstrained by structural priors, tend to produce a large proportion of invalid or low-value modifications when the accumulated inductive-bias knowledge of the field is ignored during search. Second, even when an edit succeeds, its useful
idea remains trapped inside a one-off code diff: the next generation starts again from a
codebase and a feedback transcript rather than from accumulated, validated recommender knowledge. A self-evolving recommender should not merely mutate; it should remember. 

We propose \textbf{\evoskillrec}, a framework for skill-genome evolution in recommender architecture design and discovery. First, we introduce the idea of treating \emph{\textbf{recommender modules as skills}}: \evoskillrec decomposes existing recommender models into atomic, executable modular skills, annotates each with semantic and inductive-bias knowledge, validates its functional correctness and reusability, and stores it in a library together with its provenance, empirical performance history, and implementation. Second, we design a \emph{\textbf{dual-space evolution process that jointly supports knowledge reuse and open-ended discovery}} for architecture search. In the skill space, \evoskillrec performs reliable genetic operations over known skills --- add, replace, hybridize, and specialize --- that make full use of accumulated architectural knowledge. In the code space, \evoskillrec permits open-ended LLM invention when the existing skill library is hard to use to address a diagnosed failure, producing a new module, fusion design, auxiliary objective, or routing mechanism grounded in the evolutionary trace. Successful inventions are not discarded after a single run: they are compiled back into skill cards and become part of the evolving genome, available for reuse in subsequent generations. Our contributions are as follows.
\begin{itemize}
    \item We formulate recommender self-evolution as \emph{skill-genome accumulation}: a
    model-to-skill compiler decomposes recommender code into validated, reusable
    architectural units, curated into a growing library rather than discarded after use.
    \item We design a dual-space evolution algorithm that couples type-checked mutation
    over the skill library with open-ended code-space invention, closed by an
    AutoResearch loop that distills successful inventions back into the library.
    \item We show that \evoskillrec\ outperforms strong hand-designed and automated-design
    baselines across recommendation benchmarks, and that skill-space and code-space
    evolution contribute complementary, task-dependent gains.
\end{itemize}

%% file: 2RelatedWork.tex
\section{Related Work}
\label{sec:related}

\paragraph{AutoML and NAS for recommendation.}
Automated recommender design searches over architectures, feature interactions, embeddings,
or objectives, inheriting techniques from general Neural architecture search (NAS) \citep{zoph2017nas,real2019amoebanet}.
Automated neural interaction discovery explores CTR interaction operators
\citep{song2020autonid}; NASRec amortizes recommender architecture search with weight
sharing \citep{zhang2022nasrec}; AutoLossGen generates loss functions
\citep{li2022autolossgen}; and broader automated model-design work considers co-design
across architecture and system constraints \citep{zhang2024automated}. \evoskillrec\ differs
in the \emph{object being evolved}. Rather than searching only within a predeclared operator
set, it maintains a growing library of validated skills and invents new ones when the
library proves insufficient---so the search space is an output of the procedure, not an
input to it.

\paragraph{LLM-driven program evolution.}
LLM program search has shown that code can be evolved under external evaluators. Evolution
through large models established LLMs as mutation operators over programs
\citep{lehman2023elm}; FunSearch evolves programs for mathematical discovery
\citep{romera2024funsearch}; and AlphaEvolve extends the idea into a general coding agent
for algorithmic and systems discovery \citep{novikov2025alphaevolve}. Recommender
self-evolution is a parallel emerging direction: Self-EvolveRec introduces directional
feedback for LLM-based recommender evolution \citep{kim2026selfevolverec}, and
\citet{wang2026selfevolving} describe an offline/online agent system for autonomous model
optimization in large-scale video recommendation. These systems treat each run as
self-contained: the population and the transcript carry state, but nothing durable is
extracted. \evoskillrec\ is complementary and targets exactly that gap---how an evolving
system \emph{stores, validates, and reuses} the architectural knowledge such loops produce.

%% file: 3Problem.tex
\section{Problem Setting}
\label{sec:problem}
Let a recommender training instance be $(\mathcal{D}, \mathcal{T}, \mathcal{C})$, where
$\mathcal{D}$ is the dataset, $\mathcal{T}$ specifies the task and its metrics (e.g., CTR
prediction, multi-task ranking, multi-domain ranking, or FLOPs-constrained generative
ranking), and $\mathcal{C}$ encodes deployment or research constraints such as
latency, parameter budget, or permitted feature types. A model $\model$ comprises code,
parameters, data-processing assumptions, and training logic. Conventional AutoML and NAS
solve
\begin{equation}
    \model^\star = \argmax_{\model \in \Omega}\; \eval(\model; \mathcal{D}, \mathcal{T}, \mathcal{C}),
    \label{eq:automl}
\end{equation}
where human designers fix the search space $\Omega$ in advance. \evoskillrec\
instead maintains an evolving skill library $\library_t$ at generation $t$ and
searches jointly over compositions of existing skills and newly synthesized code:
\begin{equation}
    \model_{t+1} \in \mathrm{Compose}(\library_t) \;\cup\; \mathrm{Invent}(\library_t, r_t),
    \label{eq:evoskillrec}
\end{equation}
where $r_t$ diagnoses the current failure mode (Section~\ref{sec:loop}). Whenever an
LLM-coded candidate passes validation and improves $\eval$, its reusable components are
written back into $\library_{t+1}$. Unlike Eq.~\ref{eq:automl}, the objective of
\evoskillrec\ is thus twofold: find a strong $\model^\star$ for the current task, and
improve the search distribution for future tasks by accumulating validated architectural
knowledge. Writing $\Omega_t$ for the space reachable at generation $t$, conventional
search holds $\Omega_t = \Omega$ for all $t$, whereas here $\Omega_t \subseteq \Omega_{t+1}$,
with strict inclusion exactly when a code-space invention is promoted. We call this
monotone growth of the search space skill-genome accumulation, a property that a single
run of program search lacks. Note that $\eval$ need not be a single metric:
Section~\ref{sec:exp_rq4} instantiates it as a compound objective over predictive quality
and hardware efficiency, deeming a candidate eligible only if its training MFU stays
at least $95\%$ of the seed backbone's and its validation AUC stays within $10^{-3}$ of the
incumbent, with eligibility enforced at selection time rather than folded into a scalarized
reward.

%% file: 4Method.tex
\section{\evoskillrec: Skill-Genome Evolution}
\label{sec:method}

\begin{figure}[t]
    \centering
    \includegraphics[width=\linewidth]{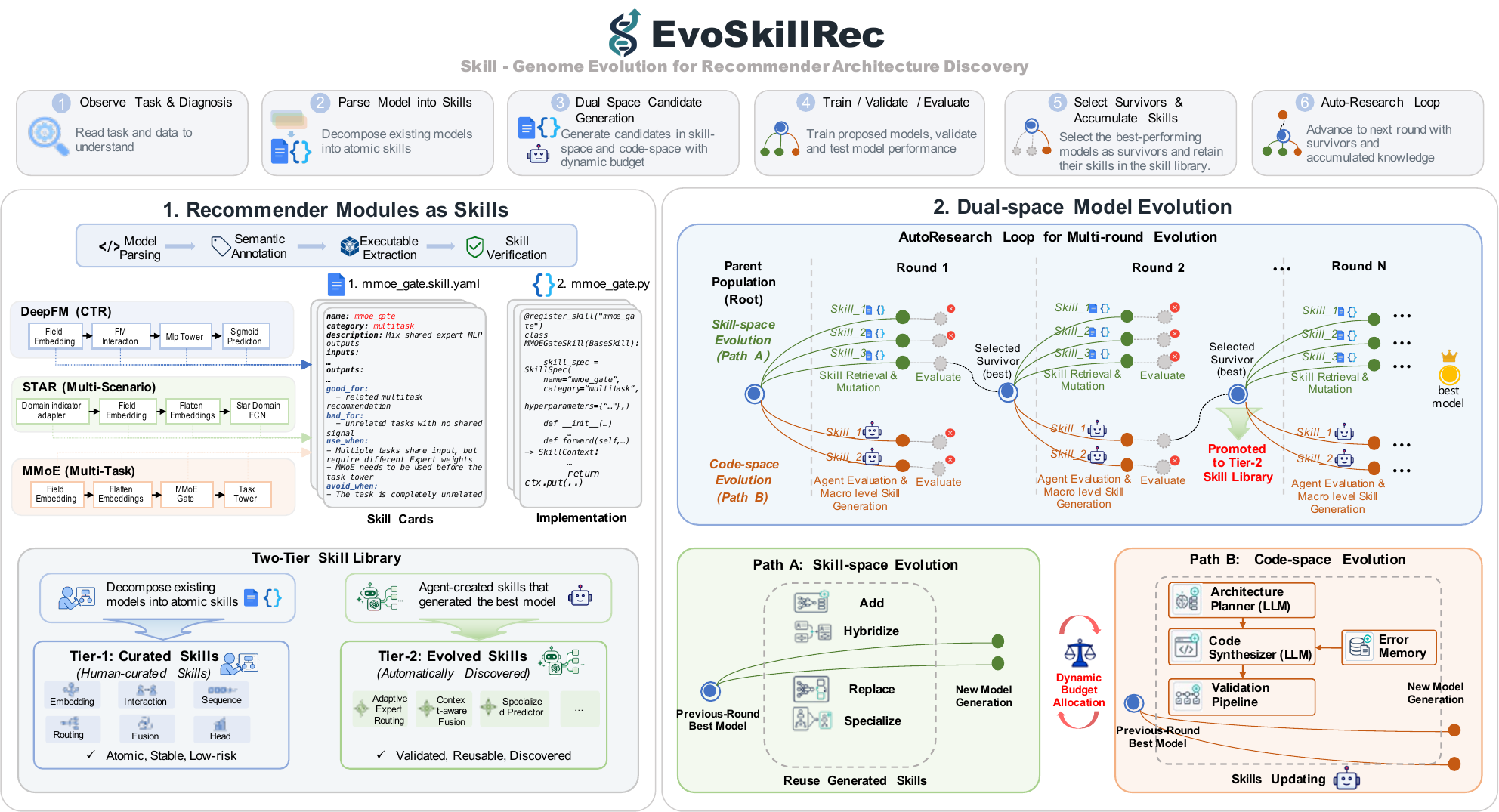}
    \caption{Overview of \evoskillrec. \textbf{(Left)} A model-to-skill compiler decomposes
    established recommender architectures into atomic, executable skills, each documented
    by a skill card with code and admitted to the curated Tier-1 library only after
    passing validation. \textbf{(Right)} A model is represented as a typed skill genome.
    Under an adaptive budget, dual-space evolution proposes candidate models via
    type-checked recombination of library skills (skill space) or LLM-based planning and
    synthesis of new modules (code space). The AutoResearch loop evaluates and diagnoses
    candidates, selects survivors, and promotes validated inventions into the Tier-2
    library, making them available to skill-space search in later generations.}
    \label{fig:overview}
\end{figure}

We now describe \evoskillrec\ in detail, and Figure~\ref{fig:overview} gives the overall picture.
Section~\ref{sec:skills} introduces the skill card abstraction together with the
model-to-skill compiler and two-tier library that populate and grow the skill genome.
Section~\ref{sec:dualspace} presents the dual-space evolution process that couples
constrained skill-space recombination with open-ended code-space invention, and the adaptive
budget that arbitrates between them. Section~\ref{sec:loop} closes the loop with the
AutoResearch controller that diagnoses failures, drives both spaces, and
consolidates validated discoveries into evolutionary memory.

\subsection{Recommender Modules as Skills}
\label{sec:skills}

Recent \emph{agent skills} frameworks package an LLM agent's capabilities as well-specified, directly invocable units that can be retrieved and composed on demand, rather than re-derived from scratch in every episode. \evoskillrec\ brings the same principle to recommender architecture design: it re-represents each established architecture as a set of atomic, executable \emph{skills}, turning the inductive-bias knowledge otherwise buried in a monolithic, task-specific implementation into explicit, validated, reusable building blocks that an evolutionary search can retrieve, recombine, and extend.

\subsubsection{The Skill Card Abstraction}
\label{sec:skillcard}

The atomic unit of \evoskillrec\ is a \emph{skill}: an executable module paired with
structured metadata,
\begin{equation}
    \skill = (c, I, O, b, a, f, v, m),
    \label{eq:skilltuple}
\end{equation}
where $c$ is the implementation code, a PyTorch module that reads from and writes to a
shared tensor context $\mathcal{X}$, a blackboard-style key--value store of named
tensors (e.g., \texttt{field\_embeddings}, \texttt{fm\_output}); $I$ and $O$ are the input
and output keys $c$ requires and produces, so two skills are composable whenever a
producer's output keys satisfy a consumer's required keys, regardless of where either sits
in the architecture; $b$ is an inductive-bias description; $a$ specifies applicable task
types and usage conditions; $f$ lists failure signatures the skill addresses or may
trigger; $v$ is a validation suite; and $m$ is accumulated empirical memory (historical
performance, mutation lineage, and reuse statistics). A skill is not a
natural-language prompt: it must be directly executable and testable, so its behavior is
verified before it enters any candidate architecture.
 
Each skill is externally described by a \textbf{skill card}, a structured record that
operationalizes Eq.~\ref{eq:skilltuple} for retrieval and validation: $I$ and $O$ become
typed tensor specifications; $b$ and $a$ become inductive-bias tags, task types, and
\texttt{good\_for}/\texttt{bad\_for} usage notes; $f$ becomes machine-checkable failure
signatures (e.g., an unexpected input rank); $v$ becomes a suite of shape, unit, and
reconstruction tests; and $m$ becomes composition statistics over compatible upstream and
downstream skills, together with a natural-language \emph{retrieval} block used for
relevance scoring (Section~\ref{sec:skillspace}). The abbreviated example below shows the
schema for
\texttt{fm\_interaction}, a second-order factorization-machine branch extracted from DeepFM
\citep{guo2017deepfm}:
\begin{quote}\small
\texttt{name}: fm\_interaction \quad \texttt{category}: interaction \\
\texttt{inputs}: field\_embeddings~$\in\mathbb{R}^{B\times F\times D}$ \quad
\texttt{outputs}: fm\_output~$\in\mathbb{R}^{B\times 1}$ \\
\texttt{inductive\_bias}: second-order feature interactions \quad
\texttt{failure\_signatures}: input rank~$\neq 3$ \\
\texttt{good\_for}: low-cost pairwise categorical interactions \quad
\texttt{bad\_for}: high-order interactions without an additional tower \\
\texttt{compatible\_with}: upstream $=$ \{field\_embedding\}, downstream $=$
\{concat\_fusion, binary\_ctr\_head\} \\
\texttt{retrieval.use\_when}: ``unflattened field embeddings are available and pairwise
interaction signal is needed without enumerating all field pairs.''
\end{quote}
Every card is paired with a code implementation that specifies, at the implementation
level, exactly how the skill turns its declared inputs into its declared outputs. This is
not a bare excerpt of the original open-source code: it inherits from \texttt{BaseSkill},
carries all the information declared in its card, and is thereby guaranteed to be reusable
and to compose into a genome without special-casing. The snippet below is the
implementation backing the \texttt{fm\_interaction} card above, where the shape assertion
enforces exactly the input contract the card declares:
\begin{quote}\small
\texttt{class FMInteractionSkill(BaseSkill):} \\
\hspace*{1.2em}\texttt{def \_\_init\_\_(self):} \\
\hspace*{2.4em}\texttt{...} \\
\hspace*{1.2em}\texttt{def forward(self, ctx: SkillContext) -> SkillContext:} \\
\hspace*{2.4em}\texttt{x = ctx.get\_required("field\_embeddings")} \\
\hspace*{2.4em}\texttt{assert x.dim() == 3\quad\# shape contract from the skill card} \\
\hspace*{2.4em}\texttt{return ctx.put("fm\_output", self.fm(x))}
\end{quote}
The full schema and worked examples across the interaction, sequence, and fusion categories
are given in Appendix~\ref{app:skill-cards}.

\subsubsection{Model-to-Skill Compiler}
\label{sec:compiler}

To bootstrap the library, \evoskillrec\ compiles a set of established open-source recommender
architectures, including DeepFM \citep{guo2017deepfm}, Wide \& Deep \citep{cheng2016wide}, DCN
\citep{wang2017dcn}, DIN \citep{zhou2018din}, DIEN \citep{zhou2019dien}, SASRec
\citep{kang2018sasrec}, MMoE \citep{ma2018mmoe}, and PLE \citep{tang2020ple}, among
other, into reusable skills through four stages.

\textbf{Parsing.} The compiler extracts candidate modules from open-source model code\footnote{We use the open-source implementations from Torch-RecHub \citep{torch_rechub} and UniRank \citep{li2026unirank}.}---embedding tables, field encoders, feature-crossing blocks, sequence encoders, attention aggregators, fusion layers, towers, gates, experts, losses, and samplers---combining static code analysis with LLM-assisted semantic grouping. A module is accepted as a candidate only if its inputs, outputs, parameters, and call sites can be cleanly isolated from the rest of the model.

\textbf{Semantic annotation.} An LLM writes the skill card for each candidate: function,
inductive bias, applicable tasks, failure modes, assumptions, and nearest known model family.
% Annotation is constrained by code evidence---if the source does not support an assumption, the
% card marks it \emph{unknown} rather than hallucinating semantics.

\textbf{Executable extraction.} The candidate's code is rewritten into a concrete, executable Python file structured as a typed \texttt{BaseSkill} module, exposing exactly the inputs, outputs, and core execution logic identified during parsing. This rewriting is what makes the result a minimal, independent execution unit and can instead be executed, tested, and reused on its own.

\textbf{Validation.} Every skill must pass validation before it is finally produced. General tests cover importability, deterministic construction, shape preservation, differentiability, and serialization, ensuring practical usability. The strongest test is reconstruction: \evoskillrec\ rebuilds each seed model from its extracted skills and verifies that end-to-end behavior matches the original implementation. When exact equivalence cannot be established, the skill construction is flagged and the extraction process is re-evaluated.

\subsubsection{Two-Tier Skill Library}
\label{sec:library}

Skill extraction (Section~\ref{sec:compiler}) only deposits knowledge that human designers
have already discovered and fixed into a published architecture---a static distillation of
prior inductive bias. Evolution itself, however, is an ongoing source of knowledge: as
\evoskillrec\ runs, LLM-driven synthesis in the code space (Section~\ref{sec:codespace}) can
turn up modules no seed architecture contained, and this dynamically discovered knowledge
likewise needs to be deposited and made reusable, or it is lost the moment the run ends.
The library $\library$ is accordingly organized into two tiers, one for each kind of
knowledge. \textbf{Tier-1} holds the curated skills
produced by the compiler of Section~\ref{sec:compiler}; it is static within a run and serves
as a reliable, well-understood foundation. \textbf{Tier-2} holds skills that originate from
code-space invention (Section~\ref{sec:codespace}) during evolution itself. A code-space
proposal is promoted into Tier-2 only if it satisfies the same extraction and
validation pipeline as any Tier-1 skill and demonstrably improves $\eval$ when
substituted into a genome: validated invention becomes evolutionary memory, and unvalidated
invention is discarded. This promotion criterion is what converts a one-off code diff into
durable, reusable architectural knowledge, addressing the trapped-edit failure mode of
unconstrained code evolution.
 
\subsection{Dual-Space Evolution}
\label{sec:dualspace}

Given the library of Section~\ref{sec:skills}, \evoskillrec\ generates candidate genomes
through two coupled search spaces---a \emph{skill space} that performs constrained,
safety-checked recombination of validated skills, and a
\emph{code space} in which an LLM invents new modules when the library cannot address a
diagnosed failure---with a per-generation budget arbitrating
how many candidates are drawn from each. The split exists
because the two spaces trade off exploitation against exploration: skill space reuses
accumulated inductive bias reliably but cannot reach a mechanism with no precedent in
$\library$, whereas code space can reach anywhere but at the cost of the invalid,
low-value edits that plague unconstrained code evolution; the
adaptive budget is what lets \evoskillrec\ shift toward whichever space the current
generation needs.

\subsubsection{Skill-Space Evolution}
\label{sec:skillspace}
Skill-space evolution treats the genome much like a genetic algorithm treats a chromosome:
at each generation, \evoskillrec\ retrieves the skills in $\library$ most likely to benefit
the current model given its task, scenario, and diagnosed failure, and composes them into
the genome through a fixed set of genetic operators---adding, replacing, hybridizing, or
specializing a branch. What makes this composition powerful is that the retrieved skills
need not all be human-designed: they may come from Tier-1, encoding an expert's inductive
bias, or from Tier-2, encoding a solution the system itself discovered and validated
earlier. Skill space is therefore not just a mutation mechanism but the channel
through which both kinds of accumulated knowledge---human-designed and self-discovered---
compound into continual improvement.

Concretely, \evoskillrec\ represents a candidate model as a genome graph $G=(V,E)$: each
node $v \in V$ is a skill instance, and each edge $e \in E$ is a tensor-context connection
that wires one skill's output into another's input, so $G$ is exactly the composition of
skills that defines the model. Skill-space evolution operationalizes mutation as four
operator categories, each a DAG-level transformation of this graph:
\begin{itemize}
    \item \textbf{Add}: insert a new skill node into the DAG, extending the pipeline with an additional skill that can make the model more effective, without modifying any existing skill.
    \item \textbf{Replace}: substitute a skill node with an interface-compatible skill, preserving its output key so downstream wiring is untouched.
    \item \textbf{Hybridize}: combine branches drawn from different architecture families
    within one candidate, e.g., pairing a DeepFM-style interaction branch with a DIN-style
    interest branch.
    \item \textbf{Specialize}: adapt an existing path to the target task, scenario, or
    capacity budget---narrowing or widening a branch, or adjusting its hyperparameters---
    without changing its \texttt{skill\_id} or wiring.
\end{itemize}
Both \emph{add} and \emph{replace} may draw their source skill from either tier
(Section~\ref{sec:library}), through a retrieval process that selects, from all of
$\library$, the skills most likely to help the current genome.\footnote{This retrieval is
rule-based rather than embedding-based: there is no vector index (e.g., FAISS, Annoy,
Milvus) over skill representations, only scoring over skill-card metadata, string matching
against the diagnosis, and tensor-key compatibility.} Retrieval proceeds in two stages.
A hard filter first discards any skill that is incompatible with the current task, whose
required inputs the parent genome cannot supply, that has no loadable implementation, that
belongs to a category excluded from direct search (e.g., losses, objectives, terminal
heads), or that would duplicate a skill already present in the genome. 
 
The skills that pass this filter are then ranked by a multi-faceted scoring function. For
skill $\skill_i$, parent genome $G$ with available tensor keys $\mathcal{X}_G$, and
diagnosis $r_t$, the score is
\begin{equation}
    \mathrm{score}(\skill_i \mid G, r_t) = w_1 \phi_{\text{task}}(\skill_i, \mathcal{T})
    + w_2 \phi_{\text{io}}(\skill_i, \mathcal{X}_G)
    + w_3 \phi_{\text{ret}}(\skill_i, r_t)
    + w_4 \phi_{\text{bias}}(\skill_i, r_t)
    + w_5 \bar{m}(\skill_i),
    \label{eq:relevance}
\end{equation}
where $\phi_{\text{task}}$ checks task-type compatibility, $\phi_{\text{io}}$ checks whether
$\skill_i$'s required input keys are present in $\mathcal{X}_G$ or producible by a one-hop
upstream skill, $\phi_{\text{ret}}$ measures textual similarity between $\skill_i$'s
retrieval description and the diagnosis, $\phi_{\text{bias}}$ rewards skills whose inductive
bias is known to counter the failure signature in $r_t$, and $\bar{m}(\skill_i)$ is
$\skill_i$'s historical average validated improvement. The top-$n$ scoring skills are
retrieved as the pool from which the genome is evolved, subject to a per-category cap that
keeps the pool from collapsing onto a single skill category.
 
\subsubsection{Code-Space Evolution}
\label{sec:codespace}
Skill space can only exploit: it recombines skills already in $\library$, so its reach is
bounded by what the library contains. Escaping that bound requires searching an open space,
in which a module is written rather than selected. Such a search is expensive---every
proposal costs an LLM call and a full training run---so what it discovers should be kept
rather than discarded once the run ends. Code space therefore comprises a disciplined
invention pipeline and a sedimentation mechanism that deposits validated inventions as
reusable evolutionary memory, the Tier-2 skills of Section~\ref{sec:library}.
 
\textbf{Invention pipeline.} Code space runs in every generation rather than only when
skill-space retrieval comes up empty: it always holds a share of the candidate budget, set
adaptively (Section~\ref{sec:budget}) and smallest in early generations, where recombining
library skills still yields most of the available gain. What the LLM receives is not the
training data but a structured \emph{design brief}
\begin{equation}
    \mathcal{B}_t = \bigl(G_t,\; r_t,\; \mathcal{M}_{t-\Delta:t},\;
    \mathcal{X}_{G_t},\; \mathcal{C}\bigr),
    \label{eq:designbrief}
\end{equation}
combining the parent genome $G_t$ with its nodes, edges, and per-node interfaces; the
diagnosis $r_t$, which names the observed failure modes and the node they are attributed
to; the most recent $\Delta$ entries of evolutionary memory $\mathcal{M}$; the tensor keys $\mathcal{X}_{G_t}$ the genome currently exposes;
and the constraints $\mathcal{C}$. Memory is read in both directions: recently promoted
skills show which structural directions are working, while proposals that failed validation
or trained poorly are supplied as negative examples, so the same dead end is not explored
twice. The constraints are also concrete: a proposal must be a macro-level change, wired
in one of four ways (replacing a node, inserting between two nodes, branching into the
fusion layer, or replacing that layer), and may read only keys already present in
$\mathcal{X}_{G_t}$, so a synthesized module consumes intermediate tensors produced upstream
in the genome rather than raw features. Generation then proceeds in two stages:
\begin{enumerate}
    \item \emph{Architecture planner}: one LLM call turns $\mathcal{B}_t$ into a set of
    architectural sketches, which describe structure rather than code. A sketch
    states the mechanism the model is missing, the innovation lane it belongs to
    (interaction, routing, fusion, embedding, or sequence), the failure mode it targets,
    the nodes it touches, the input and output keys it needs, its wiring, and its
    parameter budget.
    \item \emph{Code synthesizer}: a second LLM call turns each selected sketch into a
    proposal carrying an executable \texttt{nn.Module}, its skill identifier, and its
    declared input and output. 
\end{enumerate}
Both steps are preceded by screening, so that neither the synthesis call nor the training
run is spent on a proposal that cannot work. A sketch is discarded if $G_t$ cannot supply
its inputs, if its wiring is not one of the permitted macro-level transformations, or if its structure
duplicates one already tried. Synthesized code is checked next: static analysis rejects
non-importable code and any file, network, subprocess, or dynamic-evaluation capability;
the declared skill-card contract must match the realized
interface and tensor shapes; the code must realize the macro-level change its sketch
declared; and the module must complete forward and backward passes with valid gradients on
synthetic inputs. These checks establish that a candidate is safe, executable, and
structurally compatible, not that it will improve $\eval$, which only training can decide. A
final portability test, whether the module stays executable as the number and type of input
fields change. Candidates clearing these checks are wired into the genome, compiled, trained, and evaluated.
 
\textbf{Sedimentation into Tier-2.} Section~\ref{sec:library} introduced Tier-2 as the half
of $\library$ that accumulates knowledge acquired during evolution; we now describe how a
skill reaches it. Clearing the checks above establishes that a proposal runs, not that it
helps, so the validated skill is trained and evaluated inside its host genome alongside the
generation's other candidates, and promotion is decided by how that genome performs. A
candidate enters the survivor set $\Sigma_t$ if it stays within a tolerance $\delta$ of the
generation mean on the primary metric and ranks in the top-$K$ after sorting, and only
generated skills carried by a survivor are promoted. What is sedimented is therefore what
measurably helped the strongest models of its generation, under the task, data, and training
budget at hand.
 
A promoted skill card records not only the skill itself but the evidence that earned it a
place in $\library$: the candidate metrics and status that triggered promotion, the
architecture details, and the originating proposal and parent node, among others. Promotion
also registers the skill in the index the genome compiler reads, so from the next generation
onward a Tier-2 skill is retrieved by skill-space search exactly like a Tier-1 skill, under
the same scoring and filters as in Section~\ref{sec:skillspace}, and no longer consumes
code-space budget.

\subsubsection{Adaptive Budget Allocation}
\label{sec:budget}
 
Skill-space recombination is cheap and reliable but bounded by the library's coverage;
code-space invention escapes that bound but is more expensive and less reliable per
proposal. We therefore balance exploitation against exploration through dynamic budget
allocation. \evoskillrec\ allocates a fixed per-generation budget of $N$ candidates, $8$ by
default, between the two spaces as $(n_{\text{skill}}, n_{\text{code}})$, by default
$(6,2)$, an explore--exploit split biased toward exploiting accumulated knowledge. The
split adapts: early generations keep $n_{\text{skill}}$ high, exploiting the existing
library as far as it goes; if the best validated score has not improved by more than
$\epsilon$ over the last $k$ generations---a performance plateau, indicating that the
current library cannot express a better genome---the budget shifts toward code space,
raising $n_{\text{code}}$ so the planner can attribute the plateau's likely cause via
failure diagnosis over recent bad cases and propose new seed skills. Once a new skill is
validated and promoted, the allocation relaxes back toward the default, since the enlarged
library can now be exploited through the cheaper skill-space operators.

\subsection{AutoResearch Loop}
\label{sec:loop}
 
The AutoResearch loop turns the evolution process into one iterative procedure over
executable genomes: diagnose the current models, propose candidates in both spaces, validate
and train them, keep the survivors, and record what happened. \evoskillrec\ first trains and
evaluates the seed genome under fixed data splits. Each later generation conditions both
spaces on the diagnosis of its parents and on the evolutionary memory $\mathcal{M}$, which
they read so that a structure already tried is not proposed again. Candidates are screened
by the checks of Sections~\ref{sec:skillspace} and~\ref{sec:codespace} and de-duplicated by
architecture details, so the training budget is spent only on architectures that are executable and
new; those that remain are trained under one protocol per generation, the same splits,
optimizer, and stopping rule, which makes their scores comparable. Survivors are selected by
$\eval$, instantiated as validation AUC in our CTR experiments, and become the next
generation's parents; the generated skills they carry sediment into $\library$, and their
metrics, lineage, diagnoses, and failure records are appended to $\mathcal{M}$. When the
best validated score stops improving, the allocation shifts toward code space
(Section~\ref{sec:budget}). Algorithm~\ref{alg:loop} states the procedure in full.

\renewcommand{\algorithmiccomment}[1]{\hskip 0.8em $\triangleright$ \textit{#1}}
\begin{algorithm}[t]
\caption{AutoResearch loop. $\mathcal{O} = \{\textsc{add}, \textsc{replace},
\textsc{hybridize}, \textsc{specialize}\}$ is the skill-space operator set
(Section~\ref{sec:skillspace}), $\mathcal{P}_t$ the skills synthesized in code space,
$G_\skill$ the candidate genome hosting skill $\skill$, and $\mathrm{Top}_K$ the $K$
highest-scoring elements of a set; $\ell(G)$ and $o(G)$ record a candidate's lineage and
validation outcome.}
\label{alg:loop}
\begin{algorithmic}[1]
\REQUIRE seed genome $G_0$; initial skill library $\library_0$; instance
$(\mathcal{D}, \mathcal{T}, \mathcal{C})$; generations $T$; budget $N$; survivor size $K$
and tolerance $\delta$; plateau window $k$ and threshold $\epsilon$
\ENSURE best genome $G^\star$; grown library $\library_T$; memory $\mathcal{M}_T$
\STATE $s(G_0) \leftarrow \eval(G_0; \mathcal{D}, \mathcal{T}, \mathcal{C})$;\;
$\Sigma_0 \leftarrow \{G_0\}$;\; $\mathcal{M}_0 \leftarrow \{(G_0, s(G_0))\}$;\;
$G^\star \leftarrow G_0$
\FOR{$t = 1$ \TO $T$}
    \STATE $r_t \leftarrow \textsc{Diagnose}(\Sigma_{t-1}, \mathcal{M}_{t-1})$
    \COMMENT{failure modes attributed to genome nodes}
    \STATE $(n_{\text{skill}}, n_{\text{code}}) \leftarrow
    \textsc{Allocate}(N, \mathcal{M}_{t-1}, \epsilon, k)$
    \COMMENT{shift toward code space on a plateau}
    \STATE $\mathcal{S}_t \leftarrow
    \textsc{Retrieve}(\library_{t-1}, \Sigma_{t-1}, r_t)$
    \COMMENT{top-$n$ skills under Eq.~\ref{eq:relevance}}
    \STATE $\mathcal{Q}^{\text{skill}}_t \leftarrow \bigl\{\,\mathrm{op}(G, \skill)\;:\;
    G \in \Sigma_{t-1},\; \skill \in \mathcal{S}_t,\; \mathrm{op} \in \mathcal{O}\,\bigr\}$,
    \; $\lvert \mathcal{Q}^{\text{skill}}_t \rvert = n_{\text{skill}}$
    \STATE $\bigl(\mathcal{P}_t, \mathcal{Q}^{\text{code}}_t\bigr) \leftarrow
    \textsc{Synthesize}\bigl(\textsc{Plan}(\mathcal{B}_t)\bigr)$,
    \; $\lvert \mathcal{Q}^{\text{code}}_t \rvert = n_{\text{code}}$
    \COMMENT{$\mathcal{B}_t$ from Eq.~\ref{eq:designbrief}}
    \STATE $\mathcal{Q}_t \leftarrow \textsc{Validate}\bigl(\mathcal{Q}^{\text{skill}}_t
    \cup \mathcal{Q}^{\text{code}}_t\bigr)$
    \COMMENT{validation and fingerprint de-duplication}
    \STATE $s(G) \leftarrow \eval(G; \mathcal{D}, \mathcal{T}, \mathcal{C})$ for every
    $G \in \mathcal{Q}_t$ \COMMENT{one training protocol per generation}
    \STATE $\bar{s}_t \leftarrow \mathrm{mean}_{\,G \in \mathcal{Q}_t}\, s(G)$;\;
    $\Sigma_t \leftarrow \mathrm{Top}_K \bigl\{\, G \in \mathcal{Q}_t \;:\;
    s(G) \ge \bar{s}_t - \delta \,\bigr\}$ \COMMENT{survivor selection}
    \STATE $\library_t \leftarrow \library_{t-1} \cup \bigl\{\, \skill \in \mathcal{P}_t
    \;:\; G_\skill \in \Sigma_t \,\bigr\}$ \COMMENT{sedimentation into Tier-2}
    \STATE $\mathcal{M}_t \leftarrow \mathcal{M}_{t-1} \cup \bigl\{\,\bigl(G, s(G), r_t,
    \ell(G), o(G)\bigr) : G \in \mathcal{Q}^{\text{skill}}_t \cup
    \mathcal{Q}^{\text{code}}_t \,\bigr\}$
    \STATE $G^\star \leftarrow \argmax_{G \in \{G^\star\} \cup \Sigma_t} s(G)$
\ENDFOR
\RETURN $G^\star,\; \library_T,\; \mathcal{M}_T$
\end{algorithmic}
\end{algorithm}

%% file: 5Experiment.tex
\section{Experiments}
\label{sec:experiments}

We evaluate \evoskillrec\ through five research questions. \textbf{RQ1 (Effectiveness)} tests whether skill-genome evolution consistently outperforms hand-crafted architectures and automated-design baselines across diverse recommendation tasks. \textbf{RQ2 (Evolution dynamics)} traces how architectures improve over evolution and which mechanisms drive the gains at each stage. \textbf{RQ3 (Mechanism ablation)} compares single-space variants with the full model and analyzes how candidates from the two spaces perform within full runs. \textbf{RQ4 (Multi-objective evolution)} examines whether \evoskillrec\ jointly optimizes objectives beyond predictive accuracy, particularly model FLOPs utilization (MFU), increasingly important for efficient generative ranking models. \textbf{RQ5 (LLM backbone)} compares state-of-the-art LLMs as the invention engine driving evolution.

\subsection{Experimental Settings}
\label{sec:exp_setup}

\textbf{Tasks and datasets.}
We evaluate \evoskillrec\ on three supervised recommendation tasks: \textbf{click-through rate (CTR) prediction} on MovieLens, Amazon Beauty, and Amazon Books; \textbf{multi-task learning (MTL)} on Census-Income; and \textbf{multi-domain learning (MDL)} on Amazon and MovieLens. For \textbf{multi-objective evolution}, where AUC is co-optimized with MFU, we use the industrial-scale QK-Video corpus.

\textbf{Baselines.}
We compare against three families of methods:
(i) \textbf{manually designed models}, comprising 7 CTR, 5 MTL, and 11 MDL architectures;
(ii) \textbf{automated architecture search}, represented by NASRec,
a weight-sharing search method \citep{zhang2022nasrec}; and
(iii) \textbf{LLM-driven code evolution}, represented by OpenEvolve,
an open-source framework inspired by AlphaEvolve \citep{novikov2025alphaevolve}\footnote{\url{https://github.com/algorithmicsuperintelligence/openevolve}}.
Appendix~\ref{app:baseline-implementation} provides implementation details,
including data and training protocols, search spaces, validation-based
selection, and candidate budgets.

\textbf{Metrics and protocol.}
We report test AUC for CTR, mean test AUC over the two tasks for MTL, and overall test AUC across domains for MDL. Candidate and checkpoint selection uses the corresponding validation metric; held-out test metrics are used for reporting, not selection. For multi-objective evolution, selection uses validation AUC and training MFU jointly, and both validation and test AUC are reported alongside MFU. Further details are provided in Appendix~\ref{app:implementation}.

\textbf{Implementation.}
Unless stated otherwise, all \evoskillrec\ code-space proposals in RQ1--RQ4 are generated by GPT-5.5 through the Codex CLI; RQ5 replaces this backbone with other LLMs. Training configurations, search budgets, and LLM settings are given in Appendix~\ref{app:implementation}.

\subsection{RQ1: Effectiveness Across Recommendation Tasks}
\label{sec:exp_rq1}

Table~\ref{tab:rq1} compares \evoskillrec\ with the best manual design, the NASRec
search baseline, and the OpenEvolve code-evolution baseline across all three task families.

Under the reported settings, the NASRec baseline scores below the strongest
manual model on all six benchmarks. The OpenEvolve baseline exceeds the best
manual result on all three CTR benchmarks but scores below it on MTL and MDL;
the largest shortfall is $0.0194$ AUC on MDL-MovieLens. \evoskillrec\ has the
highest reported test AUC in each column, with gains of $0.0001$--$0.0085$ over
the strongest alternative in that column. These results compare the evaluated
implementations under the protocols in Appendix~\ref{app:baseline-implementation}.
The methods use different search spaces and proposal mechanisms, so the aggregate
comparison alone does not identify which design choice accounts for the differences.

\begin{table}[t]
\centering
\caption{\textbf{Test AUC across CTR, MTL, and MDL benchmarks.} \emph{Best Manual} is
the strongest manually designed model on each dataset. $\Delta$ is the gain
of \evoskillrec\ over the strongest non-\evoskillrec\ baseline in that column.
Baseline implementations and search budgets are specified in
Appendix~\ref{app:baseline-implementation}.}
\label{tab:rq1}
\small
\begin{threeparttable}
\setlength{\tabcolsep}{2pt}
\renewcommand{\arraystretch}{1.2}
\begin{tabular}{@{}l|cccccc@{}}
\toprule
& \multicolumn{3}{c}{\textbf{CTR}} & \textbf{MTL} & \multicolumn{2}{c}{\textbf{MDL}} \\
\cmidrule(lr){2-4} \cmidrule(lr){5-5} \cmidrule(lr){6-7}
\textbf{Method} & \textbf{MovieLens} & \makecell{\textbf{Amazon}\\\textbf{Books}}
& \makecell{\textbf{Amazon}\\\textbf{Beauty}} & \textbf{Census} & \textbf{Amazon}
& \textbf{MovieLens} \\
\midrule
Best Manual   & 0.7487\tnote{a} & 0.6587\tnote{b} & 0.6666\tnote{c} & 0.9726\tnote{d} & 0.7012\tnote{e} & 0.7924\tnote{e} \\
NASRec  & 0.7361 & 0.6487 & 0.6632 & 0.9711 & 0.6870 & 0.7619 \\
OpenEvolve & 0.7521 & 0.6654 & 0.6760 & 0.9722 & 0.6903 & 0.7730 \\
\midrule
\rowcolor{gray!10}
\textbf{\evoskillrec} & \textbf{0.7522} & \textbf{0.6708} & \textbf{0.6845} & \textbf{0.9748} & \textbf{0.7033} & \textbf{0.7932} \\
\rowcolor{gray!10}
\hspace{1em}{\scriptsize $\Delta$ vs.\ best baseline} &
{\scriptsize $\uparrow$0.0001\,(+0.01\%)} &
{\scriptsize $\uparrow$0.0054\,(+0.81\%)} &
{\scriptsize $\uparrow$0.0085\,(+1.26\%)} &
{\scriptsize $\uparrow$0.0022\,(+0.23\%)} &
{\scriptsize $\uparrow$0.0021\,(+0.30\%)} &
{\scriptsize $\uparrow$0.0008\,(+0.10\%)} \\
\bottomrule
\end{tabular}
\begin{tablenotes}
\footnotesize
\item[] Best manual models: $^{a}$Wide \& Deep, $^{b}$Wide \& Deep, $^{c}$AutoInt, $^{d}$AITM,
$^{e}$SAR-Net.
\end{tablenotes}
\end{threeparttable}
\end{table}

\subsection{RQ2: How Do Architectures Evolve?}
\label{sec:exp_rq2}

\begin{figure}[!t]
\centering
\setlength{\abovecaptionskip}{4pt}
\setlength{\belowcaptionskip}{0pt}
\begin{subfigure}[b]{0.45\textwidth}
    \centering
    \includegraphics[width=\textwidth]{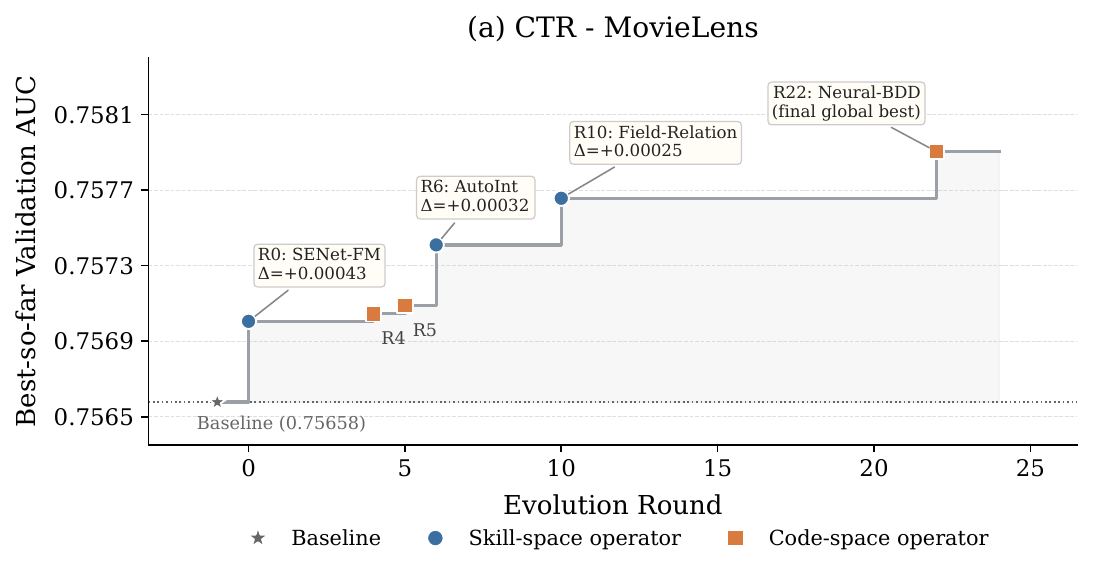}%
\end{subfigure}\hfill
\begin{subfigure}[b]{0.45\textwidth}
    \centering
    \includegraphics[width=\textwidth]{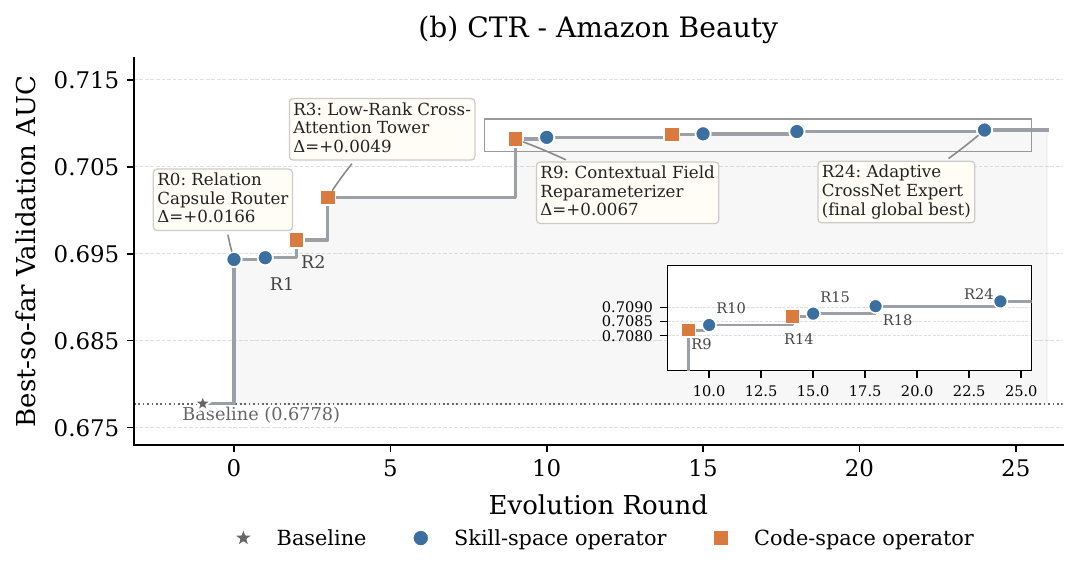}%
\end{subfigure}%
\\[0.4ex]
\begin{subfigure}[b]{0.45\textwidth}
    \centering
    \includegraphics[width=\textwidth]{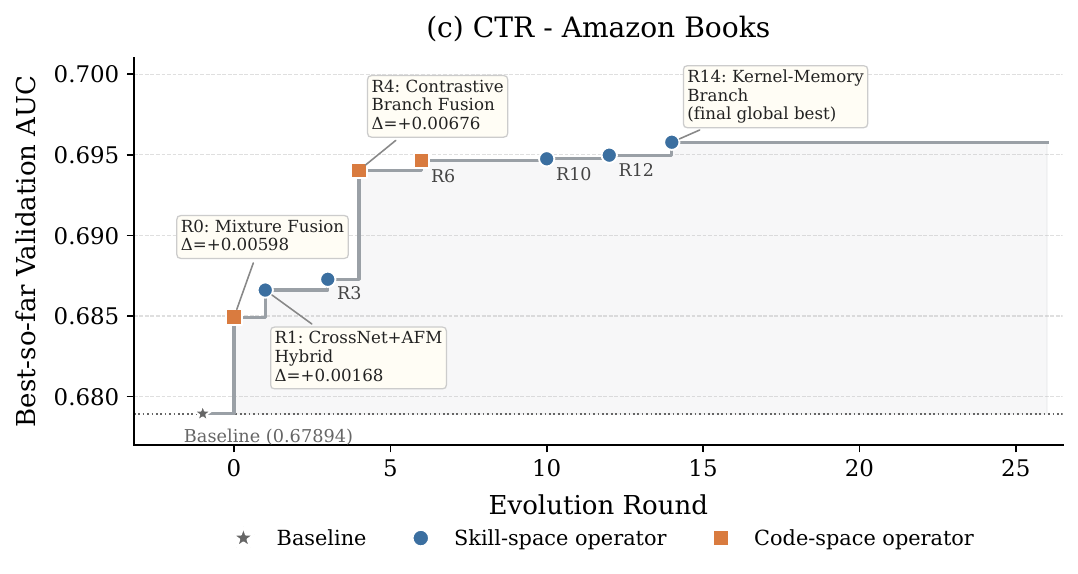}%
\end{subfigure}\hfill
\begin{subfigure}[b]{0.45\textwidth}
    \centering
    \includegraphics[width=\textwidth]{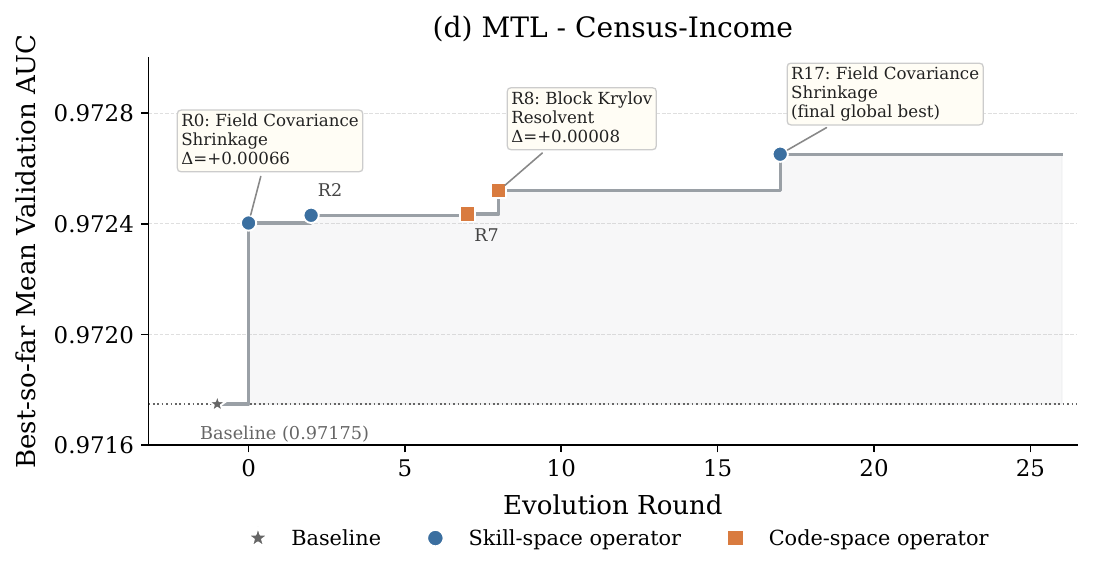}%
\end{subfigure}%
\\[0.4ex]
\begin{subfigure}[b]{0.45\textwidth}
    \centering
    \includegraphics[width=\textwidth]{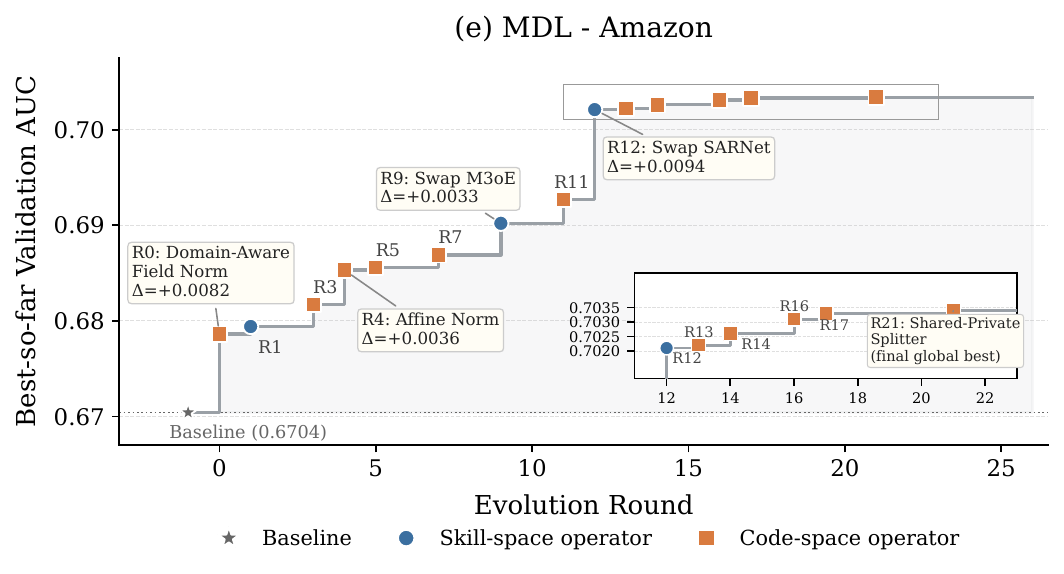}%
\end{subfigure}\hfill
\begin{subfigure}[b]{0.45\textwidth}
    \centering
    \includegraphics[width=\textwidth]{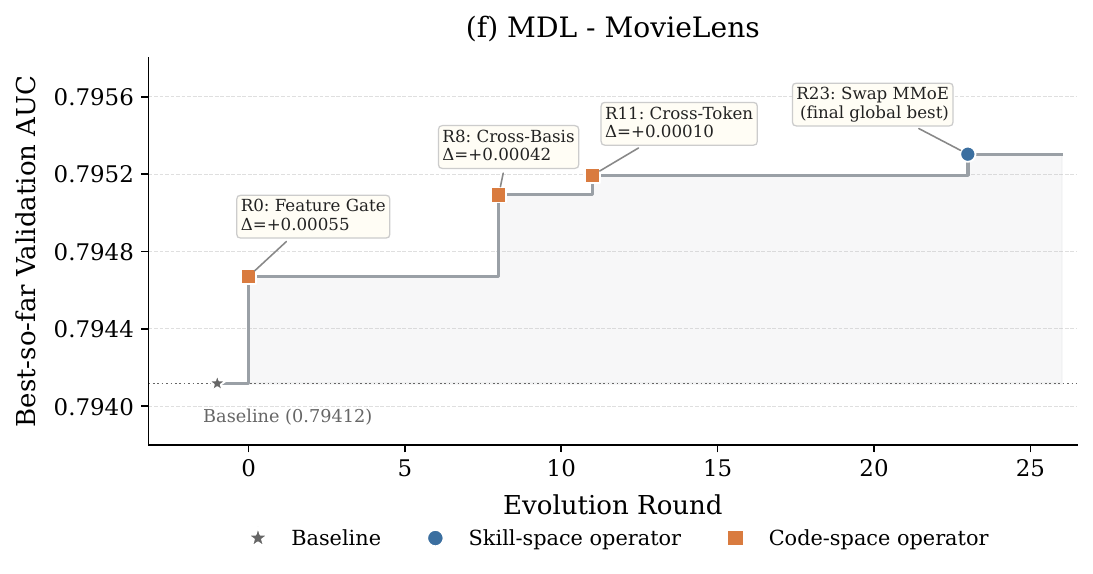}%
\end{subfigure}%
\caption{Best-so-far validation AUC over evolution rounds. Each annotated improvement reports
its operator and AUC gain, with color indicating its origin in skill space or code space.}
\label{fig:evolution_curves}
\end{figure}

To examine how improvements accumulate during evolution, Figure~\ref{fig:evolution_curves}
plots the best-so-far validation AUC against evolution round for each benchmark. Each
annotation marks an improvement with the corresponding operator and its origin in
skill space or code space.

The trajectories show that skill-space reuse and code-space synthesis play complementary
roles, with contributions varying across datasets and evolution stages. On CTR,
CTR-MovieLens benefits from both spaces: skill-space operators contribute at several
rounds, while code-space edits also improve the trajectory, including in later stages.
CTR-Books similarly combines both mechanisms. Code-space operators such as Mixture Fusion
and Contrastive Branch Fusion produce substantial improvements, while multiple skill-space
transitions also advance the best-so-far performance, including the final best-so-far model.
Thus, the effectiveness of the two search spaces can vary considerably even within the same
task family. On MDL-Amazon, code-space proposals introduce
domain-specific structures such as Domain-Aware Field Norm and Shared-Private Splitter,
whereas skill-space \emph{replace} steps that insert the M3oE or SAR-Net module in place of an
existing module also improve the trajectory. MDL-MovieLens likewise contains improvements from both mechanisms. Overall,
Figure~\ref{fig:evolution_curves} should be read as illustrating how skill reuse and code synthesis interact over individual evolution trajectories, rather than as evidence that either space universally dominates. We quantify their aggregate contributions separately in RQ3.

\subsection{RQ3: What Does Each Mechanism Contribute?}
\label{sec:exp_rq3}

\begin{table}[t]
\centering
\caption{Ablation of the two search spaces and generated-skill reuse. All values are test AUC;
MTL reports the mean over tasks and MDL the overall AUC across domains.}
\label{tab:rq3}
\small
\renewcommand{\arraystretch}{1.15}
\setlength{\tabcolsep}{2pt}
\begin{tabular}{@{}l | ccc | ccc c cc@{}}
\toprule
\multirow{2}{*}{\textbf{Method}} & \multicolumn{3}{c|}{\textbf{Search space}} & \multicolumn{3}{c}{\textbf{CTR}} & \textbf{MTL} & \multicolumn{2}{c}{\textbf{MDL}} \\
\cmidrule(lr){2-4} \cmidrule(lr){5-7} \cmidrule(lr){8-8} \cmidrule(lr){9-10}
& Skill & Code & Reuse & MovieLens & Books & Beauty & Census & Amazon & MovieLens \\
\midrule
Skill-space only & \cmark & \xmark & \xmark & 0.7512 & 0.6647 & 0.6639 & 0.9737 & 0.7019 & 0.7924 \\
Code-space only  & \xmark & \cmark & \xmark & 0.7520 & 0.6689 & 0.6840 & 0.9730 & 0.6807 & 0.7926 \\
\midrule
\rowcolor{gray!10}
Full model (\textbf{\evoskillrec}) & \cmark & \cmark & \cmark & \textbf{0.7522} & \textbf{0.6708} & \textbf{0.6845} & \textbf{0.9748} & \textbf{0.7033} & \textbf{0.7932} \\
\bottomrule
\end{tabular}
\end{table}

We conduct two complementary analyses. The first compares \evoskillrec\ with single-space
variants (Table~\ref{tab:rq3}): Skill-space only recombines existing library skills, Code-space
only invents at the code level without consolidating results back into the library, and the full
model couples both with generated-skill reuse. We do not run a variant that combines the two
spaces without reuse, so this comparison does not isolate the contribution of reuse from that of
combining the spaces. The second keeps the full model intact and groups the candidates of each
run by the space that proposed them (Table~\ref{tab:skill_vs_code}). There, the positive-gain
rate is the fraction of evaluated candidates whose validation AUC exceeds that of their parent,
$\Delta_{\max}$ is the largest such gain, and the survivor yield is the fraction of proposed
candidates that are selected as survivors, with duplicates and failed candidates counted as
non-survivors.

Table~\ref{tab:rq3} shows that the full model attains the highest test AUC on all six
benchmarks, although several margins over the stronger single-space variant are small (e.g.,
$+0.0002$ on CTR-MovieLens and $+0.0005$ on CTR-Beauty) and, like all results here, come from
single runs (Section~\ref{sec:limitations}). Neither single-space variant dominates: Code-space
only is the stronger ablation on the three CTR benchmarks, Skill-space only on MTL-Census and
MDL-Amazon, and the two are nearly tied on MDL-MovieLens (0.7924 vs.\ 0.7926). This is
consistent with the intended division of labor, in which skill space exploits primitives already
in the library and code space explores beyond them, so that combining the two does not require
choosing a space in advance.

Within the full model (Table~\ref{tab:skill_vs_code}), the positive-gain rates of the two spaces
are similar on CTR-Beauty and CTR-Books, higher for code space on CTR-MovieLens (25.7\% vs.\
17.7\%) and MTL-Census (35.1\% vs.\ 29.0\%), and higher for skill space on both multi-domain
benchmarks (56.2\% vs.\ 45.8\% and 31.6\% vs.\ 28.6\%). Skill-space candidates reach the larger
$\Delta_{\max}$ on five of six benchmarks, CTR-Books being the exception. This comparison
attributes each gain to the space that proposed the candidate; because skill-space operators can
also insert Tier-2 skills that were originally generated in code space, it does not show that
recombining human-designed modules alone yields the largest improvements. Survivor yield differs
more sharply between the spaces. Skill space has the higher yield on all CTR and MTL benchmarks
but only 3.1\% and 6.2\% on the two multi-domain benchmarks, against 62.5\% and 50.6\% for code
space; we attribute this mainly to domain-conditioned compatibility constraints, under which many
skill-space recombinations reduce to previously evaluated architectures and are removed as
duplicates before training. Finally, the number of promoted code skills ranges from 16
(CTR-Beauty) to 45 (both multi-domain benchmarks). Promoted skills become available to skill-space
search in later generations; on CTR-MovieLens, for example, the module invented in round~3
(\texttt{autoint\_\allowbreak senet\_\allowbreak bridge\_\allowbreak branch\_\allowbreak r3}) is reused in round~6 and improves the best-so-far
validation AUC (Figure~\ref{fig:evolution_curves} and Appendix~\ref{app:evolution_details}).

\begin{table*}[t]
\centering
\caption{Skill space vs.\ code space within the full model's runs, grouped by the space that
proposed each candidate. $\Delta$ is the validation AUC gain of a candidate over its parent. The
positive-gain rate is the fraction of evaluated candidates with $\Delta>0$, $\Delta_{\max}$ is the
largest $\Delta$, and survivor yield is the fraction of proposed candidates that are selected as
survivors.}
\label{tab:skill_vs_code}
\begin{threeparttable}
\resizebox{\textwidth}{!}{%
\begin{tabular}{l| cc cc cc c}
\toprule
\multirow{2}{*}{\textbf{Benchmark}} & \multicolumn{2}{c}{\textbf{Positive-gain rate} $P(\Delta>0)$ (\%)} & \multicolumn{2}{c}{\textbf{Max.\ validation gain} $\Delta_{\max}$} & \multicolumn{2}{c}{\textbf{Survivor yield} (\%)} & \textbf{Code skills} \\
\cmidrule(lr){2-3} \cmidrule(lr){4-5} \cmidrule(lr){6-7}
 & Skill & Code & Skill & Code & Skill & Code & promoted (\#) \\
\midrule
CTR-MovieLens          & 17.7          & \textbf{25.7} & \textbf{+0.001470} & +0.001171          & \textbf{40.9} & 38.1          & 43 \\
CTR-Beauty             & \textbf{19.0} & 18.9          & \textbf{+0.016589} & +0.006988          & \textbf{45.2} & 21.6          & 16 \\
CTR-Books     & 17.3          & \textbf{18.1} & +0.005369          & \textbf{+0.008282} & \textbf{44.1} & 26.0          & 19 \\
MTL-Census             & 29.0          & \textbf{35.1} & \textbf{+0.000655} & +0.000229          & \textbf{44.3} & 35.1          & 18 \\
MDL-Amazon             & \textbf{56.2} & 45.8          & \textbf{+0.011990} & +0.009539          & 3.1           & \textbf{62.5} & 45 \\
MDL-MovieLens          & \textbf{31.6} & 28.6          & \textbf{+0.004594} & +0.001665          & 6.2           & \textbf{50.6} & 45 \\
\bottomrule
\end{tabular}%
}
\end{threeparttable}
\end{table*}

\subsection{RQ4: Multi-Objective Evolution}
\label{sec:exp_rq4}

\begin{figure*}[t]
\centering
\begin{minipage}[c]{0.48\textwidth}
    \centering
    \includegraphics[width=\textwidth]{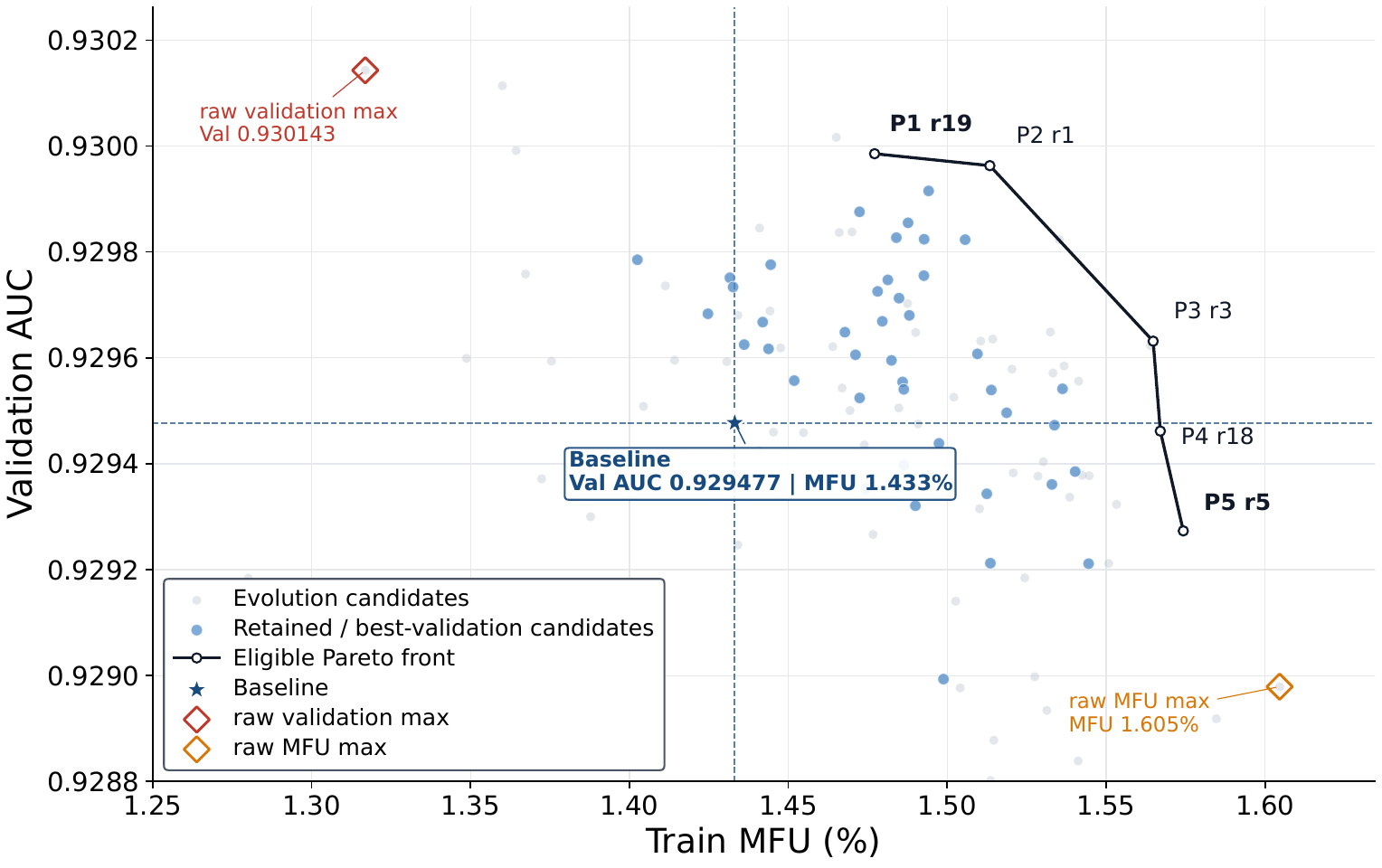}
    \captionof{figure}{Accuracy--MFU trade-off on QK-Video. Grey points are all evolution
    candidates, blue points retained candidates, and the black curve the eligible
    Pareto front. Diamonds mark the raw validation-AUC and MFU maxima.}
    \label{fig:mfu-pareto}
\end{minipage}
\hfill
\begin{minipage}[c]{0.48\textwidth}
    \centering
    \scriptsize
    \begin{threeparttable}
    \resizebox{\linewidth}{!}{%
    \begin{tabular}{@{} l |
                    S[table-format=1.5, round-precision=5]
                    S[table-format=1.5, round-precision=5] S[table-format=+1.1, round-precision=1]
                    S[table-format=1.3, round-precision=3] S[table-format=+1.3, round-precision=3] @{}}
      \toprule
      \multirow{2}{*}{\textbf{Model}} & {\multirow{2}{*}{\textbf{Val.\ AUC}}} & \multicolumn{2}{c}{\textbf{Test}}
        & \multicolumn{2}{c}{\textbf{Training MFU}} \\
      \cmidrule(lr){3-4} \cmidrule(lr){5-6}
      & & {AUC} & {$\Delta$AUC ($\times10^{-4}$)} & {MFU (\%)} & {$\Delta$MFU (pp)} \\
      \midrule
      Baseline & 0.92948 & 0.93580 & 0.0  & 1.433 & 0.000 \\
      \midrule
      P1 (r19) & \bfseries 0.92999 & \bfseries 0.93598 & \bfseries +1.7 & 1.477 & +0.044 \\
      P2 (r1)  & 0.92996 & 0.93573 & -0.7 & 1.513 & +0.080 \\
      P3 (r3)  & 0.92963 & 0.93570 & -1.0 & 1.565 & +0.132 \\
      P4 (r18) & 0.92946 & 0.93540 & -4.0 & 1.567 & +0.134 \\
      P5 (r5)  & 0.92927 & 0.93561 & -1.9 & \bfseries 1.574 & \bfseries +0.141 \\
      \bottomrule
    \end{tabular}%
    }

    \end{threeparttable}
    \captionof{table}{Performance and training efficiency of the baseline (RankMixer) and eligible
    Pareto-front candidates. All $\Delta$ values are relative to the baseline; MFU deltas are in percentage points (pp). Architectural details for P1--P5 are shown in Figure~\ref{fig:mfu-arch}; r$_N$ denotes the round that produced each candidate.}
    \label{tab:pareto_candidates}
\end{minipage}
\centering
\includegraphics[width=0.95\textwidth]{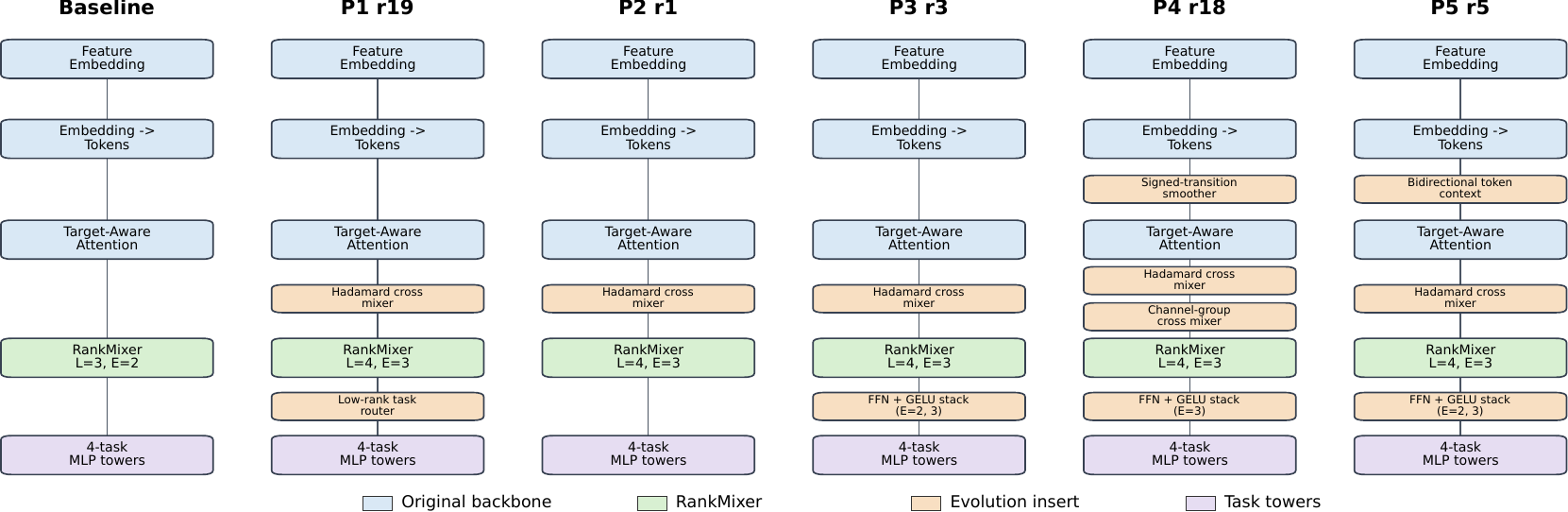}
\captionof{figure}{Architectural differences between the baseline (RankMixer) and the
Pareto-optimal models on QK-Video.}
\label{fig:mfu-arch}
\end{figure*}

Industrial recommenders are shifting towards generative ranking models
\citep{zhou2025onerec, zhu2025rankmixer}, with parameter counts growing from tens of
millions to billions. At this scale, hardware efficiency becomes a design constraint, and model
FLOPs utilization (MFU), the ratio of achieved training FLOP/s to the GPU's
theoretical peak FLOP/s, is now reported alongside accuracy
\citep{chowdhery2023palm, zhou2025onerec, zhang2026onetrans}. We therefore test whether
\evoskillrec\ can optimize a compound objective. On QK-Video, we seed evolution from a RankMixer
backbone \citep{zhu2025rankmixer} and run 30 rounds\footnote{Our pipeline builds on the public
UniRank codebase: \url{https://github.com/salmon1802/UniRank}.} with an objective jointly
rewarding AUC and MFU, subject to the eligibility criterion of Section~\ref{sec:problem}.
Figure~\ref{fig:mfu-pareto} shows the resulting accuracy--efficiency trade-off and the Pareto
front with its five candidates P1--P5, whose validation AUC, test AUC, and training MFU appear
in Table~\ref{tab:pareto_candidates}; Figure~\ref{fig:mfu-arch} shows how the evolved
architectures differ from the backbone.

In Figure~\ref{fig:mfu-pareto}, a large fraction of evolved candidates lie in the upper-right
region and dominate the RankMixer seed on both axes, so the two objectives improve together
rather than trade off. Five candidates, P1--P5, lie on the Pareto front, but only P1 (round 19)
retains both improvements on the test set. Headroom on either objective alone is also
substantial, with the best validation AUC reaching 0.9301 and the best MFU 1.605\%, so evolution
remains effective for compound objectives, not only for single metrics. As
Figure~\ref{fig:mfu-arch} shows, although produced in different rounds (r1, r3, r5, r18, r19),
all five share two modifications to the seed backbone: scaling RankMixer from $L{=}3, E{=}2$ to
$L{=}4, E{=}3$, and inserting a Hadamard cross mixer directly after DIN attention. Beyond this
shared core, their differences follow the front's geometry. Higher-MFU candidates stack
compute-dense operators: P3 and P5 add an FFN+GELU block, P4 a second channel-group cross mixer,
and P5 bidirectional token context. P1, at the accuracy-preserving end, instead adds a low-rank
task router at the tower interface and has the lowest MFU of the five ($1.477\%$). Dense
operators thus raise utilization, while low-rank routing preserves accuracy.

\subsection{RQ5: Effect of the LLM Backbone}
\label{sec:exp_rq5}

\begin{figure*}[t]
    \centering
    \includegraphics[width=0.32\textwidth]{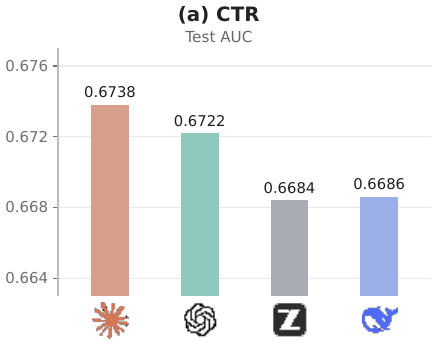}
    \hfill
    \includegraphics[width=0.32\textwidth]{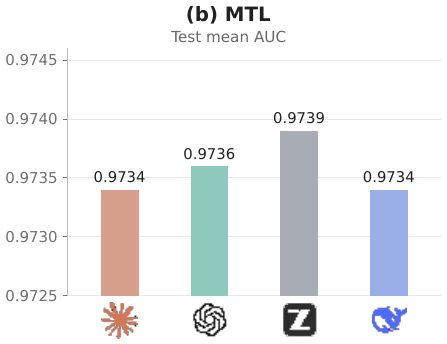}
    \hfill
    \includegraphics[width=0.32\textwidth]{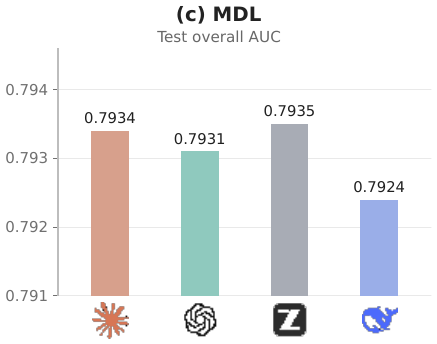}
    \caption{Effect of the LLM backbone. AUC of Claude Opus 4.8, GPT-5.6 (SOL), DeepSeek-V4-Pro
    and GLM-5.2 on CTR (Amazon-Books), multi-task learning (MTL, Census-Income) and multi-domain
    learning (MDL, MovieLens).}
    \label{fig:llm_comparison}
\end{figure*}

Because an LLM drives evolution in \evoskillrec, the backbone choice is a design variable, and
models that perform comparably on public benchmarks need not behave alike here. We compare four
models, Claude Opus 4.8, GPT-5.6 (SOL), DeepSeek-V4-Pro and GLM-5.2, each at its highest
reasoning setting and run with identical initial architectures, environment configurations and
round budgets; the main experiments (RQ1--RQ4) instead use GPT-5.5. Figure~\ref{fig:llm_comparison}
reports test AUC on CTR (Amazon-Books), MTL (Census-Income) and MDL (MovieLens).

No single backbone leads on all three tasks. Claude Opus 4.8 is strongest on CTR by a clear margin
(0.6738 test AUC), whereas GLM-5.2 leads on MTL (0.9739 mean test AUC) and MDL (0.7935), though its
$0.0001$ advantage over Claude Opus 4.8 on MDL should be read as a tie. On MDL, GLM-5.2 is also on
par with, and slightly above, the GPT-5.5 run reported in Table~\ref{tab:rq1} (0.7932), indicating
that comparable MDL results can be obtained with an open-weight backbone. Given list token prices,
GLM-5.2 reaches this performance at lower cost than Claude Opus 4.8, making it the most economical
backbone across this task suite. The gap between open- and closed-weight backbones is
model-specific rather than categorical: GLM-5.2 is competitive with both closed-weight models and
ahead of them on two tasks, whereas DeepSeek-V4-Pro is among the weaker backbones on all three
tasks (lowest on MDL, tied for lowest on MTL, and second lowest on CTR). DeepSeek's proposals also
succeed at a consistently lower rate during evolution, so its deficit lies in how reliably it
produces valid candidates, not only in their quality. The preferred backbone therefore depends on
the task family. One plausible explanation is uneven pretraining coverage: CTR architectures are
far more widely documented than multi-task and multi-domain ones, and the two closed-weight models
lead precisely on CTR while ceding the two less common families to GLM-5.2. We cannot verify this
externally, but the practical implication is stable across our runs: GLM-5.2 is the reasonable
default for MTL and MDL evolution.

%% file: 6Conclusion.tex
\section{Discussion and Limitations}
\label{sec:limitations}

The experiments above support our central claim, that evolution should accumulate reusable
architectural knowledge rather than rediscover it each run. Several questions remain open.

\textbf{Statistical resolution.}
Some of the margins we report are on the order of $10^{-4}$ AUC, which a single evolution run
cannot reliably separate from run-to-run variation. Where we draw conclusions we rely on the
larger effects, such as the ablation reversals in Section~\ref{sec:exp_rq3}; the smaller
orderings should be read as observations from one run rather than as established results.
Repeating each run under several seeds would settle this, but every cell in
Table~\ref{tab:rq1} is a full evolution run rather than a single model fit, and the compute this
would require was beyond what we could afford here.

\textbf{Cost and dependence on the LLM.}
Code-space invention is the expensive branch. Each proposal consumes two LLM calls followed by a
full training and validation cycle, and a large share of proposals are rejected before training
begins. The adaptive budget of Section~\ref{sec:budget} contains this cost without removing it.
The API bill is not trivial either: across all backbones and tasks, the experiments reported
here cost roughly \$7{,}000. This limits how much of the design space an academic group can
explore, and it pushes such work toward cheaper backbones, even though Section~\ref{sec:exp_rq5}
shows that the choice of backbone affects which architectures a run discovers. Wider academic
access programs from model providers would make this line of work easier to pursue outside
industry. Until then, our numbers should be read as characterizing \evoskillrec\ together with
the model driving it, rather than the framework in isolation.

\textbf{Scope of transfer.}
Two questions remain untested. The first is cross-task transfer: every library in our
experiments is grown and used within a single task family, so we do not know whether skills
discovered on one task would benefit evolution on other tasks. This is the stronger version of
our claim, and answering it would require a different experimental design from the one used
here. The second is industrial validation. The library is seeded from published academic
architectures and evaluated mostly on public benchmarks, and QK-Video is our only contact with
an industrial-scale setting. That result is encouraging, since evolution there improved on a
strong production backbone under a compound objective, but whether the same decomposition into
reusable skills holds for proprietary backbones and industrial oneline production traffic remains to be seen.
We see this as the direction with the most practical value.

\textbf{Outlook.}
The longer-term motivation for this line of work is to let an agent carry out architecture design
that currently requires an experienced engineer, and to leave the resulting design decisions
inspectable rather than buried in a search trajectory. Our results indicate that this is not yet
within reach: backbones differ substantially in how reliably they propose viable architectures,
and the surrounding loop, from budget allocation to promotion criteria and evaluation, still
needs careful design. We expect the balance to shift as models improve at long-horizon,
tool-using work, and we regard the skill library as the component most likely to carry over,
since it stores what a run has learned in a form that neither the model nor the next task has to
reconstruct.

\section{Conclusion}
\label{sec:conclusion}

We presented EvoSkillRec, a skill-genome framework for recommender architecture discovery. It breaks established recommenders down into executable skills, evolves them through coupled skill-space recombination and code-space invention, and promotes validated discoveries back into a growing library. EvoSkillRec is the only method we compare that improves on the strongest hand-designed baseline on all six benchmarks, and it extends to accuracy–efficiency co-optimization in generative ranking models. The two search spaces are complementary, and future work will study cross-task skill transfer and industrial deployment.

%% file: 7Appendix.tex
\clearpage
\appendix
\startcontents[appendix]
\section*{Contents of the Appendix}
\printcontents[appendix]{l}{1}{\setcounter{tocdepth}{2}}

\section{Skill Library and Skill Cards}
\label{app:skill-cards}
 
Every reusable module in the skill library comes with a skill card:
a machine-readable YAML manifest that records what the module computes,
which named tensors it consumes and produces, and how it can be retrieved
and composed with other skills. The library has two tiers. Tier-1 consists
of curated components obtained by decomposing existing recommendation
models, together with supporting adapters and training objectives. Tier-2
consists of modules synthesized by the LLM during Code Space exploration
and retained for later reuse. Both tiers use the same named-tensor
interface, so the planner can treat them uniformly; Tier-2 cards
additionally store the validation and promotion evidence collected when the
module was admitted. Below, we first specify the card schema
(\S\ref{app:skill-card-schema}), then list the Tier-1 library by category
(\S\ref{app:tier1-inventory}), and finally show one card from each tier
(\S\ref{app:skill-card-examples}).
 
\subsection{Skill Card Fields and Tensor Interfaces}
\label{app:skill-card-schema}
 
Table~\ref{tab:app-skill-schema} groups the fields of a skill card by
function. Two serialization conventions coexist in the implementation.
Curated cards usually identify a skill by \texttt{name} and declare its
interface with \texttt{inputs} and \texttt{outputs}. Generated cards also
carry a \texttt{skill\_id} and declare the interface with
\texttt{input\_signature} and \texttt{output\_signature}. The two
conventions encode the same identity and interface information. Not every
field is mandatory: the table lists the field groups that the library
supports, and a compact curated card may leave its full tensor contract to
the \texttt{SkillSpec} registered in Python.
 
\begin{table}[!htbp]
\centering
\small
\caption{Fields of a skill card, grouped by function. The Validation and
Promotion groups appear only in Tier-2 (generated) cards.}
\label{tab:app-skill-schema}
\begin{tabular}{@{}>{\raggedright\arraybackslash}p{0.17\linewidth}
                >{\raggedright\arraybackslash}p{0.38\linewidth}
                >{\raggedright\arraybackslash}p{0.38\linewidth}@{}}
\toprule
\textbf{Field group} & \textbf{Serialized fields} & \textbf{Content and purpose} \\
\midrule
Identity and role &
\texttt{name}, \texttt{skill\_id}, \texttt{skill\_name}, \texttt{category},
\texttt{description} / \texttt{function\_description} &
Stable identifier, functional category, and a natural-language
description of the module. \\
\hline
Tensor interface &
\texttt{inputs}, \texttt{outputs}; or
\texttt{input\_signature}, \texttt{output\_signature} &
Name, shape, and dtype of each input and output tensor; generated
signatures may add a semantic annotation. \\
\hline
Modeling context &
\texttt{task\_types} / \texttt{applicable\_tasks},
\texttt{inductive\_bias}, \texttt{failure\_signatures},
\texttt{failure\_modes\_addressed} &
Applicable tasks, modeling assumptions, known interface failures, and the
modeling weaknesses the skill is meant to address. \\
\hline
Retrieval &
\texttt{retrieval}: \texttt{summary}, \texttt{use\_when},
\texttt{avoid\_when}, \texttt{task\_types}, \texttt{architecture\_roles},
\texttt{input\_modalities}, \texttt{output\_semantics},
\texttt{objectives}, \texttt{model\_families}, \texttt{aliases},
\texttt{query\_examples} &
Descriptions and search terms used to match a modeling need to a
reusable component. \\
\hline
Composition &
\texttt{composition}: \texttt{requires}, \texttt{produces},
\texttt{common\_upstream}, \texttt{common\_downstream},
\texttt{example\_genome\_fragment} &
Context keys read and written, typical neighboring skills, and an example
invocation with constructor parameters. \\
\hline
Usage and cost &
\texttt{constraints}, \texttt{cost}, \texttt{good\_for},
\texttt{bad\_for}, \texttt{compatible\_with}, \texttt{mutation\_roles} &
Optional restrictions, computational cost, compatibility hints, and
suggested mutation roles. \\
\hline
Implementation and provenance &
\texttt{source}; or \texttt{implementation\_path}, \texttt{class\_name},
\texttt{source\_proposal\_id}, \texttt{parent\_skills},
\texttt{mutation\_lineage} &
Tier-1: original implementation and extraction type.
Tier-2: location of the generated code and its proposal lineage. \\
\hline
Validation (Tier-2) &
\texttt{validation\_status}, \texttt{validation\_results},
\texttt{portability} &
Results of static, forward/backward, and shape checks; where available,
tests under other feature schemas and the resulting reuse scope. \\
\hline
Promotion (Tier-2) &
\texttt{promotion\_status}, \texttt{promoted\_at},
\texttt{candidate\_id}, \texttt{candidate\_status},
\texttt{candidate\_metrics}, \texttt{architecture\_fingerprint},
\texttt{code\_hash} &
Evidence from the candidate genome that contained the skill, plus
identifiers that trace the retained architecture and code. \\
\bottomrule
\end{tabular}
\end{table}
 
\paragraph{Tensor shapes and compatibility requirements.}
A tensor contract binds a name to a shape and a dtype. For example,
\texttt{field\_embeddings} has shape $[B,F,D]$, where $B$ is the batch
size, $F$ the number of feature fields, and $D$ the embedding dimension.
Constructor arguments are supplied by the genome node that invokes the
skill; the example invocation in a card may use placeholders such as
\texttt{\$\{embedding\_dim\}}. Because dimensions are symbolic, a skill can
be re-instantiated under a different feature schema, provided its declared
constraints hold. This concerns the module definition only: trained
weights are not assumed to transfer, and compatibility is checked for each
concrete schema rather than guaranteed for all of them. Similarly, the
upstream and downstream neighbors listed in a card serve only as retrieval
hints. A connection in an actual genome is valid only when the
instantiated tensor shapes and context keys match.
 
\subsection{Curated Skills by Category}
\label{app:tier1-inventory}
 
The curated library contains \textbf{94 skills in 11 categories}.
Table~\ref{tab:app-tier1-inventory} reports the full size of each category and
shows up to four skill identifiers per category; ellipses indicate omitted entries.
Skills are grouped by the \texttt{category} field of their cards, and each
skill is counted once even if it serves several tasks or model families. Categories are defined by
functional role, and two of them need a brief note. \texttt{scenario}
collects multi-domain components, and \texttt{generative} collects
semantic-ID and generative ranking components. The name
\texttt{generative} refers to the modeling paradigm, not to how a skill was
obtained: every skill in this table is a curated Tier-1 skill. Fusion
adapters are listed under \texttt{utility}.
 
\input{appendix/appendix_skill_inventory}

\subsection{Examples of Curated and Generated Skill Cards}
\label{app:skill-card-examples}
 
We show one card from each tier: a component decomposed from an existing
model (Tier-1) and a module generated by the Code Space LLM pipeline
(Tier-2). Both listings are excerpts of stored manifests. Some fields are
omitted for space; the retained fields keep their original values and
nesting.
 
\lstdefinestyle{skillcard}{basicstyle=\ttfamily\footnotesize,
  columns=fullflexible,keepspaces=true,breaklines=true,
  showstringspaces=false,frame=single,captionpos=b}
 
\paragraph{Curated skill example: factorization-machine interaction.}
The \texttt{fm\_interaction} skill isolates the second-order interaction
branch of DeepFM by wrapping the existing FM layer. 
The card (Listing~\ref{lst:app-tier1-card}) records the source location and
extraction type, the second-order inductive bias, the tensor interface, and
a minimal invocation.
 
\begin{lstlisting}[style=skillcard,caption={Tier-1 card example: FM interaction extracted from an existing implementation.},label={lst:app-tier1-card}]
name: fm_interaction
category: interaction
description: Compute second-order factorization-machine interactions over field embeddings.
inputs:
  - name: field_embeddings
    shape: "[batch_size, num_fields, embedding_dim]"
    dtype: float32
outputs:
  - name: fm_output
    shape: "[batch_size, 1]"
    dtype: float32
task_types: [ctr, ranking]
inductive_bias: [second_order_feature_interactions]
failure_signatures: [rank_not_3]
constraints:
  input_rank: 3
cost:
  parameters: 0
  compute: "O(batch_size * num_fields * embedding_dim)"
source:
  file: torch_rechub/basic/layers.py
  class_or_function: FM
  extraction_type: wrapped
retrieval:
  summary: "Use this for low-cost second-order feature interactions over field embeddings in CTR/ranking models."
  task_types: [ctr, ranking]
  architecture_roles: [interaction, second_order, factorization_machine]
  input_modalities: [field_embeddings]
  output_semantics: [interaction_logit, fm_output]
composition:
  requires: [field_embeddings]
  produces: [fm_output]
  common_upstream: [field_embedding]
  common_downstream: [concat_fusion, binary_ctr_head]
  example_genome_fragment:
    skill: fm_interaction
    params:
      input_key: field_embeddings
      output_key: fm_output
\end{lstlisting}
 
\paragraph{Generated skill example: covariance-based field mixing.}
The generated skill \texttt{macro\_field\_covariance\_shrinkage\_v1} was
proposed while exploring a PLE-based multi-task genome. The design
hypothesis was that suppressing field correlations that conflict across
tasks, before the representation is shared, would reduce negative
transfer. The module first normalizes the field embeddings and estimates
cross-field correlations from the current batch. Two small networks map
these statistics to the factors of a field-routing matrix, and a learned
gate interpolates between this matrix and the identity. Channel
projections and a residual connection complete the update. Both the input
and the output are \texttt{field\_embeddings} with shape $[B,F,D]$, so the
module can be inserted between the feature path and the flattening step
without changing any downstream interface. The card (Listing~\ref{lst:app-tier2-card}) is shown as follows: 
 
\begin{lstlisting}[style=skillcard,caption={Tier-2 card example: a generated and promoted field-mixing module.},label={lst:app-tier2-card}]
skill_id: macro_field_covariance_shrinkage_v1
name: macro_field_covariance_shrinkage_v1
category: routing
task_types: [multitask]
failure_modes_addressed: [negative_transfer_between_tasks]
input_signature:
  - name: field_embeddings
    shape: [batch_size, num_fields, embedding_dim]
    dtype: float32
output_signature:
  - name: field_embeddings
    shape: [batch_size, num_fields, embedding_dim]
    dtype: float32
implementation_path: recskill/generated_skills/macro_field_covariance_shrinkage_v1.py
class_name: FieldCovarianceShrinkage
source_proposal_id: proposal_field_covariance_shrinkage_v1
validation_status: passed
validation_results:
  static_check: {passed: true, issues: []}
  unit_tests:
    passed: true
    checks: {import: true, instantiate: true, forward: true, backward: true}
    issues: []
  shape_tests:
    passed: true
    output_shapes: {field_embeddings: [2, 3, 16]}
    issues: []
  portability_tests:
    passed: true
portability:
  scope: schema_agnostic
  reuse_enabled: true
  constraints: {min_num_fields: 1}
composition:
  requires: [field_embeddings]
  produces: [field_embeddings]
  example_genome_fragment:
    skill: macro_field_covariance_shrinkage_v1
    params:
      covariance_rank: 8
      embedding_dim: ${embedding_dim}
      num_fields: ${num_fields}
      shrinkage_hidden_dim: 16
promotion_status: promoted
promoted_at: '2026-08-04T17:21:03+00:00'
candidate_id: round1_code_macro_field_covariance_shrinkage_v1_p0
candidate_status: survivor
candidate_metrics:
  validation_mean_auc: 0.9720098642670008
architecture_fingerprint: 1835d961a9028a03a224
\end{lstlisting}
 
Because the two cards follow the same tensor contract, the planner can wire
either skill to any compatible neighbor. The tiers differ in provenance and
in the evidence attached to it. A Tier-1 card records the source component
and how it was extracted; a Tier-2 card additionally records the generated
implementation, its verification results, and the candidate through which
it entered the library.

\FloatBarrier

\section{Skill-Space Operators and Compatibility Checks}
\label{app:operators}

Table~\ref{tab:operators} specifies the four skill-space operator categories of
Section~\ref{sec:skillspace} with their preconditions. Every precondition is checked
statically against the genome DAG and the skill cards' typed interfaces before a candidate is
constructed, so a proposal that violates one is rejected without consuming any training
budget.

\begin{table}[!htbp]
\caption{Typed mutation operators used by skill-space evolution.}
\label{tab:operators}
\centering
\small
\begin{tabular}{@{}p{0.16\linewidth}p{0.40\linewidth}p{0.36\linewidth}@{}}
\toprule
\textbf{Operator} & \textbf{Preconditions} & \textbf{Example} \\
\midrule
Add & The new skill's required input keys are present in $\mathcal{X}_G$ or producible by one
upstream hop, and its output key is accepted by the fusion layer. & Insert a target-aware
attention branch before the tower. \\
\hline
Replace & Input and output signatures are compatible, and the output key is preserved so
downstream wiring is untouched. & Replace mean pooling with target-aware attention; replace
the FM branch with identity (removal). \\
\hline
Hybridize & Two branches expose compatible fusion points and the merged graph remains acyclic
and typed. & Combine a DeepFM interaction branch with a DIN interest branch. \\
\hline
Specialize & Scenario or task metadata is available, and the change alters only
hyperparameters, not \texttt{skill\_id} or wiring. & Attach a domain-specific adapter; widen a
branch under a larger capacity budget. \\
\bottomrule
\end{tabular}
\end{table}

\FloatBarrier
\section{Evaluation Tasks, Datasets, and Models}
\label{app:task-data-baseline}

\subsection{Tasks and Evaluation Objectives}
\label{app:tasks}

We evaluate our framework on four representative recommendation tasks, covering
single-objective prediction, multi-task learning, multi-domain learning, and
joint quality--efficiency optimization.

\begin{itemize}
    \item \textbf{Click-Through Rate (CTR) Prediction.}
    Given a user--item pair with its contextual features, CTR prediction estimates
    the probability that the user interacts with (e.g., clicks) the item, which is
    formulated as binary classification. The model is trained with binary
    cross-entropy (BCE) loss and evaluated by AUC and LogLoss, following
    common CTR benchmarks~\citep{zhu2021open}.

    \item \textbf{Multi-Task Learning (MTL).}
    MTL jointly optimizes several related prediction objectives with a single model,
    where the tasks share part of the model's parameters so that knowledge transfers
    between them~\citep{caruana1997multitask,ma2018mmoe}. The key challenge is to
    balance useful sharing against negative transfer between tasks. The overall loss
    is the (weighted) sum of per-task BCE losses, and we report the AUC of each task.

    \item \textbf{Multi-Domain Learning (MDL).}
    Also known as multi-scenario recommendation, MDL serves samples from several
    domains (scenarios) with one unified model. The domains share the same feature
    space and label, but their data distributions differ~\citep{sheng2021one}.
    The model must capture both what the domains share and what is specific to each
    domain. Following~\citet{li2025scenario}, we report AUC for each domain and
    overall.

    \item \textbf{Multi-Objective Optimization: AUC Co-optimized with MFU.}
    Beyond predictive quality, industrial recommenders must also use hardware
    efficiently. In this task, a searched architecture is judged jointly by
    (i) its predictive quality (AUC) and (ii) its Model FLOPs Utilization
    (MFU)~\citep{chowdhery2023palm}. MFU is the ratio of achieved training
    FLOP/s to the hardware's theoretical peak FLOP/s.
    The goal is to find architectures that improve AUC without lowering MFU (or
    that reach a better AUC--MFU Pareto front).
\end{itemize}

\begin{table}[!htbp]
    \centering
    \footnotesize
    \renewcommand{\arraystretch}{1.15}
    \caption{Dataset statistics. For the six accuracy benchmarks, \emph{\# models} is the number
    of manually designed architectures in the comparison set. For QK-Video, it is the number of
    UniRank seed models used to compile the generative ranking skill library. Split sizes are
    numbers of interactions.}
    \label{tab:experimental-settings}
    \begin{tabular}{@{}lclccc@{}}
        \toprule
        \multirow{2}{*}{\textbf{Task}} &
        \multirow{2}{*}{\makecell{\textbf{\# Models}\\\textbf{or seeds}}} &
        \multirow{2}{*}{\textbf{Dataset}} &
        \multicolumn{3}{c}{\textbf{\# Interactions}} \\
        \cmidrule(lr){4-6}
        & & & \textbf{Train} & \textbf{Val} & \textbf{Test} \\
        \midrule
        \multirow{3}{*}{CTR}
            & \multirow{3}{*}{7} & MovieLens & 700.15K & 100.02K & 200.04K \\
            & & Beauty    & 1.20M   & 171.26K & 342.51K \\
            & & Books     & 13.76M  & 1.97M   & 3.93M   \\
        \midrule
        MTL & 5 & Census & 199.52K & 49.88K & 49.88K \\
        \midrule
        \multirow{2}{*}{MDL}
            & \multirow{2}{*}{11} & Amazon    & 658.83K & 82.35K & 82.35K \\
            & & MovieLens & 800.17K & 100.02K & 100.02K \\
        \midrule
        \makecell[l]{Multi-\\objective} & 15 & QK-Video & 394.65M & 50.08M & 48.58M \\
        \bottomrule
    \end{tabular}
\end{table}

\subsection{Datasets and Data Splits}
\label{app:datasets}

Table~\ref{tab:experimental-settings} summarizes the statistics of all datasets used
in the four tasks. The details are as follows.

\begin{itemize}
    \item \textbf{CTR task.}
    \begin{itemize}
        \item \textbf{MovieLens}~\citep{harper2015movielens}: a movie-rating dataset
        collected by GroupLens, containing user demographic features and movie
        attributes. We use the MovieLens-1M version.
        \item \textbf{Amazon Beauty} and \textbf{Amazon Books}~\citep{mcauley2015image,he2016ups}:
        two category subsets of the Amazon product review corpus, containing user
        ratings together with item metadata. Books is the largest CTR dataset in our
        study (over 19M interactions), so it tests whether searched architectures
        scale to large data. All three datasets are split into training, validation, and test sets
        with a ratio of 7:1:2.
    \end{itemize}

    \item \textbf{MTL task.}
    \begin{itemize}
        \item \textbf{Census-Income (KDD)}~\citep{census_income_kdd}: a UCI benchmark
        extracted from the 1994--1995 U.S. Current Population Surveys, with about
        300K records and 40 demographic and employment attributes. It is widely used
        to evaluate multi-task recommenders~\citep{ma2018mmoe,tang2020ple}.
    \end{itemize}

    \item \textbf{MDL task.}
    Both datasets follow the preprocessing protocol of the Scenario-Wise Rec
    benchmark~\citep{li2025scenario}, with an 8:1:1 split.
    \begin{itemize}
        \item \textbf{Amazon}~\citep{ni2019justifying}: three product categories
        (\emph{Clothing}, \emph{Beauty}, and \emph{Health}) are treated as three domains.
        \item \textbf{MovieLens}~\citep{harper2015movielens}: interactions are divided
        into three domains by user age group (\emph{1--24}, \emph{25--34}, and \emph{35+}).
    \end{itemize}

    \item \textbf{Multi-objective task.}
    \begin{itemize}
        \item \textbf{QK-Video}~\citep{yuan2022tenrec}: the largest sub-dataset of the
        Tenrec benchmark, collected from Tencent's QQ Kandian video-recommendation
        platform. It contains about 493M user--video interactions. Its industrial
        scale makes it well suited to measuring training efficiency (MFU) together
        with predictive quality.
    \end{itemize}
\end{itemize}

\subsection{Baseline and Seed Models}
\label{app:baselines}

For each task, the manually designed architectures listed below serve as manual baselines
(Table~\ref{tab:experimental-settings}).

\paragraph{CTR baselines (7 models).}
\begin{itemize}
    \item \textbf{Wide\&Deep}~\citep{cheng2016wide}: jointly trains a wide linear model
    (for memorization) and a deep MLP (for generalization).
    \item \textbf{DeepFM}~\citep{guo2017deepfm}: replaces the wide part of Wide\&Deep
    with a factorization machine (FM). The FM and the deep network share one embedding
    layer, so no manual feature engineering is needed.
    \item \textbf{DCN}~\citep{wang2017dcn}: adds a cross network that explicitly models
    bounded-degree feature interactions, in parallel with a deep network.
    \item \textbf{DCN-V2}~\citep{wang2021dcn}: improves DCN with a more expressive cross
    layer, and supports both stacked and parallel structures.
    \item \textbf{EDCN}~\citep{chen2021enhancing}: strengthens parallel CTR models by
    sharing information between the explicit and implicit interaction branches through
    bridge and regulation modules.
    \item \textbf{AFM}~\citep{xiao2017attentional}: extends FM with an attention network
    that learns a separate importance weight for each second-order feature interaction.
    \item \textbf{AutoInt}~\citep{song2019autoint}: learns high-order feature
    interactions automatically with stacked multi-head self-attention layers.
\end{itemize}

\paragraph{MTL baselines (5 models).}
\begin{itemize}
    \item \textbf{Shared-Bottom}~\citep{caruana1997multitask}: all tasks share the bottom
    layers, and each task has its own tower on top.
    \item \textbf{MMoE}~\citep{ma2018mmoe}: replaces the shared bottom with a set of
    shared experts, combined by task-specific gating networks.
    \item \textbf{PLE}~\citep{tang2020ple}: separates task-specific experts from
    shared experts and extracts representations progressively across levels, which
    reduces negative transfer and the seesaw phenomenon.
    \item \textbf{ESMM}~\citep{ma2018entire}: models post-view CTR and CTCVR over the
    entire sample space, which addresses sample selection bias and data sparsity in
    conversion-rate estimation.
    \item \textbf{AITM}~\citep{xi2021modeling}: models the sequential dependence among
    multi-step conversions with an adaptive information-transfer module based on
    attention.
\end{itemize}

\paragraph{MDL baselines (11 models).}
\begin{itemize}
    \item \textbf{Shared-Bottom}~\citep{caruana1997multitask}, \textbf{MMoE}~\citep{ma2018mmoe},
    and \textbf{PLE}~\citep{tang2020ple}: multi-task architectures adapted to
    MDL by treating each domain as a task.
    \item \textbf{STAR}~\citep{sheng2021one}: a star-topology network whose shared
    centered parameters are combined element-wise with domain-specific parameters.
    It also uses partitioned normalization and an auxiliary network.
    \item \textbf{SAR-Net}~\citep{shen2021sar}: a scenario-aware ranking network that
    transfers users' cross-scenario interests through scenario-specific attention
    and mixes scenario-specific and shared experts.
    \item \textbf{AdaSparse}~\citep{yang2022adasparse}: learns adaptively sparse network
    structures for each domain with lightweight domain-aware neuron-level weighting
    factors.
    \item \textbf{AdaptDHM}~\citep{li2022adaptdhm}: an adaptive distribution hierarchical
    model that clusters samples into distributions dynamically and routes each sample
    to the matching expert, without relying on predefined domain labels.
    \item \textbf{EPNet} and \textbf{PPNet}~\citep{chang2023pepnet}: the two components of
    PEPNet. EPNet personalizes embeddings with domain-side prior information, and PPNet
    personalizes the DNN parameters of each layer with user/item-side prior information
    via gate networks.
    \item \textbf{HAMUR}~\citep{li2023hamur}: a hyper adapter framework. A shared hyper-network
    dynamically generates the parameters of domain-specific adapters that plug into
    backbone models. We include a large and a small variant, which differ in model
    capacity.
    \item \textbf{M3oE}~\citep{zhang2024m3oe}: a multi-domain, multi-task mixture-of-experts
    framework with shared, domain-specific, and task-specific experts, fused through
    a two-level structure whose fusion weights are found by AutoML.
\end{itemize}

\paragraph{UniRank seed models for generative ranking (15 models).}
For this setting, the seed-model set comprises the 15 generative ranking
architectures listed by the UniRank benchmark~\citep{li2026unirank}
\footnote{\url{https://github.com/salmon1802/UniRank}}: OneTrans, RankMixer,
Zenith, HyFormer, MixFormer, TokenMixer, HiFormer, INFNet, EST, LONGER, HeMix,
UniMixer, TokenFormer, UltraHSTU, and SSR. These architectures are the source
models from which the generative ranking skill library is compiled, and each
is described in Appendix~\ref{app:auc_mfu}. RankMixer~\citep{zhu2025rankmixer}
is the seed backbone used to initialize the reported co-evolution run; the
other 14 seed models are source architectures rather than separately evaluated
baselines in the AUC--MFU table.

\FloatBarrier
\section{Training, Search, and Hardware Settings}
\label{app:implementation}
 
The \evoskillrec\ experiments fall into two groups. The first group runs accuracy-oriented
architecture search on the CTR, multi-task (MTL), and multi-domain (MDL)
benchmarks, where every candidate is \textbf{trained on a single GPU} and selected by
validation AUC. The second group runs AUC--MFU co-evolution on the
industrial-scale QK-Video corpus, where every candidate is \textbf{trained with multi-GPU} data parallelism and selected jointly on validation AUC and measured hardware
efficiency. The two groups share the search procedure and the LLM pipeline,
but differ in data scale, training recipe, and selection rule. All settings
are fixed within a search run and are identical for every candidate in that
run. The two automated baselines are described separately in
Section~\ref{app:baseline-implementation}.
 
% ==================================================================
\subsection{Model Training and Initial Architectures}
\label{app:training_config}
 
\textbf{Shared recipe}
The \evoskillrec\ candidate models are implemented in PyTorch and trained from scratch with Adam at a learning rate of $10^{-3}$, without learning-rate schedules or warmup. Each
prediction target is trained with binary cross-entropy; for multi-task
datasets the losses of all tasks are summed with equal weights.
Table~\ref{tab:app_train_shared} compares the remaining settings of the two
experiment groups.
 
\begin{table}[!htbp]
\centering
\small
\caption{Training settings shared by \evoskillrec\ candidates in each experiment group.
MDL: multi-domain learning.}
\label{tab:app_train_shared}
\setlength{\tabcolsep}{4pt}
\begin{tabular}{lll}
\toprule
Setting & CTR / MTL / MDL benchmarks & AUC--MFU co-evolution \\
\midrule
Optimizer & Adam, lr $10^{-3}$ & Adam, lr $10^{-3}$ \\
Weight decay & dataset-specific (Table~\ref{tab:app_train_bench}) & 0 \\
Batch size & dataset-specific (Table~\ref{tab:app_train_bench}) & 8,192 per GPU \\
Epochs & dataset-specific, with early stopping & 1 \\
Gradient clipping & - & global norm 10 \\
Precision & FP32 & BF16 autocast \\
Embedding dimension & 16 (CTR, MDL); 8 (MTL) & 16 \\
Model selection & best validation AUC & validation mean AUC over tasks \\
Reported metrics & AUC, logloss & AUC, gAUC, logloss \\
Random seed & 2022 & 2026 \\
\bottomrule
\end{tabular}
\end{table}
 
\textbf{CTR, MTL, and MDL benchmarks.}
Table~\ref{tab:app_train_bench} lists the dataset-specific settings.
Training stops when the selection metric does not improve for the given
number of epochs, and the best checkpoint is restored before test evaluation.
The selection metric is validation AUC for CTR, the mean validation AUC of
the income and marital-status tasks for Census-Income, and the global
validation AUC over all domains for MDL (per-domain AUC is reported
separately). CTR datasets are split chronologically into 70\%/10\%/20\% for
training, validation, and test. Census-Income uses the standard prepared
splits with 32 categorical and 7 continuous features. The MDL datasets use a
random 80\%/10\%/10\% split with three domains.
 
\begin{table}[!htbp]
\centering
\small
\caption{Dataset-specific training settings on the CTR, MTL, and MDL
benchmarks. Epochs is the maximum number of epochs; patience is the
early-stopping patience in epochs.}
\label{tab:app_train_bench}
\setlength{\tabcolsep}{5pt}
\begin{tabular}{llrrcr}
\toprule
Task & Dataset & Batch size & Epochs & Weight decay & Patience \\
\midrule
CTR & MovieLens-1M  & 1,024 & 5 & $3\times10^{-4}$ & 3 \\
CTR & Amazon-Beauty & 8,192 & 2 & $3\times10^{-4}$ & 2 \\
CTR & Amazon-Books  & 8,192 & 2 & $3\times10^{-4}$ & 2 \\
MTL & Census-Income & 2,048 & 5 & $10^{-4}$ & 3 \\
MDL & MovieLens     & 4,096 & 1 & $10^{-5}$ & 3 \\
MDL & Amazon        & 4,096 & 1 & $10^{-5}$ & 3 \\
\bottomrule
\end{tabular}
\end{table}
 
The \evoskillrec\ search starts from a standard model for each task (Table~\ref{tab:seeds}):
DeepFM for CTR, Shared-Bottom for MTL, and SAR-Net for MDL on both MovieLens and Amazon.
The CTR seed uses hidden widths $(256,128,64)$, ReLU, and dropout 0.2. The
MTL seed uses a shared backbone of $(128,64)$, task heads with one hidden
layer of width 32, ReLU, and dropout 0.1. On both MDL datasets, the SAR-Net seed
uses eight shared experts and two domain-specific experts of dimension 16,
with no dropout. These values define only the starting point; evolution is
free to change the structure and the architecture-specific parameters.
 
\textbf{AUC--MFU co-evolution.}
Table~\ref{tab:app_train_industrial} summarizes the industrial-scale dataset
used here. The user ID is the grouping key for gAUC. Every candidate is
trained for one epoch and evaluated once at the end of the epoch. The search
starts from RankMixer, which on QK-Video uses three RankMixer blocks with
expansion ratio 2, four non-sequential tokens of dimension 128, and task
towers of widths $(128,64)$. User behavior sequences are truncated or padded
to length 100 and processed by target attention with hidden widths
$(256,128)$, Dice activation, and no dropout or batch normalization. The
resulting model has 141.7M parameters, of which 140.0M are embeddings and
1.66M are dense.
 
\begin{table}[!htbp]
\centering
\small
\caption{The industrial-scale dataset used for AUC--MFU co-evolution, together
with the RankMixer token dimension used on it. Sample counts are
taken from the parquet metadata.}
\label{tab:app_train_industrial}
\begin{tabular}{lr}
\toprule
Dataset        & QK-Video \\
\midrule
Train          & 394,647,318 \\
Validation     & 50,079,027 \\
Test           & 48,579,958 \\
Tasks          & 4 \\
Feature fields & 10 \\
Token dim      & 128 \\
\bottomrule
\end{tabular}
\end{table}
 
% ==================================================================
\subsection{Search Budgets and Survivor Selection}
\label{app:evolution_config}
 
Each \evoskillrec\ generation evaluates a fixed budget of $N$ candidates, split between
skill-space and code-space proposals. Skill-space proposals apply one of four
operators (add, replace, hybridize, specialize) to retrieved skill cards.
Code-space proposals are macro-level modules produced by the LLM and wired
into the parent genome either between two existing nodes or in place of one
node, with input and output shapes preserved. Before training, each generated
module is checked for syntax, importability, tensor-shape compatibility, and
genome validity. Candidates whose architecture fingerprint matches a
previously evaluated architecture, including those stored in the evolution
memory, are skipped. A code-space module becomes a reusable skill only after
its candidate survives selection and passes the portability check.
 
\textbf{Budget controller.}
The split of $N$ between the two spaces is adjusted online. If the best
validation score has not improved by at least $10^{-3}$ for two consecutive
generations, the controller moves budget from skill space to code space, up
to the maximum code quota; any qualifying improvement resets it to the
initial split. $N$ itself never changes. When the LLM returns fewer valid
proposals than its quota, skill-space candidates fill the remaining slots.
Table~\ref{tab:app_evo} lists the per-run values.
 
\begin{table}[!htbp]
\centering
\small
\caption{Evolution settings for each \evoskillrec\ search run. B/B denotes Amazon-Beauty
and Amazon-Books, which share one configuration. Quotas are given as numbers
of candidates per generation.}
\label{tab:app_evo}
\setlength{\tabcolsep}{4pt}
\begin{tabular}{lrrrrrr}
\toprule
 & \multicolumn{2}{c}{CTR} & MTL & \multicolumn{2}{c}{MDL} & Co-evo. \\
\cmidrule(lr){2-3}\cmidrule(lr){5-6}
Setting & ML-1M & B/B & Census & ML & Amazon & QK-Video \\
\midrule
Generations               & 25 & 25 & 25 & 25 & 25 & 30 \\
Candidate budget $N$      &  8 &  8 &  8 &  8 &  8 &  6 \\
Survivors kept $K$        &  4 &  3 &  4 &  4 &  4 &  2 \\
Initial skill / code quota & 6/2 & 6/2 & 6/2 & 6/2 & 6/2 & 4/2 \\
Code quota increment      &  2 &  1 &  2 &  1 &  1 &  2 \\
Maximum code quota        &  6 &  6 &  6 &  6 &  6 &  4 \\
Minimum skill quota       &  2 &  2 &  2 &  2 &  2 &  2 \\
Stagnation patience       &  2 &  2 &  2 &  2 &  2 &  2 \\
\bottomrule
\end{tabular}
\end{table}
 
\textbf{Survivor selection.}
On the CTR, MTL, and MDL benchmarks, survivor selection follows
Algorithm~\ref{alg:loop}. Let $s(G)$ denote the task's validation AUC metric and
$\bar{s}_t = \mathrm{mean}_{G \in \mathcal{Q}_t} s(G)$ the generation mean.
The survivor set $\Sigma_t$ consists of the top-$K$ candidates by $s(G)$ among
those satisfying $s(G) \ge \bar{s}_t - \delta$, where $\delta$ is the tolerance
in Algorithm~\ref{alg:loop}. These survivors become the next generation's
parents, and only generated skills carried by a survivor are promoted to
Tier-2. If no candidate survives, the current parents are kept.
In co-evolution, selection is made on a constrained Pareto front of
validation mean AUC and end-to-end training MFU. A candidate is eligible only
if its validation AUC is within 0.001 of the incumbent best and its training
MFU is at least 95\% of the baseline's, with a stable MFU measurement (see
Section~\ref{app:hardware_cost}). Candidates without a valid hardware
measurement are discarded. Test metrics are never used for selection in
either setting.
 
% ==================================================================
\subsection{Hardware Setup and MFU Measurement}
\label{app:hardware_cost}
 
All experiments run on a single node with eight NVIDIA A800-SXM4 GPUs (80\,GB
each) and two Intel Xeon Platinum 8358 CPUs. On the CTR, MTL, and MDL
benchmarks, each candidate is trained on one GPU, and up to eight candidates
are evaluated in parallel. In co-evolution, each candidate is trained with
distributed data parallelism on six GPUs.
 
\paragraph{Measuring model FLOPs utilization.}
MFU is the ratio between achieved training FLOP/s and the BF16 dense peak of
the A800 (312\,TFLOP/s per GPU). FLOPs per sample are counted on a single
profiled training step after 30 warm-up steps. Step time is then measured
over 150 steps split into three windows, and in each window the slowest rank
determines the step time. The measurement is end-to-end: it includes data
loading and all training-loop overhead. We accept a measurement only if the
coefficient of variation of MFU across windows is at most 0.05.
Table~\ref{tab:app_hw_baseline} reports the resulting values for the
RankMixer baseline on QK-Video.
 
\begin{table}[!htbp]
\centering
\small
\caption{Hardware measurements of the RankMixer baseline on QK-Video
(6 GPUs, BF16, batch size 8,192 per GPU).}
\label{tab:app_hw_baseline}
\begin{tabular}{lr}
\toprule
Quantity & Value \\
\midrule
FLOPs per sample     & 58.67\,M \\
Training MFU         & 1.43\% \\
Step time            & 107.5\,ms \\
Peak memory per GPU  & 8.12\,GB \\
\bottomrule
\end{tabular}
\end{table}
 
The baseline MFU is low, which is common for embedding-heavy ranking models
whose dense part is small. This is the main reason we optimize MFU rather
than FLOPs alone.
 
% ==================================================================
\subsection{LLM Settings for Planning and Code Generation}
\label{app:llm_config}
 
Code-space proposals are generated in two stages. A planner receives the
parent genome, task and tensor dimensions, recent validation and hardware
metrics, records of failed candidates, the current budget state, and the
evolution history, and returns a set of architecture sketches. A synthesizer
turns selected sketches into JSON proposals, each containing a PyTorch
module, its tensor signature, the wiring location in the genome, and skill
metadata. The LLM is called separately for each parent that receives
code-space budget, so the total number of calls per generation depends on
the controller state. Generated modules may not read or write files, spawn
processes, or access the network. If the LLM fails to return a valid
proposal after the allowed attempts, no template-based fallback is used and
the slot is filled from skill space. See Table~\ref{tab:app_llm}
for our settings. 
\begin{table}[!htbp]
\centering
\small
\caption{LLM settings. Sketches and proposals are requested per parent that
receives code-space budget. Decoding parameters not set by our wrapper
follow the provider defaults.}
\label{tab:app_llm}
\setlength{\tabcolsep}{4pt}
\begin{tabular}{lll}
\toprule
Setting & CTR / MTL / MDL & Co-evolution \\
\midrule
Backend / model & Codex CLI, GPT-5.5 & Codex CLI, GPT-5.5 \\
Stages & planner $\rightarrow$ synthesizer & planner $\rightarrow$ synthesizer \\
Sketches requested & 8 (ML-1M CTR), 12 (others) & 8 \\
Proposals requested & 3 (B/B CTR), 2 (others) & 2 \\
Attempts per stage & 1 (CTR), 2 (MTL, MDL) & 2 \\
Timeout per call & 300\,s & 900\,s \\
Output format & one JSON object per call & one JSON object per call \\
Temperature / top-$p$ / max tokens & provider default & provider default \\
\bottomrule
\end{tabular}
\end{table}
\FloatBarrier
\input{appendix_baseline_implementation}
\FloatBarrier

\section{Detailed Experimental Results}
\label{app:results}

\subsection{Baseline Comparisons on the Six Accuracy Benchmarks}
\label{app:seed_results}

This appendix gives the full results behind Table~\ref{tab:rq1} on all six datasets.
Table~\ref{tab:ctr_baselines} covers CTR prediction, Table~\ref{tab:mtl_census} covers MTL on
Census, and Tables~\ref{tab:mdl_amazon_baselines} and~\ref{tab:mdl_movielens_baselines}
cover MDL on Amazon and MovieLens. Each table lists all manual models together with EvoSkillRec.

EvoSkillRec achieves the best performance on all six datasets, outperforming every manually designed model and every automated baseline. Relative to the strongest manual model, it improves AUC by 0.0035, 0.0121, and 0.0179 on the three CTR datasets, by 0.0022 in mean AUC on Census, and by 0.0021 and 0.0008 on the two MDL datasets. These consistent gains across all datasets show that the proposed framework can discover models that outperform expert-designed architectures.

\begin{table}[!htbp]
\centering
\caption{CTR prediction results on three datasets. AUC is higher-better ($\uparrow$) and
Logloss is lower-better ($\downarrow$). The best manual-model result in each column is in
\textbf{bold} and the second best is \underline{underlined}.}
\label{tab:ctr_baselines}
\small
\begin{tabular}{@{}lcccccc@{}}
\toprule
& \multicolumn{2}{c}{\textbf{MovieLens}} & \multicolumn{2}{c}{\textbf{Amazon Books}}
& \multicolumn{2}{c}{\textbf{Amazon Beauty}} \\
\cmidrule(lr){2-3} \cmidrule(lr){4-5} \cmidrule(lr){6-7}
\textbf{Model} & AUC~$\uparrow$ & Logloss~$\downarrow$ & AUC~$\uparrow$ & Logloss~$\downarrow$
& AUC~$\uparrow$ & Logloss~$\downarrow$ \\
\midrule
\multicolumn{7}{@{}l}{\textit{Manual models}} \\
Wide \& Deep  & \textbf{0.7487} & 0.5921             & \textbf{0.6587} & 0.5073 & 0.6565 & 0.5052 \\
DeepFM        & 0.7482 & 0.5921             & 0.6575 & 0.5024             & 0.6566 & 0.5107 \\
DCN           & 0.7484 & 0.5895             & 0.6573 & 0.5048             & 0.6570 & 0.4839 \\
DCN-V2        & 0.7479 & 0.5904             & 0.6571 & \textbf{0.4953}    & 0.6612 & 0.5404 \\
EDCN          & \underline{0.7487}& \underline{0.5880} & 0.6567 & 0.4960             & \underline{0.6639}& \textbf{0.4585} \\
AFM           & 0.7474 & \textbf{0.5869}    & 0.6562 & \underline{0.4958} & 0.6254 & \underline{0.4714} \\
AutoInt       & 0.7455 & 0.5932             & \underline{0.6579}& 0.5029             & \textbf{0.6666} & 0.5629 \\
\midrule
\textbf{EvoSkillRec}   & 0.7522    & 0.5857 & 0.6708    & 0.4952 & 0.6845    & 0.4441 \\
\bottomrule
\end{tabular}
\end{table}

\begin{table}[!htbp]
\centering
\caption{Multi-task results on Census, reported as test AUC ($\uparrow$). Income and
Marital prediction are regarded as two tasks; Mean AUC is averaged
over the two shared tasks. The best manual-model result in each column is in \textbf{bold} and the
second best is \underline{underlined}; tied entries are marked together.}
\label{tab:mtl_census}
\small
\begin{tabular}{@{}lccc@{}}
\toprule
\textbf{Model} & \textbf{Income (CVR)} & \textbf{Marital (CTR)} & \textbf{Mean AUC} \\
\midrule
\multicolumn{4}{@{}l}{\textit{Manual models}} \\
AITM           & \textbf{0.9507}    & 0.9944              & \underline{0.9726} \\
PLE            & \underline{0.9502} & \textbf{0.9946}     & 0.9724 \\
Shared-Bottom  & 0.9492             & 0.9946     & 0.9719 \\
MMoE           & 0.9485             & \underline{0.9945}  & 0.9715 \\
ESMM           & 0.8463             & 0.9912              & 0.9188 \\
\midrule
\textbf{EvoSkillRec}    & 0.9550 & 0.9946 & 0.9748 \\
\bottomrule
\end{tabular}
\end{table}

\begin{table}[!htbp]
\centering
\caption{Multi-domain results on Amazon. We report overall test AUC and Logloss (LL), as well
as per-domain test AUC. Manual models are sorted by validation AUC, which is used for
model selection. Among the manual models, the best result in each column is in
\textbf{bold} and the second best is \underline{underlined}.}
\label{tab:mdl_amazon_baselines}
\small
\setlength{\tabcolsep}{5pt}
\begin{tabular}{@{}lccccc@{}}
\toprule
& \multicolumn{2}{c}{\textbf{Overall}} & \multicolumn{3}{c}{\textbf{Per-domain AUC}~$\uparrow$} \\
\cmidrule(lr){2-3} \cmidrule(lr){4-6}
\textbf{Model} & AUC~$\uparrow$ & LL~$\downarrow$ & $D_0$ & $D_1$ & $D_2$ \\
\midrule
\multicolumn{6}{@{}l}{\textit{Manual models}} \\
SAR-Net       & \textbf{0.7012}    & \textbf{0.4632}    & \textbf{0.6970}    & \textbf{0.6768}    & \textbf{0.7204} \\
M3oE          & \underline{0.6905} & \underline{0.4678} & \underline{0.6859} & \underline{0.6752} & 0.7026 \\
AdaSparse     & 0.6876 & 0.4681 & 0.6800 & 0.6636 & \underline{0.7080} \\
MMoE          & 0.6756 & 0.4765 & 0.6669 & 0.6520 & 0.6973 \\
PPNet         & 0.6771 & 0.4720 & 0.6765 & 0.6458 & 0.6991 \\
HAMUR         & 0.6748 & 0.4778 & 0.6718 & 0.6484 & 0.6971 \\
Shared-Bottom & 0.6748 & 0.4732 & 0.6626 & 0.6498 & 0.6986 \\
EPNet         & 0.6751 & 0.4772 & 0.6690 & 0.6559 & 0.6919 \\
PLE           & 0.6719 & 0.4749 & 0.6647 & 0.6502 & 0.6969 \\
STAR          & 0.6690 & 0.4749 & 0.6680 & 0.6397 & 0.6883 \\
AdaptDHM      & 0.6305 & 0.5052 & 0.6331 & 0.6079 & 0.6504 \\
\midrule
\textbf{EvoSkillRec}   & 0.7033 & 0.4624 & 0.6963 & 0.6780 & 0.7240 \\
\bottomrule
\end{tabular}
\end{table}

\begin{table}[!htbp]
\centering
\caption{Multi-domain results on MovieLens. We report overall test AUC and Logloss (LL), as
well as per-domain test AUC. Manual models are sorted by validation AUC, which is used
for model selection. Among the manual models, the best result in each column is in
\textbf{bold} and the second best is \underline{underlined}.}
\label{tab:mdl_movielens_baselines}
\small
\setlength{\tabcolsep}{5pt}
\begin{tabular}{@{}lccccc@{}}
\toprule
& \multicolumn{2}{c}{\textbf{Overall}} & \multicolumn{3}{c}{\textbf{Per-domain AUC}~$\uparrow$} \\
\cmidrule(lr){2-3} \cmidrule(lr){4-6}
\textbf{Model} & AUC~$\uparrow$ & LL~$\downarrow$ & $D_0$ & $D_1$ & $D_2$ \\
\midrule
\multicolumn{6}{@{}l}{\textit{Manual models}} \\
AdaSparse     & \underline{0.7920} & 0.5393             & \underline{0.7974} & \textbf{0.7961}    & 0.7856 \\
SAR-Net       & \textbf{0.7924}    & \textbf{0.5384}    & 0.7970             & 0.7953             & \textbf{0.7869} \\
EPNet         & 0.7917             & \underline{0.5390} & 0.7972             & \underline{0.7960} & \underline{0.7866} \\
M3oE          & 0.7912 & 0.5419 & 0.7945 & 0.7950 & 0.7852 \\
Shared-Bottom & 0.7918 & 0.5399 & 0.7964 & 0.7952 & 0.7841 \\
MMoE          & 0.7905 & 0.5432 & \textbf{0.7975} & 0.7948 & 0.7835 \\
HAMUR         & 0.7900 & 0.5423 & 0.7965 & 0.7953 & 0.7849 \\
PLE           & 0.7890 & 0.5420 & 0.7961 & 0.7949 & 0.7851 \\
PPNet         & 0.7889 & 0.5422 & 0.7967 & 0.7955 & 0.7791 \\
STAR          & 0.7846 & 0.5470 & 0.7866 & 0.7891 & 0.7784 \\
AdaptDHM      & 0.7805 & 0.5553 & 0.7802 & 0.7851 & 0.7744 \\
\midrule
\textbf{EvoSkillRec}   & 0.7932 & 0.5371 & 0.7971 & 0.7970 & 0.7871 \\
\bottomrule
\end{tabular}
\end{table}

\subsection{Architecture Changes and Scores by Generation}
\label{app:evolution_details}
 
This appendix shows how EvoSkillRec evolves the genome, generation by generation, on all
six datasets. Table~\ref{tab:seeds} lists the seed model of each run.
Tables~\ref{tab:movielens_trace}--\ref{tab:books_trace} cover CTR prediction on MovieLens,
Amazon-Beauty, and Amazon-Books, Table~\ref{tab:census_trace} covers MTL on Census, and
Tables~\ref{tab:amazon_mdl_trace} and~\ref{tab:movielens_mdl_trace} cover MDL on Amazon and
MovieLens-1M. Each table lists the best surviving candidate of every generation, the space
it comes from, the mutation applied, and its validation and test scores. A generation is
omitted from a table when none of its mutations produced a valid candidate, so the
generation indices in these tables are not always consecutive.
 
The mutation operators are as follows. In the skill space, \emph{specialize} adapts an
existing Tier-1 path, \emph{add} inserts a library skill as a new node, \emph{replace}
substitutes one module of the genome with an interface-compatible Tier-1 module while leaving the
rest of the genome unchanged (its argument names the architecture the module comes from; e.g.,
\texttt{sarnet} denotes SAR-Net's expert-mixer module, \texttt{sarnet\_expert\_mixer}),
\emph{hybridize} recombines two Tier-1 branches, and \emph{reuse} draws a promoted Tier-2
skill. In the code space, \emph{invent} writes a new
module from scratch. Selection uses validation AUC only; test scores are reported but never
used in the search. A \cmark{} marks a generation whose validation AUC beats the best so
far, seed included.
 
Three patterns hold across the runs. First, both spaces produce large jumps in validation AUC.
Code-space inventions account for several of them (e.g., generations~3 and~9 on Amazon-Beauty
and generations~0 and~4 on Amazon-Books), and skill-space steps account for others (e.g., the
module-level \emph{replace}~\texttt{sarnet} at generation~12 on Amazon MDL). Second, skills
used early in a run often appear again in later generations.
On MovieLens, \texttt{autoint\_\allowbreak senet\_\allowbreak bridge\_\allowbreak branch\_\allowbreak r3}
is invented at generation~3 and gives the best test AUC and logloss when reused at
generation~6. On Census, \texttt{macro\_\allowbreak field\_\allowbreak covariance\_\allowbreak shrinkage\_\allowbreak v1},
used in the skill-space candidate at generation~0, appears in most later skill-space rows. Third, because selection uses
validation data only, a higher validation score does not always carry over to test.
\begin{table}[!htbp]
\centering
\caption{Seed architecture for each task and dataset. Every run initializes the genome
with an established architecture and evolves from there.}
\label{tab:seeds}
\small
\begin{tabular}{@{}lll@{}}
\toprule
\textbf{Task} & \textbf{Dataset} & \textbf{Seed architecture} \\
\midrule
CTR prediction        & MovieLens      & DeepFM \\
                      & Amazon-Beauty  & DeepFM \\
                      & Amazon-Books   & DeepFM \\
\midrule
Multi-task learning   & Census         & Shared-Bottom \\
\midrule
Multi-domain learning & Amazon (5-core) & SAR-Net \\
                      & MovieLens-1M   & SAR-Net \\
\bottomrule
\end{tabular}
\end{table}
 
\begin{table}[!htbp]
\centering
\caption{Evolution trace for CTR prediction on MovieLens, seeded with DeepFM. Each row is the best surviving candidate of one generation. Operators and markers are defined in Appendix~\ref{app:evolution_details}.}
\label{tab:movielens_trace}
\scriptsize
\setlength{\tabcolsep}{4pt}
\begin{tabular}{@{}cl>{\raggedright\arraybackslash}p{6.2cm}cccc@{}}
\toprule
 &  &  & \multicolumn{1}{c}{\textbf{Val}} & \multicolumn{2}{c}{\textbf{Test}} &  \\
\cmidrule(lr){4-4} \cmidrule(lr){5-6}
\textbf{Gen.} & \textbf{Space} & \textbf{Mutation} & AUC & AUC & LL & \textbf{Imp.} \\
\midrule
0 & Skill & \textit{specialize} \texttt{senet\_\allowbreak fm} & 0.7570 & 0.7513 & 0.5843 & \cmark \\
1 & Skill & \textit{specialize} \texttt{senet\_\allowbreak fm} & 0.7570 & 0.7515 & 0.5845 &  \\
2 & Skill & \textit{add} \texttt{crossnet\_\allowbreak v2} & 0.7568 & 0.7519 & 0.5838 &  \\
3 & Code & \textit{invent} \texttt{autoint\_\allowbreak senet\_\allowbreak bridge\_\allowbreak branch\_\allowbreak r3} & 0.7566 & 0.7518 & 0.5836 &  \\
4 & Code & \textit{invent} \texttt{contextual\_\allowbreak subspace\_\allowbreak moe\_\allowbreak branch\_\allowbreak r4} & 0.7570 & 0.7520 & 0.5825 & \cmark \\
5 & Code & \textit{invent} \texttt{bilinear\_\allowbreak spline\_\allowbreak interaction\_\allowbreak branch\_\allowbreak r5} & 0.7571 & 0.7519 & 0.5828 & \cmark \\
6 & Skill & \textit{reuse} \texttt{autoint\_\allowbreak senet\_\allowbreak bridge\_\allowbreak branch\_\allowbreak r3} & 0.7574 & 0.7529 & 0.5817 & \cmark \\
7 & Skill & \textit{reuse} \texttt{autoint\_\allowbreak senet\_\allowbreak bridge\_\allowbreak branch\_\allowbreak r3} & 0.7566 & 0.7520 & 0.5831 &  \\
8 & Code & \textit{invent} \texttt{cin\_\allowbreak cross\_\allowbreak tower\_\allowbreak replace\_\allowbreak r9} & 0.7574 & 0.7515 & 0.5862 &  \\
9 & Code & \textit{invent} \texttt{field\_\allowbreak cumulant\_\allowbreak moment\_\allowbreak branch\_\allowbreak r10} & 0.7568 & 0.7516 & 0.5849 &  \\
10 & Skill & \textit{reuse} \texttt{field\_\allowbreak relation\_\allowbreak capsule\_\allowbreak branch\_\allowbreak r3} & 0.7577 & 0.7512 & 0.5863 & \cmark \\
11 & Skill & \textit{reuse} \texttt{autoint\_\allowbreak senet\_\allowbreak bridge\_\allowbreak branch\_\allowbreak r3} & 0.7572 & 0.7513 & 0.5864 &  \\
12 & Skill & \textit{reuse} \texttt{field\_\allowbreak relation\_\allowbreak capsule\_\allowbreak branch\_\allowbreak r3} & 0.7574 & 0.7523 & 0.5853 &  \\
13 & Skill & \textit{reuse} \texttt{oblivious\_\allowbreak cross\_\allowbreak tree\_\allowbreak branch\_\allowbreak r6} & 0.7570 & 0.7514 & 0.5855 &  \\
14 & Skill & \textit{reuse} \texttt{autoint\_\allowbreak senet\_\allowbreak bridge\_\allowbreak branch\_\allowbreak r3} & 0.7571 & 0.7515 & 0.5849 &  \\
15 & Skill & \textit{reuse} \texttt{autoint\_\allowbreak senet\_\allowbreak bridge\_\allowbreak branch\_\allowbreak r3} & 0.7572 & 0.7515 & 0.5850 &  \\
16 & Skill & \textit{reuse} \texttt{oblivious\_\allowbreak cross\_\allowbreak tree\_\allowbreak branch\_\allowbreak r6} & 0.7572 & 0.7519 & 0.5858 &  \\
17 & Skill & \textit{reuse} \texttt{oblivious\_\allowbreak cross\_\allowbreak tree\_\allowbreak branch\_\allowbreak r6} & 0.7576 & 0.7517 & 0.5859 &  \\
18 & Code & \textit{invent} \texttt{contrastive\_\allowbreak hyperedge\_\allowbreak bank\_\allowbreak replace\_\allowbreak r20} & 0.7576 & 0.7517 & 0.5859 &  \\
19 & Skill & \textit{reuse} \texttt{oblivious\_\allowbreak cross\_\allowbreak tree\_\allowbreak branch\_\allowbreak r6} & 0.7569 & 0.7516 & 0.5856 &  \\
20 & Skill & \textit{reuse} \texttt{oblivious\_\allowbreak cross\_\allowbreak tree\_\allowbreak branch\_\allowbreak r6} & 0.7575 & 0.7519 & 0.5857 &  \\
21 & Code & \textit{invent} \texttt{field\_\allowbreak tuple\_\allowbreak contrastive\_\allowbreak memory\_\allowbreak branch\_\allowbreak r22} & 0.7576 & 0.7517 & 0.5859 &  \\
22 & Code & \textit{invent} \texttt{neural\_\allowbreak bdd\_\allowbreak clause\_\allowbreak lattice\_\allowbreak replace\_\allowbreak r23} & 0.7579 & 0.7522 & 0.5857 & \cmark \\
23 & Skill & \textit{reuse} \texttt{field\_\allowbreak product\_\allowbreak kernel\_\allowbreak branch\_\allowbreak r11} & 0.7572 & 0.7511 & 0.5855 &  \\
\bottomrule
\end{tabular}
\end{table}
 
\begin{table}[!htbp]
\centering
\caption{Evolution trace for CTR prediction on Amazon-Beauty, seeded with DeepFM. Operators and markers are defined in Appendix~\ref{app:evolution_details}.}
\label{tab:beauty_trace}
\scriptsize
\setlength{\tabcolsep}{4pt}
\begin{tabular}{@{}cl>{\raggedright\arraybackslash}p{6.2cm}cccc@{}}
\toprule
 &  &  & \multicolumn{1}{c}{\textbf{Val}} & \multicolumn{2}{c}{\textbf{Test}} &  \\
\cmidrule(lr){4-4} \cmidrule(lr){5-6}
\textbf{Gen.} & \textbf{Space} & \textbf{Mutation} & AUC & AUC & LL & \textbf{Imp.} \\
\midrule
0 & Skill & \texttt{relation\_\allowbreak capsule\_\allowbreak router\_\allowbreak branch} & 0.6943 & 0.6755 & 0.3893 & \cmark \\
1 & Skill & \textit{reuse} \texttt{fourier\_\allowbreak pair\_\allowbreak kernel\_\allowbreak branch} & 0.6945 & 0.6747 & 0.3869 & \cmark \\
2 & Code & \textit{invent} \texttt{tensor\_\allowbreak train\_\allowbreak field\_\allowbreak interaction} & 0.6966 & 0.6761 & 0.3915 & \cmark \\
3 & Code & \textit{invent} \texttt{lowrank\_\allowbreak cross\_\allowbreak attention\_\allowbreak tower} & 0.7015 & 0.6809 & 0.4256 & \cmark \\
4 & Code & \textit{invent} \texttt{fourier\_\allowbreak volterra\_\allowbreak kernel\_\allowbreak branch} & 0.7012 & 0.6807 & 0.4264 &  \\
5 & Code & \textit{invent} \texttt{contextual\_\allowbreak branch\_\allowbreak token\_\allowbreak transformer\_\allowbreak fusion} & 0.7013 & 0.6803 & 0.4255 &  \\
6 & Skill & \textit{reuse} \texttt{contextual\_\allowbreak branch\_\allowbreak token\_\allowbreak transformer\_\allowbreak fusion} & 0.7014 & 0.6806 & 0.4260 &  \\
7 & Skill & \textit{hybridize} \texttt{crossnet\_\allowbreak afm} & 0.7014 & 0.6806 & 0.4260 &  \\
9 & Code & \textit{invent} \texttt{contextual\_\allowbreak field\_\allowbreak reparameterizer} & 0.7082 & 0.6843 & 0.4451 & \cmark \\
10 & Skill & \textit{reuse} \texttt{contextual\_\allowbreak branch\_\allowbreak token\_\allowbreak transformer\_\allowbreak fusion} & 0.7084 & 0.6847 & 0.4465 & \cmark \\
11 & Skill & \textit{hybridize} \texttt{crossnet\_\allowbreak afm} & 0.7082 & 0.6843 & 0.4451 &  \\
14 & Code & \textit{invent} \texttt{sobol\_\allowbreak tensor\_\allowbreak mixture\_\allowbreak cross\_\allowbreak branch} & 0.7087 & 0.6844 & 0.4423 & \cmark \\
15 & Skill & \textit{reuse} \texttt{symmetric\_\allowbreak bilinear\_\allowbreak crossnet\_\allowbreak branch} & 0.7088 & 0.6842 & 0.4420 & \cmark \\
16 & Skill & \textit{hybridize} \texttt{crossnet\_\allowbreak afm} & 0.7088 & 0.6842 & 0.4420 &  \\
18 & Skill & \textit{reuse} \texttt{adaptive\_\allowbreak crossnet\_\allowbreak expert\_\allowbreak branch\_\allowbreak 1} & 0.7090 & 0.6844 & 0.4463 & \cmark \\
19 & Skill & \textit{hybridize} \texttt{crossnet\_\allowbreak afm} & 0.7090 & 0.6844 & 0.4463 &  \\
24 & Skill & \textit{reuse} \texttt{adaptive\_\allowbreak crossnet\_\allowbreak expert\_\allowbreak branch\_\allowbreak 1} & 0.7092 & 0.6845 & 0.4442 & \cmark \\
\bottomrule
\end{tabular}
\end{table}

\begin{table}[!htbp]
\centering
\caption{Evolution trace for CTR prediction on Amazon-Books, seeded with DeepFM. Operators and markers are defined in Appendix~\ref{app:evolution_details}.}
\label{tab:books_trace}
\scriptsize
\setlength{\tabcolsep}{4pt}
\begin{tabular}{@{}cl>{\raggedright\arraybackslash}p{6.2cm}cccc@{}}
\toprule
 &  &  & \multicolumn{1}{c}{\textbf{Val}} & \multicolumn{2}{c}{\textbf{Test}} &  \\
\cmidrule(lr){4-4} \cmidrule(lr){5-6}
\textbf{Gen.} & \textbf{Space} & \textbf{Mutation} & AUC & AUC & LL & \textbf{Imp.} \\
\midrule
0 & Code & \textit{invent} \texttt{contextual\_\allowbreak mixture\_\allowbreak fusion} & 0.6849 & 0.6634 & 0.5088 & \cmark \\
1 & Skill & \textit{hybridize} \texttt{crossnet\_\allowbreak afm} & 0.6866 & 0.6579 & 0.5777 & \cmark \\
2 & Skill & \textit{hybridize} \texttt{crossnet\_\allowbreak afm} & 0.6862 & 0.6578 & 0.5756 &  \\
3 & Skill & \textit{hybridize} \texttt{crossnet\_\allowbreak afm} & 0.6873 & 0.6593 & 0.5849 & \cmark \\
4 & Code & \textit{invent} \texttt{contrastive\_\allowbreak branch\_\allowbreak consensus\_\allowbreak fusion} & 0.6940 & 0.6682 & 0.5599 & \cmark \\
5 & Skill & \textit{reuse} \texttt{field\_\allowbreak wise\_\allowbreak monomial\_\allowbreak lattice\_\allowbreak branch} & 0.6939 & 0.6663 & 0.5601 &  \\
6 & Code & \textit{invent} \texttt{branch\_\allowbreak field\_\allowbreak coattention\_\allowbreak fusion} & 0.6946 & 0.6699 & 0.5482 & \cmark \\
7 & Skill & \textit{hybridize} \texttt{crossnet\_\allowbreak afm} & 0.6946 & 0.6699 & 0.5482 &  \\
10 & Skill & \textit{reuse} \texttt{complex\_\allowbreak phase\_\allowbreak correlation\_\allowbreak branch} & 0.6947 & 0.6696 & 0.5488 & \cmark \\
11 & Skill & \textit{hybridize} \texttt{crossnet\_\allowbreak afm} & 0.6947 & 0.6696 & 0.5488 &  \\
12 & Skill & \textit{reuse} \texttt{hyperbolic\_\allowbreak user\_\allowbreak item\_\allowbreak metric\_\allowbreak branch} & 0.6950 & 0.6693 & 0.5490 & \cmark \\
13 & Skill & \textit{reuse} \texttt{hyperbolic\_\allowbreak user\_\allowbreak item\_\allowbreak metric\_\allowbreak branch} & 0.6950 & 0.6693 & 0.5490 &  \\
14 & Skill & \textit{reuse} \texttt{kernel\_\allowbreak memory\_\allowbreak pair\_\allowbreak compatibility\_\allowbreak branch} & 0.6958 & 0.6708 & 0.4952 & \cmark \\
15 & Skill & \textit{hybridize} \texttt{crossnet\_\allowbreak afm} & 0.6958 & 0.6705 & 0.4955 &  \\
16 & Skill & \textit{hybridize} \texttt{crossnet\_\allowbreak afm} & 0.6955 & 0.6702 & 0.5428 &  \\
19 & Skill & \textit{reuse} \texttt{neural\_\allowbreak outer\_\allowbreak product\_\allowbreak sketch\_\allowbreak branch} & 0.6958 & 0.6701 & 0.4967 &  \\
20 & Skill & \textit{reuse} \texttt{relation\_\allowbreak capsule\_\allowbreak router\_\allowbreak branch} & 0.6958 & 0.6698 & 0.5009 &  \\
21 & Skill & \textit{hybridize} \texttt{crossnet\_\allowbreak afm} & 0.6958 & 0.6700 & 0.5003 &  \\
\bottomrule
\end{tabular}
\end{table}
 
\begin{table}[!htbp]
\centering
\caption{Evolution trace for multi-task learning on Census, seeded with Shared-Bottom. Selection uses validation mean AUC over the two tasks; test scores are reported only. Per-task validation scores are omitted for space, and values are given to five decimals because differences are small. Operators and markers are defined in Appendix~\ref{app:evolution_details}.}
\label{tab:census_trace}
\scriptsize
\setlength{\tabcolsep}{3pt}
\begin{tabular}{@{}cl>{\raggedright\arraybackslash}p{4.4cm}ccccc@{}}
\toprule
 &  &  & \multicolumn{1}{c}{\textbf{Val}} & \multicolumn{3}{c}{\textbf{Test}} &  \\
\cmidrule(lr){4-4} \cmidrule(lr){5-7}
\textbf{Gen.} & \textbf{Space} & \textbf{Mutation} & Mean AUC & Mean AUC & Income AUC & Marital AUC & \textbf{Imp.} \\
\midrule
0 & Skill & \texttt{macro\_\allowbreak field\_\allowbreak covariance\_\allowbreak shrinkage\_\allowbreak v1} & 0.97240 & 0.97473 & 0.95460 & 0.99486 & \cmark \\
1 & Skill & \textit{specialize} \texttt{wider\_\allowbreak shared\_\allowbreak bottom} & 0.97228 & 0.97425 & 0.95367 & 0.99483 &  \\
2 & Skill & \textit{reuse} \texttt{macro\_\allowbreak field\_\allowbreak covariance\_\allowbreak shrinkage\_\allowbreak v1} & 0.97243 & 0.97477 & 0.95473 & 0.99482 & \cmark \\
3 & Skill & \textit{reuse} \texttt{macro\_\allowbreak field\_\allowbreak covariance\_\allowbreak shrinkage\_\allowbreak v1} & 0.97224 & 0.97455 & 0.95427 & 0.99483 &  \\
4 & Code & \textit{invent} \texttt{macro\_\allowbreak flat\_\allowbreak subspace\_\allowbreak expert\_\allowbreak router\_\allowbreak v1} & 0.97223 & 0.97438 & 0.95392 & 0.99485 &  \\
5 & Code & \textit{invent} \texttt{macro\_\allowbreak flat\_\allowbreak cross\_\allowbreak stage\_\allowbreak lattice\_\allowbreak v1} & 0.97231 & 0.97416 & 0.95357 & 0.99475 &  \\
6 & Skill & \textit{reuse} \texttt{macro\_\allowbreak field\_\allowbreak covariance\_\allowbreak shrinkage\_\allowbreak v1} & 0.97225 & 0.97449 & 0.95410 & 0.99489 &  \\
7 & Code & \textit{invent} \texttt{macro\_\allowbreak flat\_\allowbreak multiscale\_\allowbreak dilated\_\allowbreak mixer\_\allowbreak v1} & 0.97244 & 0.97427 & 0.95366 & 0.99488 & \cmark \\
8 & Code & \textit{invent} \texttt{macro\_\allowbreak flat\_\allowbreak block\_\allowbreak krylov\_\allowbreak resolvent\_\allowbreak v1} & 0.97252 & 0.97463 & 0.95437 & 0.99489 & \cmark \\
9 & Code & \textit{invent} \texttt{macro\_\allowbreak flat\_\allowbreak virtual\_\allowbreak task\_\allowbreak view\_\allowbreak fusion\_\allowbreak v1} & 0.97230 & 0.97452 & 0.95418 & 0.99486 &  \\
10 & Skill & \textit{reuse} \texttt{macro\_\allowbreak field\_\allowbreak covariance\_\allowbreak shrinkage\_\allowbreak v1} & 0.97237 & 0.97446 & 0.95408 & 0.99485 &  \\
11 & Code & \textit{invent} \texttt{macro\_\allowbreak flat\_\allowbreak butterfly\_\allowbreak expert\_\allowbreak lattice\_\allowbreak v1} & 0.97237 & 0.97442 & 0.95402 & 0.99482 &  \\
12 & Code & \textit{invent} \texttt{macro\_\allowbreak flat\_\allowbreak hypercomplex\_\allowbreak factor\_\allowbreak mixer\_\allowbreak v1} & 0.97238 & 0.97439 & 0.95389 & 0.99488 &  \\
13 & Code & \textit{invent} \texttt{macro\_\allowbreak flat\_\allowbreak sparse\_\allowbreak coordinate\_\allowbreak graph\_\allowbreak v1} & 0.97231 & 0.97460 & 0.95433 & 0.99486 &  \\
14 & Skill & \textit{reuse} \texttt{macro\_\allowbreak field\_\allowbreak covariance\_\allowbreak shrinkage\_\allowbreak v1} & 0.97244 & 0.97464 & 0.95441 & 0.99488 &  \\
15 & Skill & \textit{reuse} \texttt{macro\_\allowbreak field\_\allowbreak covariance\_\allowbreak shrinkage\_\allowbreak v1} & 0.97239 & 0.97449 & 0.95412 & 0.99486 &  \\
16 & Code & \textit{invent} \texttt{macro\_\allowbreak flat\_\allowbreak wavelet\_\allowbreak packet\_\allowbreak cross\_\allowbreak router\_\allowbreak v1} & 0.97241 & 0.97454 & 0.95422 & 0.99487 &  \\
17 & Skill & \textit{reuse} \texttt{macro\_\allowbreak field\_\allowbreak covariance\_\allowbreak shrinkage\_\allowbreak v1} & 0.97265 & 0.97482 & 0.95500 & 0.99464 & \cmark \\
18 & Skill & \textit{reuse} \texttt{macro\_\allowbreak field\_\allowbreak covariance\_\allowbreak shrinkage\_\allowbreak v1} & 0.97252 & 0.97465 & 0.95448 & 0.99481 &  \\
19 & Code & \textit{invent} \texttt{macro\_\allowbreak field\_\allowbreak cumulant\_\allowbreak conditioned\_\allowbreak channel\_\allowbreak mixer\_\allowbreak v1} & 0.97236 & 0.97398 & 0.95333 & 0.99463 &  \\
20 & Code & \textit{invent} \texttt{macro\_\allowbreak flat\_\allowbreak semiseparable\_\allowbreak scan\_\allowbreak bank\_\allowbreak v1} & 0.97250 & 0.97453 & 0.95415 & 0.99490 &  \\
21 & Code & \textit{invent} \texttt{macro\_\allowbreak flat\_\allowbreak oblique\_\allowbreak forest\_\allowbreak recomposer\_\allowbreak v1} & 0.97246 & 0.97447 & 0.95410 & 0.99484 &  \\
22 & Skill & \textit{reuse} \texttt{macro\_\allowbreak field\_\allowbreak covariance\_\allowbreak shrinkage\_\allowbreak v1} & 0.97243 & 0.97464 & 0.95446 & 0.99482 &  \\
23 & Skill & \textit{reuse} \texttt{macro\_\allowbreak field\_\allowbreak covariance\_\allowbreak shrinkage\_\allowbreak v1} & 0.97250 & 0.97418 & 0.95351 & 0.99486 &  \\
24 & Code & \textit{invent} \texttt{macro\_\allowbreak flat\_\allowbreak butterfly\_\allowbreak permutation\_\allowbreak fabric\_\allowbreak v1} & 0.97263 & 0.97448 & 0.95413 & 0.99483 &  \\
\bottomrule
\end{tabular}
\end{table}
 
\begin{table}[!htbp]
\centering
\caption{Evolution trace for multi-domain learning on Amazon (5-core), seeded with SAR-Net. $D_0$--$D_2$ denote test AUC on each domain; per-domain validation scores are omitted for space. Scores are displayed to four decimals, and \cmark{} denotes an update of the best-so-far validation candidate. The round-12 operation replaces the current expert mixer with the SAR-Net expert-mixer component, preserving the rest of the evolved genome. Operators and markers are defined in Appendix~\ref{app:evolution_details}.}
\label{tab:amazon_mdl_trace}
\scriptsize
\setlength{\tabcolsep}{3pt}
\begin{tabular}{@{}cl>{\raggedright\arraybackslash}p{4.3cm}ccccccc@{}}
\toprule
 &  &  & \multicolumn{1}{c}{\textbf{Val}} & \multicolumn{2}{c}{\textbf{Test}} & \multicolumn{3}{c}{\textbf{Test AUC by domain}} &  \\
\cmidrule(lr){4-4} \cmidrule(lr){5-6} \cmidrule(lr){7-9}
\textbf{Gen.} & \textbf{Space} & \textbf{Mutation} & AUC & AUC & LL & $D_0$ & $D_1$ & $D_2$ & \textbf{Imp.} \\
\midrule
0 & Code & \textit{invent} \texttt{domain\_\allowbreak aware\_\allowbreak field\_\allowbreak norm} & 0.6786 & 0.6796 & 0.4713 & 0.6740 & 0.6528 & 0.7002 & \cmark \\
1 & Skill & \textit{specialize} \texttt{ple\_\allowbreak more\_\allowbreak experts} & 0.6794 & 0.6801 & 0.4718 & 0.6779 & 0.6464 & 0.7029 & \cmark \\
2 & Code & \textit{invent} \texttt{domain\_\allowbreak modulated\_\allowbreak lowrank\_\allowbreak field\_\allowbreak rotation} & 0.6794 & 0.6801 & 0.4718 & 0.6779 & 0.6464 & 0.7029 &  \\
3 & Code & \textit{invent} \texttt{domain\_\allowbreak contrastive\_\allowbreak field\_\allowbreak residual} & 0.6817 & 0.6823 & 0.4706 & 0.6825 & 0.6477 & 0.7043 & \cmark \\
4 & Code & \textit{invent} \texttt{flat\_\allowbreak domain\_\allowbreak affine\_\allowbreak norm} & 0.6853 & 0.6846 & 0.4695 & 0.6820 & 0.6515 & 0.7074 & \cmark \\
5 & Code & \textit{invent} \texttt{field\_\allowbreak token\_\allowbreak self\_\allowbreak attention\_\allowbreak adapter} & 0.6856 & 0.6848 & 0.4692 & 0.6823 & 0.6520 & 0.7075 & \cmark \\
6 & Code & \textit{invent} \texttt{flat\_\allowbreak domain\_\allowbreak prototype\_\allowbreak residual} & 0.6846 & 0.6839 & 0.4697 & 0.6812 & 0.6508 & 0.7068 &  \\
7 & Code & \textit{invent} \texttt{flat\_\allowbreak sparse\_\allowbreak mixture\_\allowbreak adapter} & 0.6869 & 0.6872 & 0.4680 & 0.6840 & 0.6545 & 0.7102 & \cmark \\
8 & Code & \textit{invent} \texttt{rep\_\allowbreak cross\_\allowbreak domain\_\allowbreak contrastive\_\allowbreak residual} & 0.6866 & 0.6881 & 0.4676 & 0.6859 & 0.6552 & 0.7106 &  \\
9 & Skill & \textit{replace} \texttt{m3oe} & 0.6902 & 0.6894 & 0.4685 & 0.6874 & 0.6701 & 0.7041 & \cmark \\
10 & Code & \textit{invent} \texttt{field\_\allowbreak domain\_\allowbreak token\_\allowbreak affine\_\allowbreak residual} & 0.6901 & 0.6890 & 0.4687 & 0.6873 & 0.6693 & 0.7038 &  \\
11 & Code & \textit{invent} \texttt{flat\_\allowbreak domain\_\allowbreak moment\_\allowbreak normalizer} & 0.6927 & 0.6926 & 0.4672 & 0.6867 & 0.6749 & 0.7085 & \cmark \\
12 & Skill & \textit{replace} \texttt{sarnet\_\allowbreak expert\_\allowbreak mixer} & 0.7021 & 0.7012 & 0.4636 & 0.6946 & 0.6777 & 0.7233 & \cmark \\
13 & Code & \textit{invent} \texttt{field\_\allowbreak domain\_\allowbreak channel\_\allowbreak competition} & 0.7022 & 0.7013 & 0.4635 & 0.6949 & 0.6773 & 0.7233 & \cmark \\
14 & Code & \textit{invent} \texttt{flat\_\allowbreak domain\_\allowbreak hyper\_\allowbreak adapter} & 0.7026 & 0.7013 & 0.4630 & 0.6947 & 0.6775 & 0.7230 & \cmark \\
15 & Code & \textit{invent} \texttt{field\_\allowbreak domain\_\allowbreak cross\_\allowbreak stitch\_\allowbreak channel\_\allowbreak mixer} & 0.7024 & 0.7020 & 0.4631 & 0.6945 & 0.6774 & 0.7237 &  \\
16 & Code & \textit{invent} \texttt{flat\_\allowbreak domain\_\allowbreak residual\_\allowbreak mixture\_\allowbreak calibrator} & 0.7031 & 0.7018 & 0.4627 & 0.6945 & 0.6778 & 0.7231 & \cmark \\
17 & Code & \textit{invent} \texttt{flat\_\allowbreak domain\_\allowbreak squeeze\_\allowbreak excitation} & 0.7033 & 0.7019 & 0.4627 & 0.6947 & 0.6779 & 0.7232 & \cmark \\
18 & Code & \textit{invent} \texttt{flat\_\allowbreak domain\_\allowbreak sparse\_\allowbreak delta\_\allowbreak router} & 0.7032 & 0.7027 & 0.4629 & 0.6956 & 0.6782 & 0.7240 &  \\
19 & Code & \textit{invent} \texttt{field\_\allowbreak domain\_\allowbreak cross\_\allowbreak stitch\_\allowbreak residual} & 0.7032 & 0.7034 & 0.4619 & 0.6961 & 0.6783 & 0.7239 &  \\
20 & Code & \textit{invent} \texttt{field\_\allowbreak domain\_\allowbreak lowrank\_\allowbreak hyper\_\allowbreak token\_\allowbreak adapter} & 0.7031 & 0.7028 & 0.4624 & 0.6959 & 0.6780 & 0.7229 &  \\
21 & Code & \textit{invent} \texttt{field\_\allowbreak domain\_\allowbreak shared\_\allowbreak private\_\allowbreak splitter} & 0.7034 & 0.7033 & 0.4624 & 0.6963 & 0.6780 & 0.7240 & \cmark \\
\bottomrule
\end{tabular}
\end{table}
 
\begin{table}[!htbp]
\centering
\caption{Evolution trace for multi-domain learning on MovieLens-1M, seeded with SAR-Net.
$D_0$--$D_2$ denote test AUC on each domain; per-domain logloss is omitted for space.
Operators and markers are defined in
Appendix~\ref{app:evolution_details}.}
\label{tab:movielens_mdl_trace}
\scriptsize
\setlength{\tabcolsep}{3pt}
\begin{tabular}{@{}cl>{\raggedright\arraybackslash}p{4.3cm}ccccccc@{}}
\toprule
 &  &  & \multicolumn{1}{c}{\textbf{Val}} & \multicolumn{2}{c}{\textbf{Test}} & \multicolumn{3}{c}{\textbf{Test AUC by domain}} &  \\
\cmidrule(lr){4-4} \cmidrule(lr){5-6} \cmidrule(lr){7-9}
\textbf{Gen.} & \textbf{Space} & \textbf{Mutation} & AUC & AUC & LL & $D_0$ & $D_1$ & $D_2$ & \textbf{Imp.} \\
\midrule
0 & Code & \textit{invent} \texttt{flat\_\allowbreak domain\_\allowbreak feature\_\allowbreak gate} & 0.7947 & 0.7928 & 0.5379 & 0.7976 & 0.7953 & 0.7869 & \cmark \\
1 & Skill & \textit{replace} \texttt{m3oe} & 0.7930 & 0.7912 & 0.5419 & 0.7945 & 0.7950 & 0.7852 &  \\
2 & Skill & \textit{specialize} \texttt{ple\_\allowbreak more\_\allowbreak experts} & 0.7941 & 0.7923 & 0.5384 & 0.7970 & 0.7957 & 0.7864 &  \\
3 & Code & \textit{invent} \texttt{domain\_\allowbreak representation\_\allowbreak slot\_\allowbreak norm} & 0.7939 & 0.7923 & 0.5387 & 0.7962 & 0.7958 & 0.7849 &  \\
4 & Code & \textit{invent} \texttt{domain\_\allowbreak slot\_\allowbreak contextual\_\allowbreak lowrank\_\allowbreak adapter} & 0.7938 & 0.7919 & 0.5385 & 0.7961 & 0.7954 & 0.7860 &  \\
5 & Code & \textit{invent} \texttt{domain\_\allowbreak representation\_\allowbreak active\_\allowbreak slot\_\allowbreak cross\_\allowbreak attention} & 0.7942 & 0.7923 & 0.5395 & 0.7966 & 0.7957 & 0.7852 &  \\
7 & Code & \textit{invent} \texttt{flat\_\allowbreak domain\_\allowbreak basis\_\allowbreak residual} & 0.7942 & 0.7930 & 0.5384 & 0.7976 & 0.7961 & 0.7863 &  \\
8 & Code & \textit{invent} \texttt{flat\_\allowbreak domain\_\allowbreak cross\_\allowbreak basis\_\allowbreak adapter} & 0.7951 & 0.7932 & 0.5375 & 0.7970 & 0.7953 & 0.7878 & \cmark \\
9 & Code & \textit{invent} \texttt{flat\_\allowbreak domain\_\allowbreak blockwise\_\allowbreak calibrator} & 0.7949 & 0.7932 & 0.5378 & 0.7978 & 0.7959 & 0.7873 &  \\
10 & Code & \textit{invent} \texttt{field\_\allowbreak domain\_\allowbreak token\_\allowbreak fourier\_\allowbreak bias} & 0.7946 & 0.7929 & 0.5374 & 0.7968 & 0.7958 & 0.7872 &  \\
11 & Code & \textit{invent} \texttt{field\_\allowbreak domain\_\allowbreak cross\_\allowbreak token\_\allowbreak squeeze} & 0.7952 & 0.7927 & 0.5379 & 0.7969 & 0.7953 & 0.7870 & \cmark \\
12 & Code & \textit{invent} \texttt{flat\_\allowbreak domain\_\allowbreak hadamard\_\allowbreak expert\_\allowbreak residual} & 0.7950 & 0.7931 & 0.5374 & 0.7977 & 0.7959 & 0.7869 &  \\
14 & Code & \textit{invent} \texttt{flat\_\allowbreak domain\_\allowbreak residual\_\allowbreak temperature\_\allowbreak gate} & 0.7950 & 0.7933 & 0.5372 & 0.7975 & 0.7961 & 0.7867 &  \\
15 & Code & \textit{invent} \texttt{flat\_\allowbreak domain\_\allowbreak covariance\_\allowbreak residual\_\allowbreak smoother} & 0.7942 & 0.7922 & 0.5391 & 0.7981 & 0.7951 & 0.7857 &  \\
19 & Skill & \textit{replace} \texttt{adasparse} & 0.7926 & 0.7910 & 0.5395 & 0.7972 & 0.7972 & 0.7853 &  \\
22 & Skill & \textit{specialize} \texttt{deeper\_\allowbreak domain\_\allowbreak tower} & 0.7879 & 0.7881 & 0.5529 & 0.7956 & 0.7931 & 0.7810 &  \\
23 & Skill & \textit{replace} \texttt{mmoe} & 0.7953 & 0.7932 & 0.5371 & 0.7971 & 0.7970 & 0.7871 & \cmark \\
24 & Skill & \textit{replace} \texttt{shared\_\allowbreak bottom} & 0.7927 & 0.7917 & 0.5395 & 0.7971 & 0.7952 & 0.7840 &  \\
\bottomrule
\end{tabular}
\end{table}

\subsection{Generative Ranking on QK-Video}
\label{app:auc_mfu}

We evolve the RankMixer backbone to improve predictive quality (AUC) and hardware
efficiency (model FLOPs utilisation, MFU) at the same time. The skill library is
compiled (Section~\ref{sec:compiler}) from 15 generative ranking architectures:

\begin{itemize}
  \item \textbf{OneTrans}~\citep{zhang2026onetrans}: a single causal Transformer over
    one token stream of sequential and non-sequential features. Sequential tokens share
    parameters, non-sequential tokens have their own, and the KV cache is reused across
    requests.
  \item \textbf{RankMixer}~\citep{zhu2025rankmixer}: a hardware-aware design that
    replaces self-attention with parameter-free multi-head token mixing and per-token
    FFNs. A Sparse-MoE variant scales it to 1B parameters.
  \item \textbf{Zenith}~\citep{zhang2026zenith}: scales a small set of high-dimensional
    Prime Tokens with Token Fusion and Token Boost, at little extra inference latency.
  \item \textbf{HyFormer}~\citep{huang2026hyformer}: alternates Query Decoding, where
    global tokens decode long behaviour sequences, and Query Boosting, which mixes tokens
    across features.
  \item \textbf{MixFormer}~\citep{huang2026mixformer}: scales dense interaction and
    sequence modelling together in one shared-parameter block with a query mixer,
    cross-attention and output fusion.
  \item \textbf{TokenMixer}~\citep{jiang2026tokenmixer}: scales token mixing to
    15B parameters with mixing-and-reverting, inter-layer residuals, auxiliary losses and
    a sparse per-token MoE.
  \item \textbf{HiFormer}~\citep{gui2023hiformer}: heterogeneous self-attention with
    feature-specific projections, made servable by low-rank approximation and pruning.
  \item \textbf{INFNet}~\citep{li2025infnet}: per-group hub tokens gather context by
    cross-attention and broadcast it back to feature tokens, giving task-aware
    interaction at linear cost.
  \item \textbf{EST}~\citep{liu2026est}: treats all raw inputs as one sequence, with
    lightweight cross-attention for heterogeneous features and content-sparse attention
    for behaviours.
  \item \textbf{LONGER}~\citep{chai2025longer}: models ultra-long user sequences with
    global tokens, token merging through inner Transformers, and hybrid attention.
  \item \textbf{HeMix}~\citep{wang2026hemix}: extracts user interests with dynamic and
    fixed queries, and replaces self-attention with a HeteroMixer block for
    multi-granularity feature interaction.
  \item \textbf{UniMixer}~\citep{ha2026unimixer}: makes rule-based token mixing a
    learnable operator that covers attention-, TokenMixer- and FM-style interaction.
  \item \textbf{TokenFormer}~\citep{zhou2026tokenformer}: unifies multi-field and
    sequential tokens with bottom-full/top-sliding attention and multiplicative gating,
    which prevents representation collapse.
  \item \textbf{UltraHSTU}~\citep{ding2026bending}: co-designs the input sequence,
    sparse attention and topology of HSTU for better training and inference scaling.
  \item \textbf{SSR}~\citep{yu2026ssr}: replaces dense connectivity with
    filter-then-fuse blocks, which apply sparse dimension-level filtering over parallel
    views and then fuse densely within each subspace.
\end{itemize}

Table~\ref{tab:app-skill-inventory-unirank} lists the base skills obtained by
decomposing these architectures. The compiler breaks each model into reusable
components and keeps the generic modules they depend on, giving \textbf{92 skills in
10 categories}. This library is \textbf{separate from the one used in the CTR, MTL and
MDL experiments} and shares no skills with it. We build it \textbf{specifically for
generative ranking models}. These models work on unified token sequences and rely on
token mixing, attention and model-specific backbone blocks, so they need different
operators from classical feature-interaction models. \textbf{Interaction (25),
embedding (17) and attention (16)} make up \textbf{63\%} of the library. \textbf{54
skills are model-specific} and carry their source model's prefix. The other \textbf{38
are shared building blocks}, such as cross networks, pooling, activations and PyTorch
primitives.

\input{appendix/appendix_skill_inventory_unirank}

\paragraph{AUC--MFU score for comparison after search.}
To summarise the trade-off between ranking quality and hardware utilisation, we give
each evaluated architecture $c$ the score
\begin{equation}
\label{eq:compound_score}
\mathrm{Score}(c)
= \frac{1}{2}\,\frac{A(c)-A_{\min}}{\max(A_{\max}-A_{\min},\epsilon)}
+ \frac{1}{2}\,\frac{M(c)-M_{\min}}{\max(M_{\max}-M_{\min},\epsilon)},
\end{equation}
where $A(c)$ is the mean validation AUC across tasks and $M(c)$ is the measured
training MFU. For $X\in\{A,M\}$, $X_{\min}$ and $X_{\max}$ are the minimum and maximum
of $X$ over $\mathcal{C}$, the set of architectures with valid measurements in the
completed run, and $\epsilon=10^{-12}$ avoids division by zero. All candidates share
the same training budget and evaluation protocol. The bounds are fixed for the whole
run and are not recomputed for each generation, and test metrics are not used. The
score is only used for comparison after the run. Survivors are selected during
evolution by the constrained Pareto criterion.

\paragraph{Accuracy--efficiency trade-offs and search progress.}
Pushing either objective alone costs the other. The best-AUC candidates in
Table~\ref{tab:mfu_tradeoff} are all retokenizers. The signed-bucket retokenizer
(generation 14) gains $+6.7\times10^{-4}$ validation AUC but loses 8\% MFU, because
bucketing and hashing are memory-bound. The highest-MFU candidate (generation 7,
$+12\%$ MFU) falls below the baseline in AUC. Both are discarded, and every survivor
lies between 1.47\% and 1.57\% MFU. The Hadamard cross mixer appears in all five points
of the Pareto front over all eligible candidates (Table~\ref{tab:pareto_mfu}). Paired with the low-rank task-basis
router (P1), it is the only configuration that beats the baseline on validation AUC,
test AUC and MFU. At the other end, P5 raises MFU by 9.8\% over the baseline but gives
up $7\times10^{-4}$ validation AUC relative to P1.

The per-generation log (Table~\ref{tab:mfu_rounds}) shows three phases. Generations
0--5 make structural edits that raise MFU, such as per-token FFNs and Hadamard mixing.
Generations 7--16 try sequence modules placed before DIN, and retokenizers. They reach
the best AUC of the run (generation 14), but the retokenizers are discarded for their
MFU cost. From generation 17 onward the search moves toward task routing, and the
low-rank router (generation 19) improves both objectives. Of the 180 candidates, 26\%
survive and 23 skills are promoted to Tier-2. After generation 2 the budget shifts
from skill space (4/2) to code space (mostly 2/4), because the cheap recombinations
of the library have been used up. Failures concentrate late in the run (15 of 33 in
generations 23--29). Rising failures and fewer promotions could serve as a stopping
signal.

\begin{table}[!htbp]
\centering
\caption{Top candidates on the RankMixer backbone, sorted by validation AUC.
\textbf{Gen.}: generation; \textbf{St.}: outcome (S\,=\,survivor, D\,=\,discarded);
\textbf{Score}: \eqref{eq:compound_score}. Validation gAUC and logloss are omitted.
Best candidate value per column in \textbf{bold}.}
\label{tab:mfu_tradeoff}
\scriptsize
\setlength{\tabcolsep}{3pt}
\begin{tabular}{@{}lcccccccc@{}}
\toprule
& & & \textbf{Val} & \multicolumn{3}{c}{\textbf{Test}} & & \\
\cmidrule(lr){4-4} \cmidrule(lr){5-7}
\textbf{Candidate} & \textbf{Gen.} & \textbf{St.} & AUC & AUC & gAUC & Logloss
& \textbf{MFU (\%)} & \textbf{Score} \\
\midrule
Baseline (RankMixer)              & --- & --- & 0.9295 & 0.9358 & 0.7833 & 0.0744 & 1.433 & 0.793 \\
\midrule
Signed-bucket retokenizer         & 14 & D & \textbf{0.9301} & \textbf{0.9365} & 0.7822 & \textbf{0.0740} & 1.317 & 0.868 \\
Field-hash retokenizer            & 13 & D & \textbf{0.9301} & 0.9364 & 0.7815 & 0.0741 & 1.360 & 0.882 \\
Markov bucket exchange            & 11 & D & 0.9300 & 0.9362 & 0.7833 & 0.0745 & 1.465 & 0.912 \\
Field-covariance retokenizer      & 13 & D & 0.9300 & 0.9361 & 0.7831 & 0.0743 & 1.364 & 0.861 \\
Low-rank task-basis router        & 19 & S & 0.9300 & 0.9360 & 0.7844 & 0.0744 & 1.477 & 0.911 \\
Hadamard cross mixer              & 1  & S & 0.9300 & 0.9357 & 0.7825 & 0.0745 & 1.513 & \textbf{0.924} \\
Directional-decay exchange        & 12 & S & 0.9299 & 0.9360 & 0.7834 & 0.0744 & 1.494 & 0.906 \\
Lag-delta interest filter         & 7  & S & 0.9299 & 0.9359 & \textbf{0.7848} & 0.0744 & 1.472 & 0.888 \\
Field-moment retokenizer          & 26 & D & 0.9299 & 0.9360 & 0.7813 & 0.0741 & 1.206 & 0.763 \\
Orthogonal task-subspace fusion   & 22 & S & 0.9299 & 0.9360 & 0.7845 & 0.0744 & 1.488 & 0.891 \\
Per-token FFN/GELU hybrid         & 25 & D & 0.9298 & 0.9358 & 0.7828 & 0.0743 & 1.535 & 0.907 \\
Wider RankMixer block             & 21 & S & 0.9298 & 0.9362 & 0.7838 & 0.0743 & 1.506 & 0.893 \\
Bidirectional context exchange    & 5  & S & 0.9293 & 0.9356 & 0.7821 & 0.0743 & 1.574 & 0.819 \\
Per-token FFN/GELU hybrid         & 7  & D & 0.9290 & 0.9353 & 0.7828 & 0.0746 & \textbf{1.605} & 0.776 \\
\bottomrule
\end{tabular}
\end{table}

\begin{table}[!htbp]
\centering
\caption{Pareto front over all eligible candidates on validation AUC and MFU
(RankMixer backbone), using the eligibility criteria in Section~\ref{app:evolution_config}.
$\Delta$: difference from the baseline in $10^{-4}$ AUC. Rows run from highest AUC to
highest MFU. Best value per column in \textbf{bold}.}
\label{tab:pareto_mfu}
\scriptsize
\setlength{\tabcolsep}{3pt}
\begin{tabular}{@{}lc>{\raggedright\arraybackslash}p{4.4cm}ccccc@{}}
\toprule
& & & \multicolumn{2}{c}{\textbf{Val}} & \multicolumn{2}{c}{\textbf{Test}} & \\
\cmidrule(lr){4-5} \cmidrule(lr){6-7}
\textbf{Model} & \textbf{Gen.} & \textbf{Architectural change} & AUC & $\Delta$ & AUC & $\Delta$
& \textbf{MFU (\%)} \\
\midrule
Baseline (RankMixer) & --- & Original backbone
  & 0.929477 & --- & 0.935804 & --- & 1.433 \\
\midrule
P1 & 19 & Hadamard cross mixer $+$ low-rank task-basis router
  & \textbf{0.929986} & $\mathbf{+5.08}$ & \textbf{0.935978} & $\mathbf{+1.74}$ & 1.477 \\
P2 & 1  & Hadamard cross mixer
  & 0.929963 & $+4.86$ & 0.935729 & $-0.75$ & 1.513 \\
P3 & 3  & Hadamard cross mixer $+$ per-token FFN/GELU
  & 0.929632 & $+1.54$ & 0.935701 & $-1.03$ & 1.565 \\
P4 & 18 & Signed-transition smoother $+$ Hadamard $+$ channel-group mixer $+$ FFN/GELU
  & 0.929462 & $-0.16$ & 0.935401 & $-4.04$ & 1.567 \\
P5 & 5  & Bidirectional token context $+$ Hadamard $+$ FFN/GELU
  & 0.929273 & $-2.04$ & 0.935614 & $-1.90$ & \textbf{1.574} \\
\bottomrule
\end{tabular}
\end{table}

\begin{table}[!htbp]
\centering
\caption{Per-generation log on the RankMixer backbone. \textbf{Change}: the
generation's best-AUC candidate, with its outcome in \textbf{St.} \textbf{MFU}: the
generation's highest MFU, with that candidate's outcome as superscript.
\textbf{K/C}: skill-space/code-space split of the
six candidates (Section~\ref{sec:budget}). \textbf{S/D/F/U}: survived, discarded,
failed, duplicate. \textbf{P}: skills promoted to Tier-2. Best values in \textbf{bold}.}
\label{tab:mfu_rounds}
\scriptsize
\setlength{\tabcolsep}{3pt}
\begin{tabular}{@{}c>{\raggedright\arraybackslash}p{3.65cm}ccccccc@{}}
\toprule
& & & & \textbf{Val} & \textbf{Test} & & & \\
\cmidrule(lr){5-5} \cmidrule(lr){6-6}
\textbf{Gen.} & \textbf{Change} & \textbf{St.} & \textbf{K/C} & AUC & AUC
& \textbf{MFU (\%)} & \textbf{S/D/F/U} & \textbf{P} \\
\midrule
0  & token-mix $+$ per-token FFN            & S & 4/2 & 0.929776 & 0.935908 & 1.536$^{\mathrm{S}}$ & 2/2/2/0 & 0 \\
1  & Hadamard cross post-attention          & S & 4/2 & 0.929963 & 0.935729 & 1.519$^{\mathrm{S}}$ & 2/4/0/0 & 1 \\
2  & token-mix $+$ per-token FFN            & S & 4/2 & 0.929473 & 0.935687 & 1.534$^{\mathrm{S}}$ & 2/1/2/1 & 0 \\
3  & per-token FFN $+$ GELU                 & S & 2/4 & 0.929631 & 0.935701 & 1.565$^{\mathrm{S}}$ & 2/3/1/0 & 0 \\
4  & add FFN after RankMixer                & D & 2/4 & 0.929623 & 0.935608 & 1.564$^{\mathrm{D}}$ & 1/4/1/0 & 1 \\
5  & triangular token context before DIN    & S & 2/4 & 0.929625 & 0.935573 & 1.574$^{\mathrm{S}}$ & 2/4/0/0 & 2 \\
6  & consensus-disagreement fusion bridge   & D & 3/3 & 0.929662 & 0.935807 & 1.467$^{\mathrm{D}}$ & 0/6/0/0 & 0 \\
7  & lag-delta interest filter before DIN   & S & 2/4 & 0.929876 & 0.935922 & \textbf{1.605}$^{\mathrm{D}}$ & 2/3/1/0 & 1 \\
8  & multi-scale delta exchange before DIN  & S & 3/3 & 0.929747 & 0.935558 & 1.510$^{\mathrm{D}}$ & 2/4/0/0 & 2 \\
9  & spectral token filter before DIN       & S & 2/4 & 0.929713 & 0.935676 & 1.553$^{\mathrm{D}}$ & 2/3/1/0 & 1 \\
10 & field-moment retokenizer               & D & 2/4 & 0.929736 & 0.936431 & 1.542$^{\mathrm{D}}$ & 0/4/2/0 & 0 \\
11 & Markov bucket exchange before DIN      & D & 2/4 & 0.930017 & 0.936176 & 1.541$^{\mathrm{D}}$ & 2/4/0/0 & 1 \\
12 & directional-decay exchange             & S & 2/4 & 0.929915 & 0.936034 & 1.545$^{\mathrm{D}}$ & 2/3/1/0 & 2 \\
13 & field-hash retokenizer                 & D & 2/4 & 0.930114 & 0.936423 & 1.585$^{\mathrm{D}}$ & 2/3/1/0 & 0 \\
14 & signed-bucket retokenizer              & D & 2/4 & \textbf{0.930143} & \textbf{0.936516} & 1.483$^{\mathrm{S}}$ & 2/3/1/0 & 1 \\
15 & compact anchor reservoir retokenizer   & D & 2/4 & 0.929599 & 0.935658 & 1.528$^{\mathrm{D}}$ & 0/3/2/1 & 0 \\
16 & reuse multi-scale delta exchange       & S & 2/4 & 0.929785 & 0.936061 & 1.521$^{\mathrm{D}}$ & 2/4/0/0 & 1 \\
17 & linear EMA contrast                    & S & 2/4 & 0.929725 & 0.936000 & 1.540$^{\mathrm{S}}$ & 2/2/1/1 & 1 \\
18 & token-transition Laplacian             & D & 2/4 & 0.929845 & 0.935729 & 1.567$^{\mathrm{S}}$ & 2/4/0/0 & 2 \\
19 & low-rank task-basis router             & S & 2/4 & 0.929986 & 0.935977 & 1.577$^{\mathrm{D}}$ & 2/3/1/0 & 2 \\
20 & reuse directional-decay exchange       & D & 2/4 & 0.929838 & 0.936072 & 1.484$^{\mathrm{S}}$ & 1/5/0/0 & 1 \\
21 & wider RankMixer block                  & S & 3/3 & 0.929823 & 0.936163 & 1.545$^{\mathrm{S}}$ & 2/4/0/0 & 0 \\
22 & orthogonal task-subspace fusion        & S & 2/4 & 0.929855 & 0.936027 & 1.533$^{\mathrm{D}}$ & 2/3/1/0 & 1 \\
23 & per-token FFN $+$ GELU                 & S & 2/4 & 0.929397 & 0.935656 & 1.531$^{\mathrm{D}}$ & 1/1/3/1 & 0 \\
24 & rank-1 task-affinity fusion            & S & 2/4 & 0.929540 & 0.935561 & 1.504$^{\mathrm{D}}$ & 1/2/3/0 & 1 \\
25 & per-token FFN $+$ GELU                 & D & 3/3 & 0.929823 & 0.935760 & 1.535$^{\mathrm{D}}$ & 1/2/3/0 & 0 \\
26 & field-moment retokenizer               & D & 2/4 & 0.929863 & 0.935986 & 1.530$^{\mathrm{D}}$ & 2/3/1/0 & 1 \\
27 & causal token spline                    & D & 2/4 & 0.929593 & 0.935939 & 1.514$^{\mathrm{D}}$ & 2/3/0/1 & 1 \\
28 & reuse task-axis contextual router      & S & 2/4 & 0.929606 & 0.936007 & 1.471$^{\mathrm{S}}$ & 1/3/2/0 & 0 \\
29 & affine-free task-token router          & D & 2/4 & 0.929377 & 0.935614 & 1.533$^{\mathrm{S}}$ & 1/1/3/1 & 0 \\
\bottomrule
\end{tabular}
\end{table}

% Appendix section. Place after \appendix in the main file.
% Requires: \usepackage{booktabs,multirow,array}
\FloatBarrier

\section{Results with Different LLM Backbones}
\label{app:llm_backbones}

\begin{table}[!htbp]
\centering
\caption{Best candidate found by each LLM backbone. \emph{Best Round} is the evolution round in which the best-performing candidate first appeared. \emph{Operation} gives the operator type (Code: code-space generation; Skill: skill-space reuse; Template: template-level specialization) and the concrete architectural change it made. For MTL, the AUC is the mean over the two tasks. Datasets: Amazon-Books (CTR), Census-Income (MTL) and MovieLens (MDL); the runs start from DeepFM, Shared-Bottom and SAR-Net, respectively.}
\label{tab:app_llm_backbones}
\small
\setlength{\tabcolsep}{3.5pt}
\begin{tabular}{llccc >{\raggedright\arraybackslash}p{4.6cm}}
\toprule
Setting & Backbone & Val AUC & Test AUC & Best Round & Operation \\
\midrule
\multirow{4}{*}{CTR}
 & GLM-5.2         & 0.6937 & 0.6684 & 24 & Skill: add a reciprocal field-subspace router \\
 & DeepSeek-V4-Pro & 0.6952 & 0.6686 & 18 & Code: replace DCN-V2 cross fusion with sample-adaptive bilinear gated fusion (23 branches) \\
 & GPT-5.6 (SOL)   & 0.6968 & 0.6722 & 6  & Code: add a dual-anchor cross-lattice interaction module \\
 & Claude Opus 4.8 & \textbf{0.6972} & \textbf{0.6738} & 23 & Skill: add a complex Volterra resonance factor branch \\
\midrule
\multirow{4}{*}{MTL}
 & GLM-5.2         & \textbf{0.9718} & \textbf{0.9739} & 11 & Code: add hyperbolic (Poincar\'e) projection on field embeddings \\
 & DeepSeek-V4-Pro & 0.9714 & 0.9734 & 24 & Code: add dimension-wise 1D-conv gating on field embeddings (residual) \\
 & GPT-5.6 (SOL)   & 0.9714 & 0.9736 & 24$^\dagger$ & Code: add Low-rank Neural Flow  \\
 & Claude Opus 4.8 & 0.9714 & 0.9734 & 18 & Skill: add a bidirectional orthogonal task-exchange module (shared/private subspaces) \\
\midrule
\multirow{4}{*}{MDL}
 & GLM-5.2         & \textbf{0.7957} & \textbf{0.7935} & 19 & Code: add a low-rank relation mixer \\
 & DeepSeek-V4-Pro & 0.7954   & 0.7924 & 2  & Template: replace the domain module with PLE; shared experts $1\!\rightarrow\!2$, specific experts $2\!\rightarrow\!3$ \\
 & GPT-5.6 (SOL)   & 0.7954 & 0.7931 & 21 & Code: add a domain slot memory over fields \\
 & Claude Opus 4.8 & 0.7955 & 0.7934 & 5  & Code: add a domain-conditioned soft field selection \\
\bottomrule
\end{tabular}

\vspace{2pt}
\raggedright\footnotesize
$^\dagger$The round-1 candidate already reaches $0.97136$ validation AUC; round 24 adds only $6.6\times10^{-6}$.
\end{table}

We check whether the gains reported in the main text depend on the LLM that proposes mutations. To do so, we re-run the search in all three settings with four LLM backbones. Within each setting, all runs start from the same architecture and use the same operators and the same selection rule. Table~\ref{tab:app_llm_backbones} reports the results. All four backbones exceed the strongest manual model on CTR (Amazon-Books; Wide \& Deep, $0.6587$ test AUC) and on MTL (Census-Income; AITM, $0.9726$ mean test AUC). On MDL (MovieLens), three backbones exceed the strongest manual model (SAR-Net, $0.7924$), and DeepSeek-V4-Pro matches it. On MTL, three backbones reach the same validation mean AUC of $0.9714$. This suggests that the gains do not hinge on a single LLM. No backbone is best everywhere. Claude is best on CTR. GLM is best on MTL and MDL. On MTL it is the only backbone to pass the shared plateau, using a hyperbolic Poincar\'e projection of the field embeddings. On MDL, GLM, GPT and Claude each found a field-level, domain-conditioned module on their own: a low-rank relation mixer (GLM), a domain slot memory (GPT) and a soft field selection (Claude).

The backbones differ mostly in how they search. GPT converges early: it finds its best CTR model at round 6, and its best MTL result is already reached at round 1. Claude and GLM keep improving until late rounds (rounds 18--24), and most of these late gains come from reusing skills stored in the library rather than from new code. DeepSeek tends to build wide ensembles: its best CTR model fuses 23 branches. On MDL, however, it only swapped existing templates and tuned their hyperparameters (switching to PLE with more experts), and its test AUC only matches that of the strongest manual model. This is the weakest result among the twelve runs, and it suggests that generating new modules in code matters when strong manual architectures already exist. One limitation is shared by all backbones: selecting only on AUC can hurt calibration. On CTR, every best model has a higher test LogLoss ($0.533$--$0.554$) than all manual models on Amazon-Books ($0.495$--$0.507$). All results come from a single run per backbone, so differences between backbones below about $5\times10^{-4}$ AUC should be treated as ties.

\FloatBarrier

%% file: appendix/appendix_skill_inventory.tex
% Curated Tier-1 skill inventory (abbreviated for presentation).
% Category sizes refer to the full library. Show the first four identifiers
% from the supplied inventory, or all identifiers when fewer than four exist.

\begin{table*}[!htbp]
\centering
\caption{Curated Tier-1 skills by category. The Size column gives the full number
of skills in each category. Up to four identifiers are shown per category;
\(\ldots\) denotes omitted entries. Identifiers and category assignments follow
the stored skill manifests.}
\label{tab:app-tier1-inventory}

\footnotesize
\setlength{\tabcolsep}{3pt}
\renewcommand{\arraystretch}{1.08}
\begin{tabular}{@{}
  >{\raggedright\arraybackslash}p{0.15\linewidth}
  >{\centering\arraybackslash}p{0.04\linewidth}
  >{\raggedright\arraybackslash}p{0.22\linewidth}
  >{\raggedright\arraybackslash}p{\dimexpr0.59\linewidth-6\tabcolsep\relax}
@{}}
\toprule
\textbf{Category} & \textbf{Size} & \textbf{Role} & \textbf{Example skill identifiers} \\
\midrule
\texttt{embedding} & 4 &
Categorical and sequential feature embeddings. &
\skillid{field_aware_embedding_adapter},\allowbreak\space
\skillid{field_embedding},\allowbreak\space
\skillid{sequence_field_embedding},\allowbreak\space
\skillid{shared_sequence_embedding} \\

\addlinespace[4pt]
\texttt{interaction} & 11 &
Explicit feature interactions, attention, and gating. &
\skillid{afm_attention_pooling},\allowbreak\space
\skillid{autoint_attention},\allowbreak\space
\skillid{bilinear_interaction},\allowbreak\space
\skillid{cen_feature_gate},~\(\ldots\) \\

\addlinespace[4pt]
\texttt{tower} & 2 &
Dense representation transformations. &
\skillid{mlp_tower},\allowbreak\space
\skillid{shared_mlp_tower} \\

\addlinespace[4pt]
\texttt{sequence} & 19 &
Sequence encoding, interest modeling, and temporal features. &
\skillid{dien_interest_evolution},\allowbreak\space
\skillid{din_target_attention},\allowbreak\space
\skillid{exact_leakyrelu_transformer_encoder},\allowbreak\space
\skillid{full_augru_evolution},~\(\ldots\) \\

\addlinespace[4pt]
\texttt{matching} & 7 &
Candidate scoring, retrieval, and negative sampling. &
\skillid{all_item_scoring_adapter},\allowbreak\space
\skillid{candidate_dot_logits},\allowbreak\space
\skillid{dot_product_logits},\allowbreak\space
\skillid{in_batch_negative_head},~\(\ldots\) \\

\addlinespace[4pt]
\texttt{multitask} & 10 &
Shared experts, task routing, and task-specific prediction. &
\skillid{aitm_transfer},\allowbreak\space
\skillid{esmm_chain_head},\allowbreak\space
\skillid{exact_attention_transfer},\allowbreak\space
\skillid{mmoe_gate},~\(\ldots\) \\

\addlinespace[4pt]
\texttt{scenario} & 11 &
Domain conditioning, adaptation, and prediction. &
\skillid{adaptdhm_cluster_tower},\allowbreak\space
\skillid{adasparse_pruned_tower},\allowbreak\space
\skillid{domain_indicator_adapter},\allowbreak\space
\skillid{domain_select},~\(\ldots\) \\

\addlinespace[4pt]
\texttt{generative} & 8 &
Quantization, semantic IDs, and generative recommendation. &
\skillid{exact_residual_quantizer_stack},\allowbreak\space
\skillid{exact_t5_encoder_decoder},\allowbreak\space
\skillid{kmeans_sinkhorn_initialization},\allowbreak\space
\skillid{normalized_item_dot_head},~\(\ldots\) \\

\addlinespace[4pt]
\texttt{head} & 6 &
Prediction logits, probabilities, and language-model outputs. &
\skillid{binary_ctr_head},\allowbreak\space
\skillid{linear_logit},\allowbreak\space
\skillid{sasrec_pairwise_logits},\allowbreak\space
\skillid{sequence_lm_head},~\(\ldots\) \\

\addlinespace[4pt]
\texttt{loss} & 9 &
Training objectives and regularization. &
\skillid{bce_loss},\allowbreak\space
\skillid{covariance_regularizer},\allowbreak\space
\skillid{dien_auxiliary_interest_loss},\allowbreak\space
\skillid{hinge_loss},~\(\ldots\) \\

\addlinespace[4pt]
\texttt{utility} & 7 &
Tensor adaptation, normalization, and fusion. &
\skillid{additive_fusion},\allowbreak\space
\skillid{concat_fusion},\allowbreak\space
\skillid{dense_feature_path},\allowbreak\space
\skillid{field_sum},~\(\ldots\) \\
\midrule
\textbf{Total} & \textbf{94} & \multicolumn{2}{l}{11 categories; 42 identifiers shown} \\
\bottomrule
\end{tabular}
\end{table*}

%% file: appendix_baseline_implementation.tex
% Included from 7Appendix.tex under Training, Search, and Hardware Settings.

\subsection{Implementation of the Automated Search Baselines}
\label{app:baseline-implementation}

This section describes the NASRec weight-sharing search baseline and
the OpenEvolve code-evolution baseline evaluated on the six CTR, MTL,
and MDL benchmarks in Table~\ref{tab:rq1}. Our implementations adapt these
methods to the recommendation tasks considered here. The task-specific data
processing, model interfaces, search settings, and evaluation protocols
used in these experiments are detailed below. Both use standalone PyTorch models
and a shared candidate trainer, with the same preprocessing, data splits,
random seed, optimizer, and early-stopping settings. Their dataset-specific
training schedules match those in Table~\ref{tab:app_train_bench};
method-specific search procedures and budgets are detailed below. These
settings apply to the six accuracy benchmarks, separately from the QK-Video
co-evolution experiment. Each baseline candidate uses one GPU and random
seed 2022; independent workers can run across the eight-GPU node described
in Section~\ref{app:hardware_cost}.

\paragraph{Datasets, labels, and input features.}
Table~\ref{tab:baseline-data-config} records the baseline input and label
configurations. Both baselines use temporal 70\%/10\%/20\% splits for CTR,
the prepared Census-Income train/validation/test files, and seeded random
80\%/10\%/10\% splits for MDL, following the dataset protocols in
Section~\ref{app:training_config}. Amazon CTR uses the provided binary
\texttt{label} column. For Amazon MDL, Beauty, Clothing, and Health form
three domains, and ratings greater than 3 are positive. MovieLens CTR uses
ratings of at least 4 as positives; MovieLens MDL uses ratings of at least 3.
The latter groups age codes into $\{1,18\}$, $\{25\}$, and
$\{35,45,50,56\}$, corresponding to the age groups in
Section~\ref{app:datasets}. Categorical features are label-encoded with zero
reserved for missing or padding entries. Continuous features use a separate
dense-feature path; MovieLens MDL normalizes the age feature. Embedding
dimensions are 16 for CTR and MDL and 8 for MTL.

\begin{table}[!htbp]
\centering
\caption{Input configurations of the two automated baselines. CTR splits
are temporal, Census-Income uses prepared files, and MDL splits are seeded
random splits. Sparse and dense refer to categorical and continuous input
features, respectively.}
\label{tab:baseline-data-config}
\small
\setlength{\tabcolsep}{3pt}
\begin{tabular}{@{}>{\raggedright\arraybackslash}p{0.18\linewidth}r
                >{\raggedright\arraybackslash}p{0.29\linewidth}
                >{\raggedright\arraybackslash}p{0.28\linewidth}@{}}
\toprule
Dataset / task & Samples & Target & Input features \\
\midrule
MovieLens / CTR & 1,000,209 & rating $\geq 4$ & 7 sparse \\
Books / CTR & 19,651,355 & provided binary label & 2 sparse \\
Beauty / CTR & 1,712,558 & provided binary label & 2 sparse \\
Census / MTL & 299,285 & income and marital status & 32 sparse + 7 dense \\
Amazon / MDL & 823,534 & rating $>3$ & 2 sparse; 3 domains \\
MovieLens / MDL & 1,000,209 & rating $\geq 3$ & 6 sparse + 1 dense; 3 domains \\
\bottomrule
\end{tabular}
\end{table}

\paragraph{Candidate training and validation-based selection.}
Each candidate implements a PyTorch \texttt{nn.Module} and is trained with
Adam and binary cross-entropy with logits. Table~\ref{tab:baseline-training}
lists the shared baseline hyperparameters. The selection metric is validation
AUC for CTR, the arithmetic mean of the two validation AUCs for Census-Income,
and global validation AUC across domains for MDL. The corresponding test
metrics are evaluated after candidate selection and do not guide search.
For Census-Income, both baseline trainers average the two task losses with
equal weights; the \evoskillrec\ training description in
Section~\ref{app:training_config} uses their sum. Thus, the matching schedules
do not imply identical loss normalization. For MDL, global and per-domain
AUC are reported separately.

\begin{table}[!htbp]
\centering
\caption{Candidate-training settings shared by the NASRec and
OpenEvolve baselines. Epochs is the maximum number of epochs, and
patience is measured in epochs. These hyperparameters match
Table~\ref{tab:app_train_bench}; the Census loss reduction is described in
the text.}
\label{tab:baseline-training}
\small
\setlength{\tabcolsep}{4pt}
\begin{tabular}{@{}lrrrrr@{}}
\toprule
Dataset / task & Epochs & Batch & Learning rate & Weight decay & Patience \\
\midrule
MovieLens / CTR & 5 & 1,024 & $10^{-3}$ & $3\times10^{-4}$ & 3 \\
Books / CTR & 2 & 8,192 & $10^{-3}$ & $3\times10^{-4}$ & 2 \\
Beauty / CTR & 2 & 8,192 & $10^{-3}$ & $3\times10^{-4}$ & 2 \\
Census / MTL & 5 & 2,048 & $10^{-3}$ & $10^{-4}$ & 3 \\
Amazon / MDL & 1 & 4,096 & $10^{-3}$ & $10^{-5}$ & 3 \\
MovieLens / MDL & 1 & 4,096 & $10^{-3}$ & $10^{-5}$ & 3 \\
\bottomrule
\end{tabular}
\end{table}

\paragraph{Configured candidate budgets.}
Table~\ref{tab:baseline-candidate-budgets} gives the configured budgets.
Each baseline requests $B=250$ standalone candidate-training slots on
MovieLens CTR and MovieLens MDL, and $B=200$ on the other four benchmarks.
For OpenEvolve search, these correspond to 25 rounds with at most
$n_b=10$ or $8$ proposals per round, respectively. \evoskillrec\ uses 25
rounds with eight candidate slots per round on all six benchmarks
(Table~\ref{tab:app_evo}). These are configured budgets, not counts of
successfully trained models; invalid or failed proposals are not reported
as successful evaluations. NASRec search additionally trains a
weight-sharing supernet before standalone candidate training. Accordingly,
candidate budgets do not measure total search computation.

\begin{table}[!htbp]
\centering
\caption{Configured candidate budgets on the six accuracy benchmarks.
$B$ is the standalone training budget for each automated baseline.
$n_b=B/25$ is the OpenEvolve per-round proposal limit.
\evoskillrec\ uses $25\times8=200$ candidate slots on each benchmark.
Supernet training is additional to the NASRec budget $B$.}
\label{tab:baseline-candidate-budgets}
\small
\setlength{\tabcolsep}{5pt}
\begin{tabular}{@{}lrrr@{}}
\toprule
Dataset / task & $n_b$ & Each baseline: $B$ & \evoskillrec: slots \\
\midrule
MovieLens / CTR & 10 & 250 & 200 \\
Books / CTR & 8 & 200 & 200 \\
Beauty / CTR & 8 & 200 & 200 \\
Census / MTL & 8 & 200 & 200 \\
Amazon / MDL & 8 & 200 & 200 \\
MovieLens / MDL & 10 & 250 & 200 \\
\bottomrule
\end{tabular}
\end{table}

\paragraph{NASRec architecture search.}
The baseline samples a pool of $2B$ architectures from a fixed, task-specific
search space and trains a weight-sharing supernet for three epochs.
Architectures are sampled in a balanced manner during supernet training
and ranked by inherited validation AUC. The highest-ranked distinct
architectures are then instantiated as standalone models and trained
independently under Table~\ref{tab:baseline-training}, up to budget $B$.
This separates inherited-weight screening from the standalone evaluation
used to select the reported model. The search spaces are:
\begin{itemize}
    \item \textbf{CTR:} linear, FM, deep MLP, cross-network, AutoInt, and
    SENET-deep branches, with sum, gated, or concatenation-MLP fusion. The
    default configuration has embedding dimension 16, hidden widths
    $(256,128,64)$, dropout 0.2, three cross layers, and two attention heads.
    \item \textbf{MTL:} Shared-Bottom, MMoE, PLE, and task-specific towers,
    with variants using 4--8 experts and representation widths of 32--128.
    \item \textbf{MDL:} Shared-Bottom, STAR, MMoE, PLE, SAR-Net, and
    domain-specific towers, including 4--8-expert and wider STAR variants.
\end{itemize}

\paragraph{OpenEvolve code evolution.}
The proposal provider receives the current population, archive, recent
failures, and island-specific parents, and generates ordinary PyTorch model
modules. The entry point is
\path{tools/ordinary_model_llm_provider.py}. Each proposal must define
\texttt{build\_model(context)}, use the feature dimensions provided by
\texttt{context}, and return logits with the required task-specific shape.
Candidate code cannot import \evoskillrec\ genome classes or access the
filesystem or network. Before training, proposals undergo static safety,
import/instantiation, forward-shape, and output-contract checks. Valid
proposals use the same standalone trainer as the NASRec baseline.
The search maintains a population of 32, an archive of 64, four islands,
and two parents per proposal. One provider call per round returns at most
$n_b$ proposals. Invalid proposals and training failures are recorded;
they are not assigned test results.

\begin{table}[!htbp]
\centering
\caption{Method-specific search settings. $B$ and $n_b$ are defined in
Table~\ref{tab:baseline-candidate-budgets}; they do not denote completed
evaluation counts.}
\label{tab:baseline-search-config}
\small
\setlength{\tabcolsep}{3pt}
\begin{tabular}{@{}>{\raggedright\arraybackslash}p{0.23\linewidth}
                >{\raggedright\arraybackslash}p{0.32\linewidth}
                >{\raggedright\arraybackslash}p{0.38\linewidth}@{}}
\toprule
Method & Search and training budget & Search state \\
\midrule
NASRec & 3 supernet epochs; $2B$ screened architectures;
$B$ standalone training slots & Balanced sampling and inherited-AUC ranking \\
OpenEvolve & 25 provider calls; at most $n_b$ proposals per round;
$B$ configured training slots & Population 32; archive 64; 4 islands;
2 parents per proposal \\
\bottomrule
\end{tabular}
\end{table}

\paragraph{Shared settings and differences between methods.}
The two baselines share their data and candidate-training implementation,
and the optimizer settings, epoch limits, batch sizes, and validation
selection metrics align with those reported for \evoskillrec. The Census
loss reduction differs as specified above. Search spaces, candidate
representations, and proposal procedures remain method-specific; the
comparison therefore evaluates the configured search systems under the
reported protocols. Candidate records retain architecture and metric
information, elapsed time, and the failure stage where applicable, to
support inspection of individual evaluations.

%% file: appendix/appendix_skill_inventory_unirank.tex
% Monospaced identifiers with automatic line wrapping.
% Underscores can be entered directly.

\begin{table*}[h]
\centering
\setlength{\belowcaptionskip}{6pt}
\caption{Skill inventory for generative ranking (92 skills, 10 categories).
Model-specific skills are prefixed with their source model.}
\label{tab:app-skill-inventory-unirank}

\small
\setlength{\tabcolsep}{4pt}
\renewcommand{\arraystretch}{1.10}

\begin{tabular}{@{}
  >{\raggedright\arraybackslash}p{0.10\linewidth}
  >{\centering\arraybackslash}p{0.04\linewidth}
  >{\raggedright\arraybackslash}p{0.22\linewidth}
  >{\raggedright\arraybackslash}p{\dimexpr0.64\linewidth-6\tabcolsep\relax}
@{}}
\toprule
\textbf{Category} &
\textbf{\#} &
\textbf{Role} &
\textbf{Skill identifiers} \\
\midrule

\texttt{embedding} & 17 &
Feature embeddings and model-specific feature-to-token transformations. &
\skillid{est_tokenization},
\skillid{feature_embedding},
\skillid{feature_embedding_dict},
\skillid{hemix_tokenization},
\skillid{hiformer_tokenization},
\skillid{hyformer_tokenization},
\skillid{infnet_tokenization},
\skillid{longer_tokenization},
\skillid{mixformer_tokenization},
\skillid{onetrans_tokenization},
\skillid{pretrained_embedding},
\skillid{rankmixer_tokenization},
\skillid{tokenformer_tokenization},
\skillid{tokenmixer_tokenization},
\skillid{ultrahstu_tokenization},
\skillid{unimixer_tokenization},
\skillid{zenith_tokenization} \\
\hline

\addlinespace[3pt]
\texttt{interaction} & 25 &
Explicit feature interactions, token mixing, and token-wise feed-forward or expert layers. &
\skillid{bilinear_interaction},
\skillid{bilinear_interaction_v2},
\skillid{compressed_interaction_net},
\skillid{cross_interaction},
\skillid{cross_net},
\skillid{cross_net_mix},
\skillid{cross_net_v2},
\skillid{factorization_machine},
\skillid{hemix_hetero_mixer},
\skillid{hemix_token_mixing},
\skillid{holographic_interaction},
\skillid{inner_product_interaction},
\skillid{interaction_machine},
\skillid{mixformer_query_mixer},
\skillid{multi_head_token_mixing},
\skillid{per_token_ffn},
\skillid{per_token_swiglu},
\skillid{rankmixer_block},
\skillid{sparse_moe},
\skillid{swiglu},
\skillid{tokenformer_interaction_blocks},
\skillid{tokenmixer_blocks},
\skillid{unimixer_lite_blocks},
\skillid{unimixer_lite_layer},
\skillid{unimixer_siamese_ffn} \\
\hline

\addlinespace[3pt]
\texttt{attention} & 16 &
Self-, cross-, and target-aware attention, sequence transduction, and channel recalibration. &
\skillid{din_attention},
\skillid{est_cross_attention},
\skillid{hemix_hetero_attention},
\skillid{hiformer_attention},
\skillid{longer_cross_causal_attention},
\skillid{longer_self_causal_attention},
\skillid{mixformer_cross_attention},
\skillid{multi_head_target_attention},
\skillid{onetrans_mixed_mha},
\skillid{scaled_dot_product_attention},
\skillid{squeeze_excitation},
\skillid{tokenformer_gated_mha},
\skillid{transformer_encoder},
\skillid{ultrahstu_interaction_blocks},
\skillid{ultrahstu_transduction_unit},
\skillid{zenith_self_attention} \\
\hline

\addlinespace[3pt]
\texttt{tower} & 3 &
MLP towers and mixed feed-forward transformations. &
\skillid{est_mixed_ffn},
\skillid{mlp_block},
\skillid{onetrans_mixed_ffn} \\
\hline

\addlinespace[3pt]
\texttt{block} & 10 &
Composite model-specific backbone blocks. &
\skillid{est_block},
\skillid{hiformer_block},
\skillid{hyformer_block},
\skillid{infnet_block},
\skillid{longer_block},
\skillid{longer_inner_trans},
\skillid{mixformer_block},
\skillid{onetrans_block},
\skillid{sasrec_block},
\skillid{zenith_block} \\
\hline

\addlinespace[3pt]
\texttt{pooling} & 3 &
Masked sum/average aggregation and top-$k$ pooling. &
\skillid{kmax_pooling},
\skillid{masked_average_pooling},
\skillid{masked_sum_pooling} \\
\hline

\addlinespace[3pt]
\texttt{activation} & 2 &
Adaptive and smooth nonlinear activations. &
\skillid{dice},
\skillid{gelu} \\
\hline

\addlinespace[3pt]
\texttt{head} & 2 &
Linear logits and per-task output activations. &
\skillid{logistic_regression},
\skillid{output_activation_list} \\
\hline

\addlinespace[3pt]
\texttt{primitive} & 5 &
Basic PyTorch layers and containers. &
\skillid{dropout},
\skillid{embedding},
\skillid{identity},
\skillid{linear},
\skillid{sequential} \\
\hline

\addlinespace[3pt]
\texttt{utility} & 9 &
Query processing, sequence representation, aggregation, broadcast, and fusion. &
\skillid{hyformer_query_boosting},
\skillid{hyformer_query_decoding},
\skillid{hyformer_query_generation},
\skillid{hyformer_sequence_representation},
\skillid{infnet_aggregation},
\skillid{infnet_broadcast},
\skillid{infnet_broadcast_gated_unit},
\skillid{infnet_hub_aggregator},
\skillid{mixformer_output_fusion},
 \\

\midrule
\textbf{Total} & \textbf{92} &
\multicolumn{2}{l}{10 categories} \\
\bottomrule
\end{tabular}
\end{table*}